\documentclass[twocolumn,final,aps,prl,showpacs,superscriptaddress,amsmath,amssymb,amsfonts,floatfix]{revtex4-1}
\usepackage{graphicx}
\usepackage{multirow}
\usepackage{epsfig}

\def\be{\begin{equation}}
\def\ee{\end{equation}}

\usepackage{url}
\usepackage{color}
\usepackage[colorlinks,breaklinks,bookmarks=true,citecolor=blue,linkcolor=red,urlcolor=blue]{hyperref}
\definecolor{darkred}{rgb}{0.7,0.0,0}

\begin{document}

\title{Pitfalls and solutions for perovskite transparent conductors}

\author{Liang Si}
\affiliation{Key Laboratory of Magnetic Materials and Devices \& Zhejiang Province Key Laboratory of Magnetic Materials and Application Technology, Ningbo Institute of Materials Technology and Engineering (NIMTE), Chinese Academy of Sciences, Ningbo 315201, China}
\affiliation{Institute for Solid State Physics, Vienna University of Technology, 1040 Vienna, Austria}

\author{Josef Kaufmann}
\affiliation{Institute for Solid State Physics, Vienna University of Technology, 1040 Vienna, Austria}

\author{Jan M.\ Tomczak}
\affiliation{Institute for Solid State Physics, Vienna University of Technology, 1040 Vienna, Austria}

\author{Zhicheng Zhong}
\email{zhong@nimte.ac.cn}
\affiliation{Key Laboratory of Magnetic Materials and Devices \& Zhejiang Province Key Laboratory of Magnetic Materials and Application Technology, Ningbo Institute of Materials Technology and Engineering (NIMTE), Chinese Academy of Sciences, Ningbo 315201, China}

\author{Karsten Held}
\email{held@ifp.tuwien.ac.at}
\affiliation{Institute for Solid State Physics, Vienna University of Technology, 1040 Vienna, Austria}

\date{\today}

\begin{abstract}
Transparent conductors---nearly an oxymoron---are in pressing demand, as ultra-thin-film technologies become ubiquitous commodities.
As current solutions rely on non-abundant elements, perovskites such as SrVO$_3$ and SrNbO$_3$ have been suggested as next generation transparent conductors.
Our \emph{ab-initio} calculations and analytical insights show, however, that reducing the plasma frequency below the visible spectrum by strong electronic correlations---a recently proposed	 strategy---unavoidably comes at a price: an enhanced scattering and thus a substantial optical absorption above the plasma edge.
As a way out of this dilemma we identify several perovskite transparent conductors, relying on hole doping, somewhat larger bandwidths and separations to other bands.
\end{abstract}

\maketitle

Transparent conductors are highly sought after due to rapidly growing application as displays, touch screens, photovoltaics, smart windows and solid-state lighting technology \cite{Ginley2010}.
The current way of producing transparent conductors is doping carriers into transparent semiconductors, typically oxides of  the  post-transition metals Zn, Cd, In and Sn.
Electron doping makes these transparent insulators conductive, whereby strongly delocalized vacant $s$-orbitals form the conduction band with a large bandwidth and a small  mass enhancement ($m^*/m_b$). This class of materials is referred to as transparent conducting oxides (TCOs) \cite{ginley2000transparent,Ginley2010} \footnote{Amporphous graphene has also been proposed as a transparent conductor \cite{PhysRevB.84.205414}}.

However, the synthesis of thin films exhibiting both, an excellent electrical conductivity and optical transparency in the visible spectrum, is challenging.
Among all TCOs, tin-doped indium oxide (ITO) exhibits the best balance between optical transparency and electrical conductivity, leading to its wide application.
However, surging prices for indium and integration of the transparent conductors in ultra-thin-film  fabrication call for the next generation of transparent conductors.

Making a conductor transparent feels like squaring the circle.
Conventional metals such as Ag or Cu have a high carrier concentration and their small effective mass greatly enhances their conductivity.
However, this also shifts their plasma edge  $\omega_p$ above the visible spectrum, making them transparent only in the ultraviolet.
This opaqueness can be mitigated by reducing the film thickness. But when the thickness approaches nm's, which is below the electronic mean-free path, the scattering time $\tau$ is enhanced and the conductivity reduced.

Alternatively, one can push the plasma frequency $\omega_{p}$=$e$$\sqrt{ 4\pi/\epsilon_{core}}$$\sqrt{n/m^*}$ below the visible spectrum.
Here, $\epsilon_{core}$ is the core dielectric constant, $e$ the elemental charge, $n$ the charge carrier density and $m^*$ the effective mass of the electrons.
Conventional doped TCOs largely rely on a small $n$ to achieve this objective.

Instead, one may also increase $m^*$ by electronic correlations.
This second route has been proposed in Ref.\,\cite{zhang2016correlated}, where SrVO$_3$ and CaVO$_3$ films have been identified as good candidates for transparent correlated conductors. Later SrNbO$_3$ films have been suggested \cite{Park2020} as well, which have a larger inter-band optical gap and are hence more transparent in the ultraviolet.

In this Letter, we show on the basis of simple physical relations that such a many-body enhancement comes at a price: enhancing $m^*$ not only reduces the plasma frequency $\omega_p$ but at the same time inevitably enhances the electron-electron scattering.
As a consequence the optical reflectivity and absorption above the plasma frequency are finite, the conductor opaque. 
Our density-functional theory (DFT) \cite{PhysRev.136.B864} and dynamical mean-field theory (DMFT) \cite{RevModPhys.68.13,kotliar2004strongly,PhysRevLett.62.324,held2007electronic} calculations  for perovskites and double perovskites confirm this. Previous, pioneering DFT+DMFT calculations \cite{Paul2019,Park2020,PhysRevResearch.2.033156} for transparent conductors focused on the quasi-particle weight $Z=m_b/m^*$ and the one-particle spectrum, but did not calculate optical properties. Hence this inherent pitfall went unnoticed.
To overcome these difficulties we propose an alternative route to transparent conductors: transition metal oxides (TMOs) with only a mild quasiparticle renormalization of $m^*$, but employing doping and using 4$d$ and 5$d$ TMOs instead of 3$d$ TMOs.


\begin{figure*}[tb]
\begin{minipage}{0.69\linewidth}
\includegraphics[width=1.00\columnwidth]{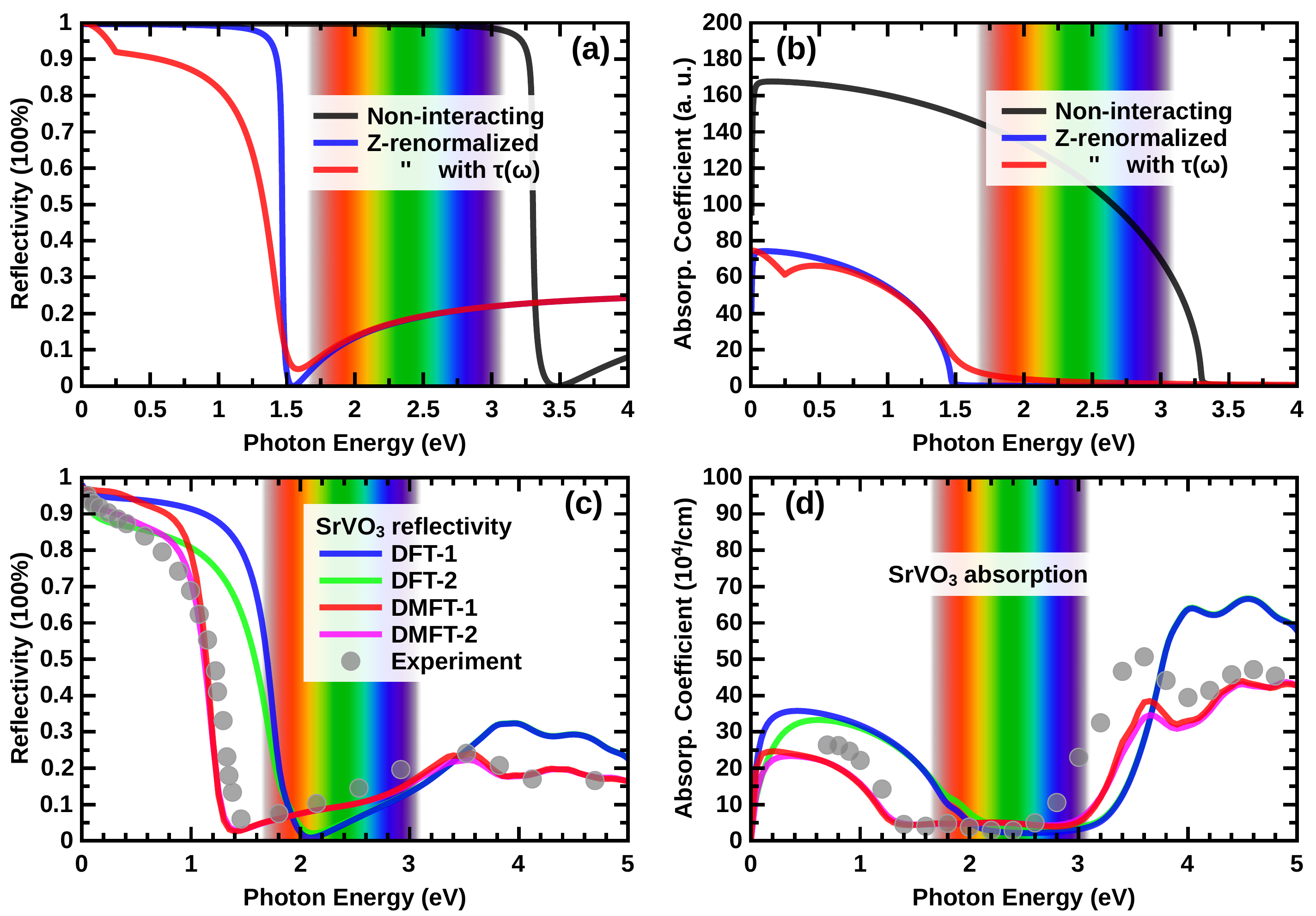}
\end{minipage}\hfill
\begin{minipage}{5cm}
\caption{(a,b) Reflectivity ($R$) and absorption coefficient ($A$) vs.\ frequency (photon energy) with and without quasiparticle renormalization $Z$ and scattering rate $\tau(\omega)^{-1}$ in the Drude model. Only considering $Z$ shifts the plasma frequency $\omega_p$ below the visible range  (rainbow) and $R=0$ at $\omega_p$ and $A\approx 0$ above. The renormalization $Z$ however also leads to an enhanced $\tau(\omega)^{-1}$, increasing $R$ and $A$. (c,d) $R$, $A$ in DFT and DFT+DMFT for SrVO$_3$. DMFT includes $Z$ and $\tau(\omega)$, and agrees with experiments. The experimental $R$ is from Ref.\,\cite{PhysRevB.58.4384}; the experimental $A$ is computed from the dielectric functions $\epsilon$ from Ref.\,\cite{zhang2016correlated}. For the DFT+DMFT optical conductivity $\sigma(\omega)$ see Supplemental Materials (SM) Section S.~6.\\ }
\label{Fig1}
\end{minipage}
\end{figure*}

\emph{Computational Details.}
The DFT-level optical properties of (undoped and doped) Sr$B$O$_3$ and Sr$_2$$B$$B'$O$_6$ are calculated by \textsc{wien2k} \cite{blaha2001wien2k,wien2k2020} within the PBE version of  the generalized gradient approximation \cite{PhysRevLett.77.3865} using the mBJ potential \cite{PhysRevLett.102.226401}. A dense $k$-mesh of 13$\times$13$\times$13 is used to guarantee convergence. Electron and hole doping is achieved in  the virtual crystal approximation (VCA) \cite{PhysRevB.61.7877} implemented in \textsc{wien2k}. All  lattice constants used  are obtained by performing structural relaxations. 

For the DMFT calculations, we firstly project the corresponding \textsc{wien2k} $t_{2g}$ band structure \footnote{For Sr$_2$TiNbO$_6$, the Ti-$e_g$ bands are also projected onto the Wannier orbitals because they are energetically comparable to the Nb-$t_{2g}$ bands. However, DMFT predicts the Ti-$e_g$ orbitals to be unoccupied.} to Wannier functions \cite{PhysRev.52.191,RevModPhys.84.1419} using \textsc{wien2wannier} \cite{mostofi2008wannier90,kunevs2010wien2wannier} and supplement it by local density-density interactions with standard values of the intra-orbital $U$=5.5\,eV, inter-orbital $U'$=3.5\,eV and Hund's exchange $J$=1.0\,eV for 3$d$ TMOs; 3.0\,eV, 2.4\,eV, and 0.3\,eV for 4$d$ TMOs; and 2.0\,eV, 1.4\,eV and 0.3\,eV for 5$d$ TMOs.
These are consistent with previous studies \cite{PhysRevB.73.155112,PhysRevLett.119.026402,PhysRevB.89.195121}. The resulting Hamiltonian is then solved at room temperature (300\,K) using continuous-time quantum Monte Carlo simulations in the hybridization expansions \cite{RevModPhys.83.349} through \textsc{w2dynamics} \cite{PhysRevB.86.155158,wallerberger2019w2dynamics}, and the maximum entropy method \cite{PhysRevB.44.6011,PhysRevB.57.10287} for an analytic continuation of the spectra.
The optical conductivity  $\sigma$($\omega$) is calculated from the DMFT self-energy $\Sigma$($\omega$) and the \textsc{wien2k}-calculated dipole matrix elements \cite{AmbroschDraxl20061} using  \textsc{woptic} \cite{assmann2016woptic}.

\emph{Pitfalls when reducing $\omega_p$ by correlations.}
A first insight is already possible on the basis of simple analytical considerations. The optical conductivity in the Drude model is given by
\begin{equation}
\label{Eq:sigma}
\sigma(\omega)= \frac{\sigma_0}{1- i\omega\tau},
\end{equation}
where $\sigma_0 =  \frac{n e^2 \tau}{m^*}$ already takes into account the mass renormalization
(of a free electron dispersion); $\tau$ is the scattering time and $n$ the electron density. 

This optical conductivity is directly connected to the dielectric function  
$\epsilon(\omega)= \epsilon_{core} + 4\pi i \sigma(\omega)/\omega$, 
where  $\epsilon_{core}$ denotes the dielectric contribution of the core electrons.
If we  assume  a constant $\tau$ we have $\omega \tau \gg 1$ 
at large frequencies typically including the visible range. 
Then, we can neglect the ``1'' in the denominator of Eq.~(\ref{Eq:sigma});
$\epsilon(\omega)$ becomes (approximately) purely real with a $1/\omega^2$ 
decay and a change of sign from negative to positive at the plasma frequency:
\begin{eqnarray}
\begin{split}
\epsilon(\omega) \approx  \epsilon_{core} \big(1-  \frac{\omega_{p}^{2}}{\omega^2} \big) & \mbox{\;\;\; with \; \; }
  \omega_p=\sqrt{\frac{ 4\pi}{\epsilon_{core}}\frac{n e^ 2}{m^*}}.
\label{Eq:omegaP}
\end{split}
\end{eqnarray}
From the real and imaginary part ($\tilde{n}$ and $k$) of 
$\sqrt{\epsilon} =\tilde{n}+ik$  we obtain the (normal incident) reflectivity $R =  | 1-\sqrt{\epsilon}| ^2/| 1+\sqrt{\epsilon}|^2$ and absorption $A={2 \omega k}/{c}$ \cite{ashcroft1976solid}.
This shows that, if Eq.~(\ref{Eq:omegaP}) holds, there is no reflectivity at and no absorption above $\omega_p$; cf.\ Supplemental Material (SM) Section S.\,1 for further details. The idea of Refs.\,\cite{zhang2016correlated,Paul2019,Park2020,PhysRevResearch.2.033156} is to reduce $\omega_p$ by a correlation-induced mass enhancement $m^*/m_b=Z^{-1}$ which is exemplified by the blue and black curves in Fig.~\ref{Fig1}(a,b) which differ by a quasiparticle renormalization factor $Z^{-1}=5.06$. Further parameters:  $\tau_0$=5.0$\times$10$^{-13}$s, $m_b=10 m_e$=9.11$\times$10$^{-27}$g,  $n$=2.0$\times$10$^{22}$ cm$^{-3}$.

The pitfall of this idea is that the mass-enhancement comes at a price: electronic correlations increase the scattering rate  $\tau^{-1}$ as well, in particular at larger frequencies, including the visible range of the spectrum for TMOs. This is unavoidable since the real part of the self-energy, Re$\Sigma (\omega) = -\gamma \omega$, where $\gamma=Z^{-1}-1$ is directly connected to the mass enhancement, necessarily leads through the Kramers-Kronig relation to an imaginary part with the same prefactor $\gamma$:
\begin{equation}
\Sigma (\omega) = -\gamma \omega - i \frac{3}{4} \frac{\pi}{2} \frac{\gamma}{\omega^*} [ (\pi T)^2 + \omega^2 ] -\frac{i}{2} \tau_0^{-1}
\label{Eq:Sigma}
\end{equation}
which holds for $\omega < {\omega^*}\approx \frac{U}{2 \sqrt{\gamma}}$. See SM Section S.\,2.1 for a derivation using the Kramers-Kronig relation and the behavior at large frequencies;  $T$ is the temperature and $\tau_0^{-1}$ an additional (impurity) scattering rate. The imaginary part of the self-energy in turn is related to the scattering rate:
\begin{eqnarray}
\!\! \!\! \tau(\omega)^{-1}\! =\!  -2  Z {\rm Im}\Sigma(\omega)\! =\! \frac{3 \pi Z}{4}  \frac{\gamma}{\omega^*} [(\pi T)^2\! + \!\omega^2 ]\! -\!Z \tau_0^{-1}\!.
\label{Eq4}
\end{eqnarray}
This enhancement due to electron-electron scattering and its frequency dependence must be taken into account in Eq.~(\ref{Eq:sigma}). Then the assumption $\omega \tau(\omega) \gg 1$  leading to  Eq.~(\ref{Eq:omegaP}) cannot be made; there is a substantial  $\tau^{-1}(\omega)$ and hence ${\rm Im }\, \epsilon(\omega)$ when ${\rm Re }\, \epsilon(\omega)$  changes sign. Consequently the reflectivity at and the absorption above the plasma edge is not zero but remains substantial, see red curves in Fig.~\ref{Fig1}(a,b) which are for $Z$$\approx$0.2, $U$=2.4\,eV, $T$=300\,K, $\tau_0^{-1}$=0.

Note that $\omega \tau(\omega)$  even decreases as  $\tau(\omega)\sim 1/\omega^2$ in Eq.~(\ref{Eq4}). However, this decrease and Eq.~(\ref{Eq4}) only hold in the low frequency regime, up to $\omega^*$. Eventually, $\tau(\omega)$ will increase again, but it will remain finite and substantial up to the point where excitations to the Hubbard bands become possible, leading to additional absorption beyond the quasiparticle model. To mimic the fact that $\tau^{-1}(\omega)$ is not growing indefinitely but is nonetheless still substantial in the relevant frequency regime, we have used a simple cut-off 
$\tau(\omega)^{-1} \rightarrow \min (\tau(\omega)^{-1},\tau_{\rm cut-off}^{-1})$ in  Fig.~\ref{Fig1}(a,b) with  $\tau_{\rm cut-off}^{-1}$=4.83$\times$10$^{13}\,$s$^{-1}$ (which corresponds to Im$\Sigma(\omega)_{\rm cut-off}$=$-$0.5\,eV). A more elaborate modeling as well as numerical renormalization group (NRG) data presented in the SM Section S.\,2.2 justify this simple cut-off picture. The cut-off also leads to a kink at the cut-off frequency of $\sim$ 0.25\,eV in Fig.~\ref{Fig1}(a,b). A feature that is also present in the better modeling and NRG data because $\tau(\omega)$ changes abruptly around $\omega^*$, see SM Section S.\,2.2.
This feature is visible as an additional shoulder also for SrVO$_3$ in Fig.~\ref{Fig1} (c) discussed below.

While the absorption in Fig.~\ref{Fig1}(b) remains low in the visible spectrum when including electronic correlations,  it has still increased by several orders of magnitude. Also the minimum of the reflectivity is increased considerably. Both enter in the figure of merit (FOM) for a transparent conductor given by the ratio of 10th power of the transmittance $T$ to sheet resistance $R_S$ \cite{Haake1976}: \begin{equation}
 \Phi_{TC}=T^{10}/R_S=\sigma(0)t\{(1-R)^2[e^{A t}-R^2e^{-A t}]^{-1}\}^{10}.
 \label{Merit}
\end{equation}
As customary \cite{zhang2016correlated,Haake1976}, we compute the FOM from $\sigma(0)$, $R$ and $A$ at $\omega \sim$2.25\,eV (550\,nm: the wavelength the human eye is most sensitive to).

\begin{figure*}[tb]
\begin{minipage}{0.78\linewidth}
\includegraphics[width=1.00\columnwidth]{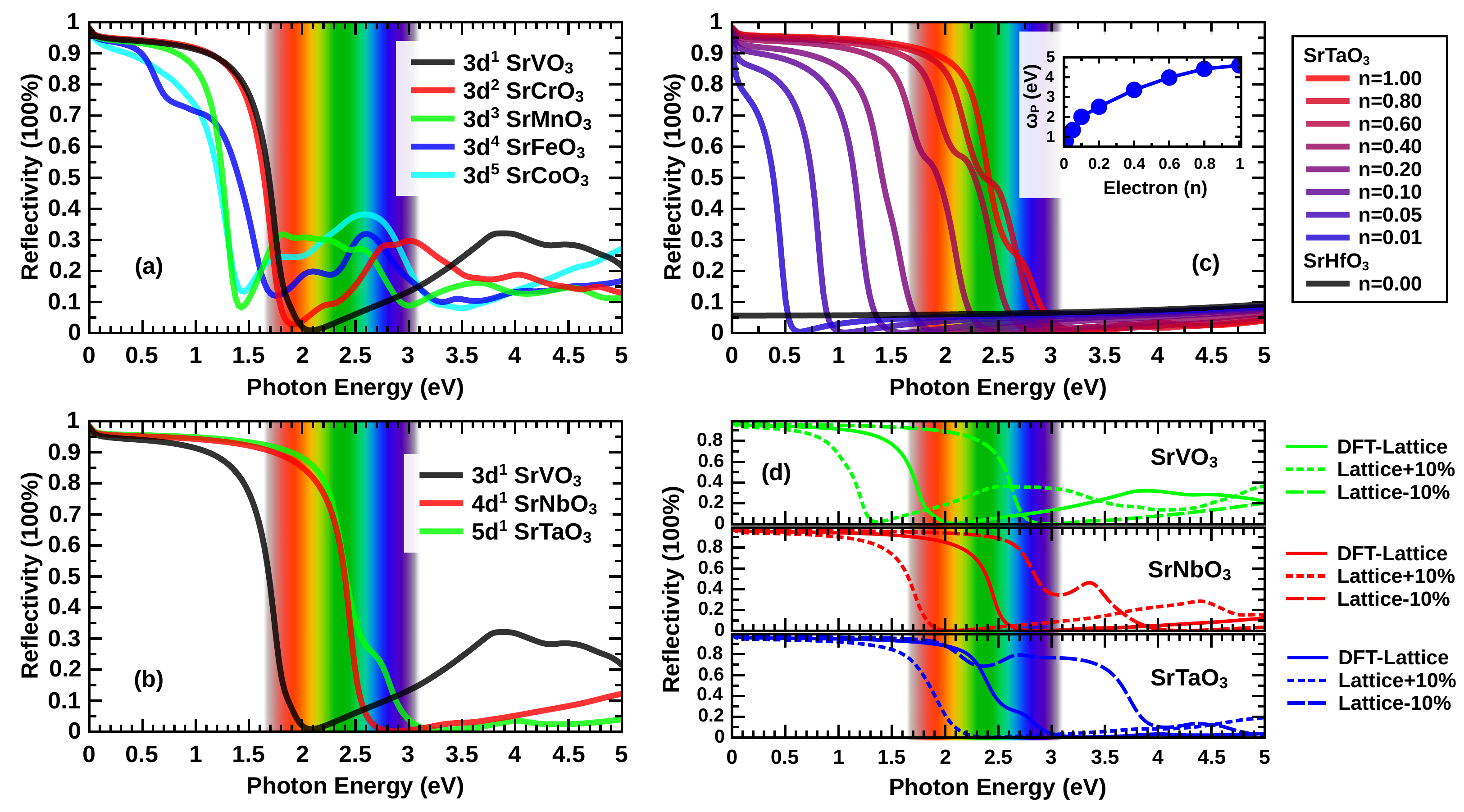}
\end{minipage}\hfill
\begin{minipage}{3.5cm}
\caption{DFT optical reflectivity of Sr$B$O$_3$ with (a) $B$=V, Cr, Mn, Fe, Co, and (b) $B$=V, Nb, Ta; (c) hole doping from 5$d^1$ SrTaO$_3$ to 5$d^0$  SrHfO$_3$; (d)  SrVO$_3$, SrNbO$_3$, and SrTaO$_3$ with DFT-relaxed lattice, lattice+10\% and lattice-10\%. Inset of (c): doping dependence of the plasma frequencies. For the absorption of doped SrTaO$_3$ and optical properties of SrVO$_3$ and SrNbO$_3$ see SM Section S.\,4. \\ }
\label{Fig2}
\end{minipage}
\end{figure*}

{\em DFT+DMFT for SrVO$_3$.} The above considerations are instructive and disclose the general difficulties to exploit electronic correlations for getting better transparent conductors. However, some aspects are not included in the simple analysis above: Hubbard bands emerge from the real part of the self-energy at larger frequencies; the free electron model considered above does not consider band edges nor multi-band effect. All these effects are now taken into account in our numerical materials calculation for SrVO$_3$ where DMFT yields not only the quasiparticle renormalizations of the  $t_{2g}$-orbitals, but also frequency-dependent scattering rates and Hubbard bands.  Our calculated $Z$$\sim$0.4 ($m^*/m_b$=2.5) is consistent with previous theoretical predictions \cite{PhysRevLett.92.176403,huang2012dynamical} and the experimental value \cite{PhysRevB.58.4384}; as is the  DMFT self-energy $\Sigma(\omega)$ for real frequencies, see SM Section S.\,3 and,  e.g.,  Ref.\,\cite{PhysRevB.73.155112}. 

Fig.~\ref{Fig1}(c,d) shows the DFT, DMFT and experimental \cite{PhysRevB.58.4384} optical reflectivity $R$ and absorption $A$ of SrVO$_3$; for the optical conductivity $\sigma(\omega)$ see SM Section S.\,3. 
DMFT describes all three quantities reasonably, especially for the energies covering the visible range (1\,eV$ \textless\omega \textless$4\,eV). Here, we have taken into account an additional impurity scattering which for DFT-1 and DFT-2
is $\tau_0^{-1}=$0.10\,eV and 0.27\,eV; 
for  DMFT-1 and DMFT-2 it is  $\tau_0^{-1}=0$\,eV and 0.13\,eV, see SM Section S.\,3 for further details. While this scattering rate is essential for DFT,
in DMFT it only slightly improves the description of the low-frequency behavior of the conductivity. For larger frequencies ($\omega$$\gtrsim$1.0\,eV) which are relevant for the transparency, the electron-electron scattering of DMFT dominates anyhow.
Indeed, already for frequencies $\omega\sim 0.2\;$eV within the Drude peak the electron-electron scattering is substantial. A proper description of the frequency behavior of the reflectivity and absorption hence requires the DMFT frequency-dependent electron-electron scattering.  In the following we will use DMFT  results without additional impurity scattering  (aka.~DMFT-1) unless specified otherwise.

The reflectivity in Fig.~\ref{Fig1}(c) shows that the DFT plasma edge ($\omega_p^{\rm DFT}\sim3.8$\,eV) is reduced by electronic correlations in DMFT ($\omega_p^{\rm DMFT}\sim2.5$\,eV), moving the minimum of $R$  ($R_{min}$) below the visible range. This agrees well with experiments  \cite{zhang2016correlated}, on the basis of which SrVO$_3$ has been suggested as a good material for transparent conductors. But we see at the same time that the reflectivity at the plasma edge is finite, not zero, again in agreement with experiment \cite{PhysRevB.58.4384}.
The same electronic correlations that reduce $\omega_p$ also lead to an enhanced scattering at finite frequencies, just as we have pointed out  in our analytical considerations above. 

The same holds for the absorption coefficient in Fig.~\ref{Fig1}(d)  which is finite in the optical range with good agreement between DMFT and experiment \footnote{Here, the experimental absorption coefficient is calculated from the measured dielectric functions of Ref.\,\cite{zhang2016correlated} via $A$=2$\omega k$/c.}.  Note the DFT absorption is also not zero because of inter-band transitions.

\emph{Trends for TMO transparent conductors.} Given the discussed difficulties to exploit correlation-induced mass enhancements for making TMOs good transparent conductors, let us now try to identify the optimal TMO. We do so by analyzing first some general trends in DFT.

In Fig.~\ref{Fig2}(a), we change the 3$d$ transition metal in  Sr$B$O$_3$ moving to the right in the periodic table from $B$=V to  Cr, Mn, Fe, and Co. Clearly, the  3$d^1$ configuration in SrVO$_3$ gives the lowest DFT reflectivity, indicating that late  3$d$ TMOs are less suitable as transparent conductors. 
Second, in Fig.~\ref{Fig2}(b), we move down in the periodic table from 3$d^1$ (SrVO$_3$) to  4$d^1$ (SrNbO$_3$) and 5$d^1$ (SrTaO$_3$). The DFT results indicate that SrNbO$_3$---and even more so SrTaO$_3$---have a wider range of low reflectivity but also require a larger shift of the plasma edge for it to be below the visible spectrum. The reason for this is that on the one hand there is a larger gap from the $t_{2g}$ bands around the Fermi level to the oxygen $p$ bands below and the $e_g$ states above. This reduces interband transitions above $\omega_p$ (see SM Section S.\,4). On the other hand, we have a wider bandwidth (smaller $m_b$) and thus a larger $\omega_p$.

In Fig.~\ref{Fig2}(c) we instead consider hole doping of SrTaO$_3$, reducing the number $n$ of electrons per Ta site. Also shown is the limiting case of SrHfO$_3$  with  $n=0$ (5$d^0$), and in the inset of Fig.~\ref{Fig2}(c) the relationship between electron filling and $\omega_p$. We see that through hole doping, we can move the range of low reflectivity into the visible spectrum. Hence our DFT calculations indicate that hole-doped SrNbO$_3$ and SrTaO$_3$ or electron doped SrZrO$_3$ and SrHfO$_3$ might be good candidates for transparent conductors (see SM Section S.\,4).

In  Fig.~\ref{Fig2}(d), we consider an additional parameter for optimizing the transparency: lattice engineering. The results indicates that also an expanding lattice (+10\%) can be employed to lower $\omega_p$, as it reduces the bandwidth [enhances $m_b$ (and hence $m^*$)  in Eq.~(\ref{Eq:omegaP})] and leads to a lower charge density $n$. 

\emph{Materials proposal.}
From our insight above 4$d^1$ and 5$d^1$ TMOs have lower DFT reflectivities and also absorptions (Fig.~\ref{Fig2} and SM Section S.\,4) in a large frequency range. To shift the plasma edge below the visible spectrum, a combination of a mild quasiparticle renormalization and hole doping appears most promising. In fact, the weaker interaction of the more extended 4$d$/5$d$ orbitals and the weaker correlations of the hole-doped system automatically reduce the renormalization and electron-electron scattering. All we need is hole doping  4$d^1$ and/or 5$d^1$ TMOs.

\begin{figure*}[tb]
\begin{minipage}{0.65\linewidth}
\includegraphics[width=1.00\columnwidth]{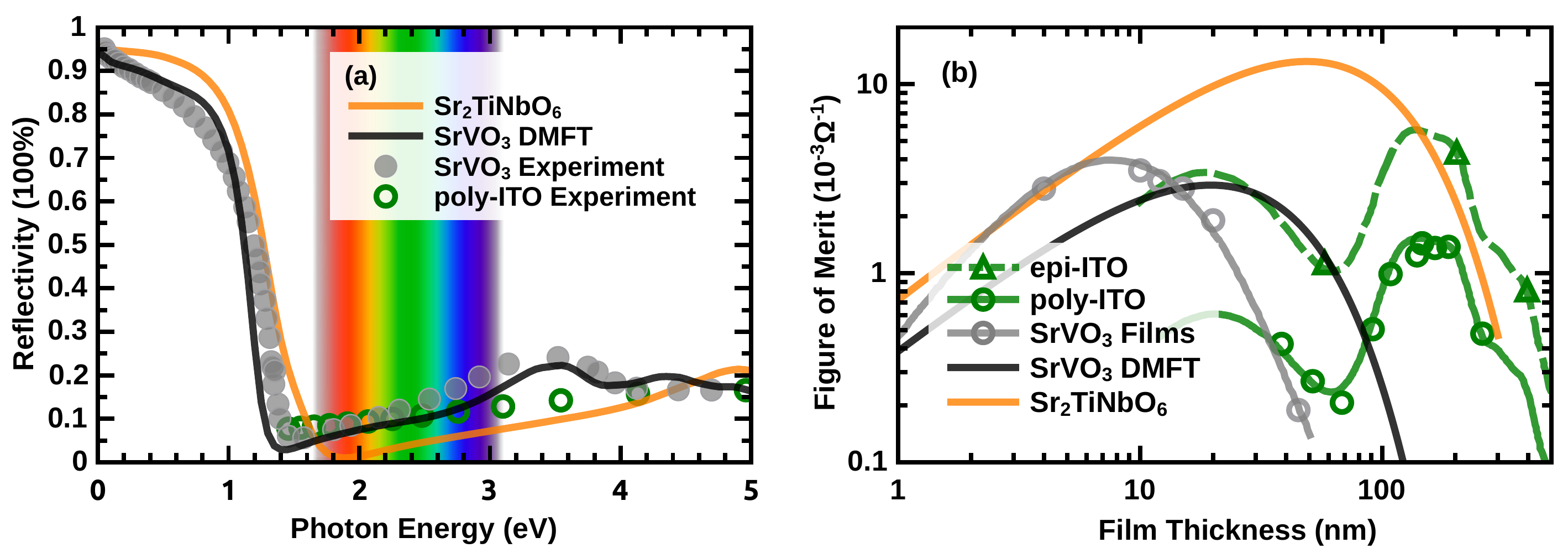}
\end{minipage}\hfill
\begin{minipage}{6.0cm}
\caption{DFT+DMFT (a) reflectivity and (b) figure of merit vs.\ film thickness of the proposed double perovskite Sr$_2$TiNbO$_6$ (orange), compared to SrVO$_3$ (grey), poly(crystalline) and epi(taxial) ITO (green). 
In (a) data for SrVO$_3$ as before; for poly-ITO from Ref.~\onlinecite{polyITO}.
In (b), data for SrVO$_3$, epi-ITO and poly-ITO from measurements \cite{ellmer2012past,polyITO,ohta2002surface,moyer2013highly,zhang2016correlated} (symbols) and calculations (lines) as presented in Ref.~\onlinecite{zhang2016correlated}.}
\label{Fig3}
\end{minipage}
\end{figure*}

To achieve this, we have studied three possible routes: A-site doping is achieved by doping holes to $d^1$ or electrons to $d^0$ materials, e.g., K-doped SrNbO$_3$ and SrTaO$_3$ and La-doped SrZr$O_3$ and SrHfO$_3$. B-site doping is achieved by SrTi$_{1-x}$Nb$_x$O$_3$, SrZr$_{1-x}$Ta$_x$O$_3$ and SrTi$_{1-x}$Ta$_x$O$_3$. However, $B$-site doping  induces a strong disorder scattering potential at the sites where the low-energy electrons reside, and thus larger reflectivities and absorptions: a problem encountered e.g.\,in Ref.\,\cite{mizoguchi2013electrical}. Here, we will hence only focus on the third route that turned out most promising: double perovskites.  This way,  $B$-site doping can be achieved without disorder scattering because of the regular periodic structure. For the two other routes, see SM Section S.\,5. 

Fig.~\ref{Fig3} shows our DFT+DMFT calculations for the proposed double perovskite Sr$_2$TiNbO$_6$ along with experimental data of SrVO$_3$. This double perovskite shows a considerably lower absorption (see SM Section S.\,6) and in particular reflectivity [see Fig.~\ref{Fig3}(a)] than SrVO$_3$, especially in the visible region. 
The electron donated by Nb$^{4+}$ is shared with Ti, leading to Ti $d^{0.58}$ and Nb $d^{0.42}$, respectively. The effective mass renormalization is $Z_{\rm Ti}\sim 0.7$ and $Z_{\rm Nb}\sim 0.9$.
Crucially, because of overall milder correlations, we have less electron-electron scattering and the reflectivity at the plasma edge is almost zero [Fig.~\ref{Fig3}(a)]. We thus find Sr$_2$TiNbO$_6$ to combine the smaller $\omega_p$ of 3$d$ TMOs with the excellent optical transmission of 4$d$/5$d$ TMOs. Altogether, Sr$_2$TiNbO$_6$ realizes a small enough $\omega_p$ (in DFT: 3.12\,eV vs. 3.82\,eV for SrVO$_3$), a low reflectivity and absorption throughout the visible spectrum and still a good conductivity (DMFT: 1.37$\times$10$^{4}$\,S/cm). 
Indeed, according to the FOM shown in Fig.~\ref{Fig3}(b), our proposed materials (for other materials see SM Section S.\,6) are better than SrVO$_3$ by a factor of 2-4, even exceeding commercial ITO \footnote{The shift of  the SrVO$_3$ FOM curve in DFT+DMFT compared to experiment is mainly because the absorption coefficient becomes larger for thin films while it is constant in our bulk calculations. Indeed ultrathin SrVO$_3$ films behave very different from bulk \cite{james2020quantum,PhysRevLett.114.246401,PhysRevLett.104.147601}}.

\emph{Conclusions.} We have pointed out the intrinsic pitfalls of using electronic correlations to reduce the plasma frequency as a design guideline in transparent conductors: 
the same quasiparticle renormalization factor that reduces the plasma frequency simultaneously enhances the finite-frequency scattering and thus enhances unwanted reflection and absorption above the plasma edge.
We have investigated routes to overcome this obstacle: we find that doping, 4$d$ and 5$d$ instead of 3$d$ transition metal oxides, as well as lattice expansion (i.e.\,tensile strain) all help in increasing performance. This strategy leads to a minimum of the reflectivity and absorption in the visible spectrum which is lower and more shallow than in the afore proposed SrVO$_3$.
As particularly promising candidates for transparent conductors, we identify strained (lattice+10\%) SrNbO$_3$ [see Fig.~\ref{Fig2}(b) and SM Section S.\,6] and the double perovskite Sr$_2$TiNbO$_6$. These optimized materials constitute a good balance between being  good conductors and having a small reflectivity and absorption in the optical range. Their FOM in Fig.~\ref{Fig3}(b) is even better than traditional ITO films.

\begin{acknowledgments}
\emph{Acknowledgments.}  The authors acknowledge discussions with P.\,Hansmann.
The research was supported by the Austrian Science Fund (FWF) through the Doctoral School W1243 Building Solids for Function and projects P 30819, P 30997 and P 32044. L.\,S. and Z.\,Z. also by the National Key R\&D Program of China (2017YFA0303602), 
Key Research Program of Frontier Sciences, CAS (Grant No.\,ZDBS-LY-SLH008),
and the National Nature Science Foundation of China (11774360, 11904373). Calculations were done on the Vienna Scientific Cluster (VSC).
\end{acknowledgments}

%


\clearpage

\setcounter{equation}{0}
\setcounter{figure}{0}


\onecolumngrid

\subsection*{\large Supplementary Material for ``Pitfalls and solutions for perovskite transparent conductors''}

\onecolumngrid
In this supplementary material, we include the following information that, while not essential for understanding the article's main results, provides more details on:
(1) Material-design rules for transparent conductors.
(2) Kramers-Kronig transformation and optical properties for different self-energy models.
(3) DFT and DMFT electronic structures of SrVO$_3$, including real frequency self-energies $\Sigma(\omega)$.
(4) DFT-calculated reflectivity $R(\omega)$ and absorption $A(\omega)$ for SrVO$_3$, SrNbO$_3$ and SrTaO$_3$ with hole doping in the virtual crystal approximation (VCA).
(5) DFT-calculated  $R(\omega)$ and  $A(\omega)$ for the actual materials proposed on the basis of (4).
(6) Additional DMFT results for the most promising materials identified in (5).
(7) Details on the analytical continuation of the self-energy.

\subsection{Section 1: Guidelines for designing transparent conductors}

In order to quantify the optical performance of transparent conductors, 
we define three characteristic properties on the basis of the reflectivity $R$ and the absorption $A$: 
\begin{enumerate}
  \item $\Delta R$ and $\Delta A$, which indicate the height of the tail 
    of the reflectivity and absorption curves at $E$$\sim$3.10\,eV (blue-light edge) (Fig.~\ref{Fig1}). 
  \item The plasma frequency $\omega_p$, which determines the photon energy at which the reflectivity 
    is minimal and above which the absorption curve levels off in the 
    Drude model [Fig.~\ref{Fig1}(a,b) of main text]. 
    In realistic materials, the minimum (valley) is not fully decided 
    by $\omega_p$ because of multi-band effects, 
    which are beyond the free carrier Drude model. 
    For instance, for SrVO$_3$ our DFT calculation predicts the 
    screened $\omega_p$ to be $\sim$3.8\,eV. However, in the DFT reflectivity
    and absorption curves [Fig.~\ref{Fig2}(c,d) of main text] the minimum is located at $\sim$2.1\,eV. 
  \item $R_{min}$ and $A_{min}$: the minimum of the reflectivity and absorption curves. 
    The photon energy realizing $R_{min}$ and $A_{min}$ is partially 
    decided by $\omega_p$ and the band structure of the materials.
\end{enumerate}

For ideal transparent conductors, $\Delta R$ and $\Delta A$ have to be as small as possible, as should be $R_{min}$ and $A_{min}$. The optimized value of $\omega_p$ is desired to be slightly below the red-edge ($\sim$1.65\,eV) of the visible-light window. As we discussed in the main text, for correlated metals $R_{min}$ and $A_{min}$ are invariably non-zero because of finite  electron-electron scattering. For comparison, we use the dashed cyan line in Fig.~\ref{Fig1} representing an ideal transparent conductor, in which $\omega_p$ puts the photon energy of $R_{min}$ and $A_{min}$ slightly below 1.65\,eV. In this ideal case,  $\Delta R$, $\Delta A$, $R_{min}$, and $A_{min}$ are all zero for $E$$>$1.65\,eV.

\begin{figure*}[h]
\includegraphics[width=0.80\textwidth]{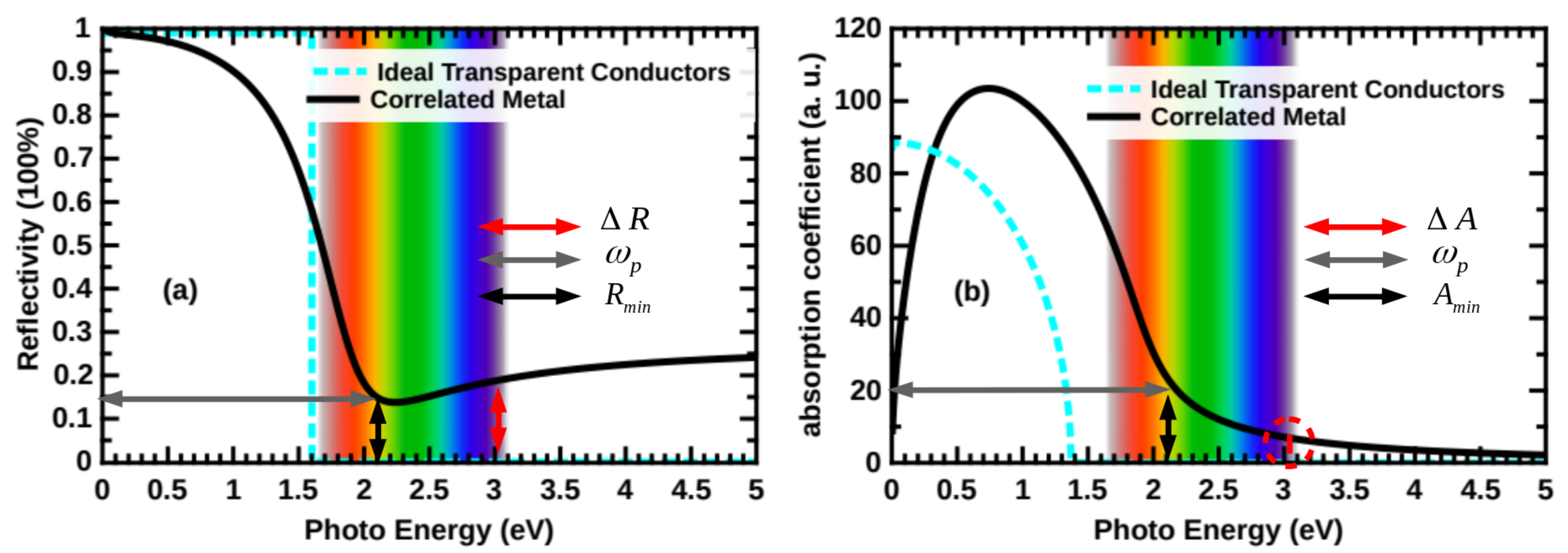}
\caption{Design rules for transparent conductors: (a) reflectivity and (b) absorption. The dashed cyan lines are the optical properties of a metal with perfect transparency; the black lines are the optical reflectivity and absorption typical for correlated metals.}
\label{Fig1}
\end{figure*}

\clearpage
\subsection{Section 2: Optical properties computed from self-energy model(s)}

\subsubsection{Section 2.1: Kramers-Kronig relation between the real and the imaginary part of the self-energy}

In the main text, we have made use of the connection between the real part of the self-energy and its imaginary part,
which necessitates that a large quasiparticle renormalization comes along with a large imaginary part of the self-energy or scattering rate. 
Here, we present a detailed derivation for this statement using the Kramers-Kronig transform,
\begin{equation}
  \label{eq:kramers-kronig}
  {\rm Re} \Sigma(\omega) = \frac{1}{\pi} \mathcal{P}\!\!\!\int_{-\infty}^\infty {\rm d}\omega' \frac{{\rm Im}\Sigma(\omega')}{\omega'-\omega}
  \hspace{2em} \text{and} \hspace{2em}
  {\rm Im} \Sigma(\omega) = -\frac{1}{\pi} \mathcal{P}\!\!\!\int_{-\infty}^\infty {\rm d}\omega' \frac{{\rm Re}\Sigma(\omega')}{\omega'-\omega},
\end{equation}
which relates the real and imaginary part of the frequency-dependent part of the self-energy $\Sigma(\omega)$.

In the main text, we have focused on the low frequency behavior, which for the 
real-part of the self-energy is linear; 
the prefactor $\gamma= Z^{-1}-1$ is directly connected to the quasiparticle renormalization factor $Z$. 
Further, the behavior at large frequency  is known to follow a $1/\omega$ behavior 
with prefactor $U^2/4$ for the symmetric one-band case at half-filling, see, e.g., Ref.~\onlinecite{Wang11}.
Altogether this yields
\begin{equation}
{\rm Re} \Sigma = 
  \left\{ \begin{array}{ll} 
    -\gamma \; \omega & {\rm \; \; for  \; \;}  \omega\rightarrow 0\\
    \frac{U^2}{4} \; \; \frac{1}{\omega} & {\rm \; \; for  \; \;}  |\omega|\rightarrow \infty\\
  \end{array}\right. \label{Eq:ReS}
\end{equation}
Due to the Kramers-Kronig relation Eq.~(\ref{eq:kramers-kronig}),
the imaginary part has then the following corresponding asymptotic behavior:
\begin{equation}
{\rm Im} \Sigma = \left\{ \begin{array}{ll} 
- \alpha \; \omega^2 & {\rm \;\; for \;\; }  \omega\rightarrow 0\\
- \beta  \; \frac{1}{\omega^2} & {\rm \;\; for \;\; }  |\omega|\rightarrow \infty\\
\end{array}\right.
\end{equation}
with real coefficients $\alpha$, $\beta$.
In a simple {\em model self-energy} we mimic the entire self-energy just by this asymptotic behavior. That is,
\begin{equation}
{\rm Im} \Sigma = \left\{ \begin{array}{ll} 
- \alpha \;  \omega^2 & {\rm \;\; for \;\; }  |\omega|< \omega^*\\
- \beta \; \frac{1}{\omega^2} & {\rm \;\; for \;\; }  |\omega|>\omega^*\\
\end{array}\right. \label{Eq:ModelS}
\end{equation}
with a cut-off frequency $\omega^*$  between the low and high-frequency behavior. Here, $\omega^*$ should be in the range where the high and low frequency asymptotics become comparably large, which for the real-part of the self-energy is at $\omega^* = U/2 \times 1/\sqrt{\gamma}$. The same criterion for the imaginary part (derived below) yields the identical  $\omega^*$. In the next Section, we will further see that this  model self-energy actually works reasonably well in comparison to numerical data.

To determine $\alpha$ and $\beta$ we will use Eq. (\ref{eq:kramers-kronig})
with the model self-energy Eq.~(\ref{Eq:ModelS})
This yields 
\begin{eqnarray}
{\rm Re} \Sigma(\omega) =   \underbrace{\frac{1}{\pi}  \int_{-\omega^*}^{\omega^*} {\rm d}\omega' \; \frac{-\alpha {\omega'}^2}{\omega'-\omega}}_{\textcircled{1}} 
+\underbrace{\frac{1}{\pi}  \left[\int_{-\infty}^{-\omega^*} {\rm d}\omega' + \int_{\omega^*}^{\infty} {\rm d}\omega'\right] \;  \frac{-\beta/\omega'^2}{\omega'-\omega}}_{\textcircled{2}}.
\end{eqnarray}
Integration gives
\begin{eqnarray}
\textcircled{1}&=&-\frac{\alpha}{\pi} \left. [
\omega^ 2 \ln|\omega'-\omega| +\omega\omega'+{\omega'}^2/2
]\right|_{-\omega^*}^{\omega^*} \nonumber \\
&&\rightarrow \left\{ \begin{array}{ll} 
- \frac{2\alpha}{\pi}\omega^* \;  \omega^2 & {\rm \;\; for \;\; }  \omega\rightarrow 0\\
 \;  \frac{2\alpha}{3\pi}{{\omega^*}^3} \;    \frac{1}{\omega} & {\rm \;\; for \;\; }  \omega\rightarrow \infty\\
\end{array}\right.
\\
\textcircled{2}&=&-\frac{\beta}{\pi}\frac{\omega'\ln\omega'-\omega|+\omega-\omega'\ln|\omega'|}{\omega^2\omega'}
\left[\Big|_{-\infty}^{-\omega^*}  + \Big|_{\omega^*}^{\infty}\right]\nonumber \\
&&\rightarrow  \left\{ \begin{array}{ll} 
- \frac{2\beta}{3\pi}\frac{1}{{\omega^*}^3} \; \omega & {\rm \;\; for \;\; }  \omega\rightarrow 0\\
\;  \;\frac{2\beta}{\pi} \frac{1}{\omega^*} \;\; \; \frac{1}{\omega} & {\rm \;\; for \;\; }  \omega\rightarrow \infty\\
\end{array}\right. \; .
\end{eqnarray}
Adding $\textcircled{1}$ and  $\textcircled{2}$ yields Eq.~(\ref{Eq:ReS}) with
$\gamma=\frac{2\alpha}{\pi}{{\omega^*}} +\frac{2\beta}{3\pi}\frac{1}{{\omega^*}^3}$ and $U^2/4=\frac{2\beta}{\pi} \frac{1}{\omega^*}+\frac{2\alpha}{3\pi} {\omega^*}^3$. Resolving this for $\alpha$ and $\beta$ we obtain the model self-energy with imaginary part 

\begin{equation}
{\rm Im} \Sigma (\omega) = \left\{ \begin{array}{ll} 
-  \frac{3\pi}{8}\frac{1}{{\omega^*}} \gamma \; \; \; \omega^2 & {\rm \;\; for \;\; }  \omega< \omega^*\\
-\frac{3\pi}{8}{{\omega^*}} \frac{U^2}{4} \; \frac{1}{\omega^2} & {\rm \;\; for \;\; }  \omega>\omega^*\\
\end{array}\right. \label{Eq:ModelS2}
\end{equation}
and real-part  Eq.~(\ref{Eq:ReS}).
This self-energy, specifically the corresponding electron-electron life time $\tau(\omega)^{-1}=-2Z{\rm Im} \Sigma(\omega)$, is 
taken in the main paper in the regime  $\omega< \omega^*$.
It describes life-time effects due to electron-electron scattering. When the self-energy is momentum-independent and there are no vertex corrections (as in DMFT \cite{Pruschke93,RevModPhys.68.13}), the optical (transport) life time is the same as this electron-electron scattering life time. We add a small impurity scattering rate $\tau_0^{-1}$, which is frequency-independent.

\subsubsection{Section 2.2: Comparison of model self-energy, cut-off self-energy and NRG data}

In this section we compare the {\em model self-energy} Eq.~(\ref{Eq:ModelS2}) 
derived in the previous Section, to (i) the numerical renormalization group (NRG) 
self-energy calculated for the one-band Hubbard model at interaction $U=2.4$
and zero temperature within DMFT for a Bethe lattice with bandwidth 
2 in Ref.~\onlinecite{Bulla1999}, 
and to (ii) the simplified form of the model self-energy used in the main paper. 
The last simply takes the low-frequency part of Eq.~\ref{Eq:ModelS2} 
which is cut off above a certain strength of the self-energy or scattering rate ($\tau^{-1}_{\rm cut-off}=-2 Z{\rm Im}\Sigma_{\rm cut-off}$):  
\begin{equation}
{\rm Im} \Sigma (\omega)= - \min\big\{
 \frac{3\pi}{8}\frac{1}{{\omega^*}} \gamma \; \omega^2, 
-{\rm Im} \Sigma_{\rm cut-off}  \big\}
 \label{Eq:CutOffS}.
\end{equation}
The reason why we have chosen this {\em cut-off} self-energy is merely a pedagogical simplification as it avoids the additional discussion of the large frequency asymptotics in the main text. As we will see next, if   ${\rm Im}\Sigma_{\rm cut-off}$ is chosen such that it is a comparable to the model or NRG self-energy in the visible range of the spectrum, also the resulting reflectivity and absorption is very similar to the NRG and the full {model self-energy}.

\begin{figure*}[tb]
\includegraphics[width=1.00\textwidth]{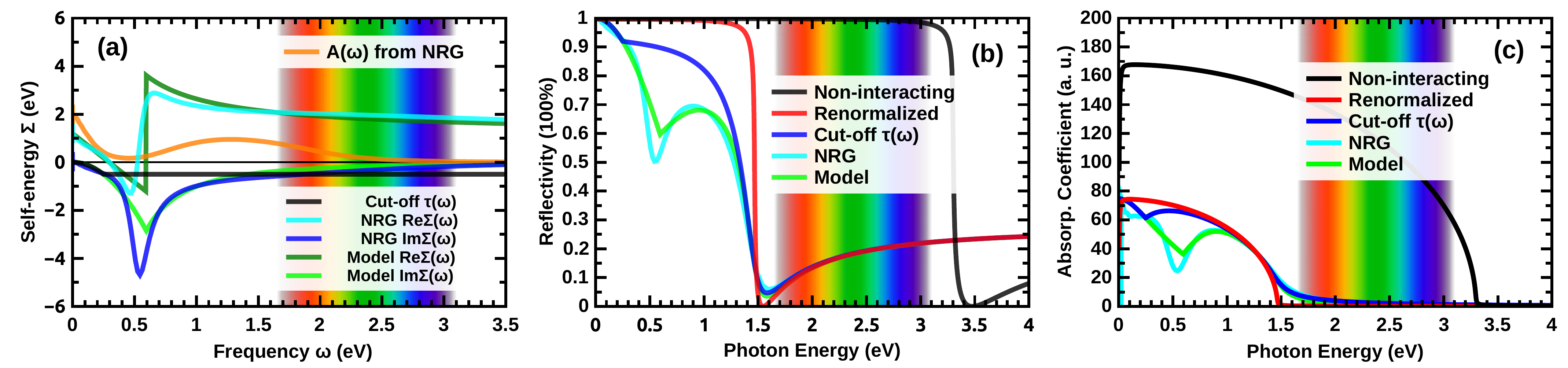}
\caption{Model and NRG self-energies for optical calculations.\\
(a) Real and imaginary part of the model self-energy Eqs.~(\ref{Eq:ReS}),~(\ref{Eq:ModelS2}); the cut-off self-energy Eq.~(\ref{Eq:CutOffS}) used in Fig.~\ref{Fig1}(a,b) of the main text (denoted there and here by the related cut-off $\tau(\omega)$); and the NRG self-energy and spectral function $A(\omega)$ taken from Ref.~\onlinecite{Bulla1999}.\\
(b) Reflectivity and (c) absorption calculated  with $\tau(\omega)^{-1}=-2Z{\rm Im} \Sigma(\omega)$ and the quasiparticle renormalization $Z=\sim$0.2 for the three self energies from panel (a); as well as for ${\rm Im }\Sigma=0$ (denoted as ``renormalized'') and $Z=1$, ${\rm Im }\Sigma=0$ (``non-interacting''). In the non-interacting and renormalized cases we have added a small impurity scattering $\tau_0^{-1}\sim$0.01\,eV.
}
\label{Fig2}
\end{figure*}

Fig.~\ref{Fig2} (a) shows these three self-energies. As we can see, the  full {\em model self-energy} derived in the previous Section agrees with the NRG data quite well, with a single free parameter $\gamma$ fitted to the low frequency real-part of the self-energy ($U=2.4$ as in NRG).
For low energies, 0\,eV$<$$\omega$$<$0.3\,eV, Im$\Sigma$ of all three self-energies is very similar. 
The most remarkable difference is that the imaginary part of the model self-energy 
deviates from the NRG data in so-far as in the crossover regime 
between low-frequency and high-frequency asymptotics it shows a sharp kink, 
whereas it rather has the form of an even more pronounced hump in  NRG. 
The cut-off self-energy deviates even stronger with a kink at 
smaller energies where the self-energy or $\tau(\omega)$ is cut off.
These kinks, respectively the hump, are however not a problem,
because they are located in the not particularly relevant \emph{in-between} region:
On the one hand the important low-frequency optical conductivity
is unchanged. On the other hand the optical range of the spectrum,
where we are interested in reflectivity and absorption, 
starts at higher energies and thus remaines untouched by such kinks.
Let us note that they are also located in between the quasi-particle peak
and the upper (lower) Hubbard band.
For large frequencies, the constant $\Sigma_{\rm cut-off}$ of the {cut-off} self-energy
deviates from the correct $1/\omega^2$ behavior. But we choose the 
cut-off $\Sigma_{\rm cut-off}$ such that it yields very 
comparable scattering rates in the visible range of the spectrum.

Note that there is a further (hardly visible) feature in the NRG data, 
namely a kink in the real part of the self-energy, 
which reflects as a change of curvature for the imaginary part. 
This kink emerges from electronic correlations \cite{Nekrasov05a,Byczuk2007} 
and can be traced back to the Kondo effect in a DMFT bath \cite{Held13}. 
Again, as shown below, it is neither relevant for $\sigma(0)$
nor for $R$ and $A$ in the visible range of the spectrum.

Fig.~\ref{Fig2} (b,c) compares the reflectivity $R$ and absorption $A$ 
obtained by these three self-energies of Fig.~\ref{Fig2} (a), 
using the corresponding $\tau(\omega)$ and $Z$ in the Drude model 
[Eq.~(1) of the main text; further parameters as in the main text]. 
As one can see, qualitatively all of them yield a similar reflectivity and absorption. 
Quantitatively, there are some minor deviations:
The hump in the NRG self-energy and the kink in the model self-energy 
lead to a dip in the reflectivity and absorption around 0.5 eV;  
the cut-off in the cut-off self-energy similarly leads to a dip but already at 0.25 eV. 
This feature is also observed in experiment and DFT+DMFT calculations 
for the reflectivity of SrVO$_3$, see Fig.~1~(c) of the main text.

The aforementioned electronic kink due to the Kondo effect additionally leads 
to deviations of the NRG absorption at small frequencies, 
since the quasiparticle renormalization changes.  
However, these are relatively minor difference far from the visible range of the spectrum.

In the visible range of the spectrum all three self-energies (model, cut-off, NRG) show essentially the same reflectivity and absorption. In contrast, if the imaginary part of the self-energy is not taken into account, but only the quasiparticle renormalization, we grossly underestimate the reflectivity and absorption.
As we discussed in the main text, the non-zero Im$\Sigma(\omega)$ in the 
visible light window leads to a non-zero $R$ and $A$ at and above the plasma frequency, respectively. This is the pitfall of using electronic correlations to push the plasma frequency below the visible range of the spectrum, but neglecting intrinsically linked finite lifetimes.

\clearpage

\subsection{Section 3: Further DFT and DMFT optical properties of SrVO$_3$}

In this Section we provide further details of our DFT+DMFT calculation for SrVO$_3$. 
That is, in  Fig.~\ref{Fig3}~(a) we show the perovskite crystal structure, 
in  Fig.~\ref{Fig3}~(b) the DFT and Wannier-projected bandstructure, 
and in Fig.~\ref{Fig3}~(c) the DFT density of states (DOS). 
Essential to reproduce the DMFT calculation is the self-energy 
shown in  Fig.~\ref{Fig3}~(d). From this self-energy and the 
DFT(\textsc{Wien2K})-calculated dipole matrix elements, 
we calculate the optical conductivity $\sigma(\omega)$ shown in Fig.~\ref{Fig4} (a,b) 
using \textsc{Woptic} \cite{assmann2016woptic}. 
From this optical conductivity in turn we determine the dielectric function 
$\epsilon(\omega)= \epsilon_{core} + 4\pi i \sigma(\omega)/\omega$, 
whose real and imaginary part yield the reflectivity and absorption shown in Fig.~\ref{Fig4} (c) and (d), respectively. 

\begin{figure*}[h]
\includegraphics[width=0.70\textwidth]{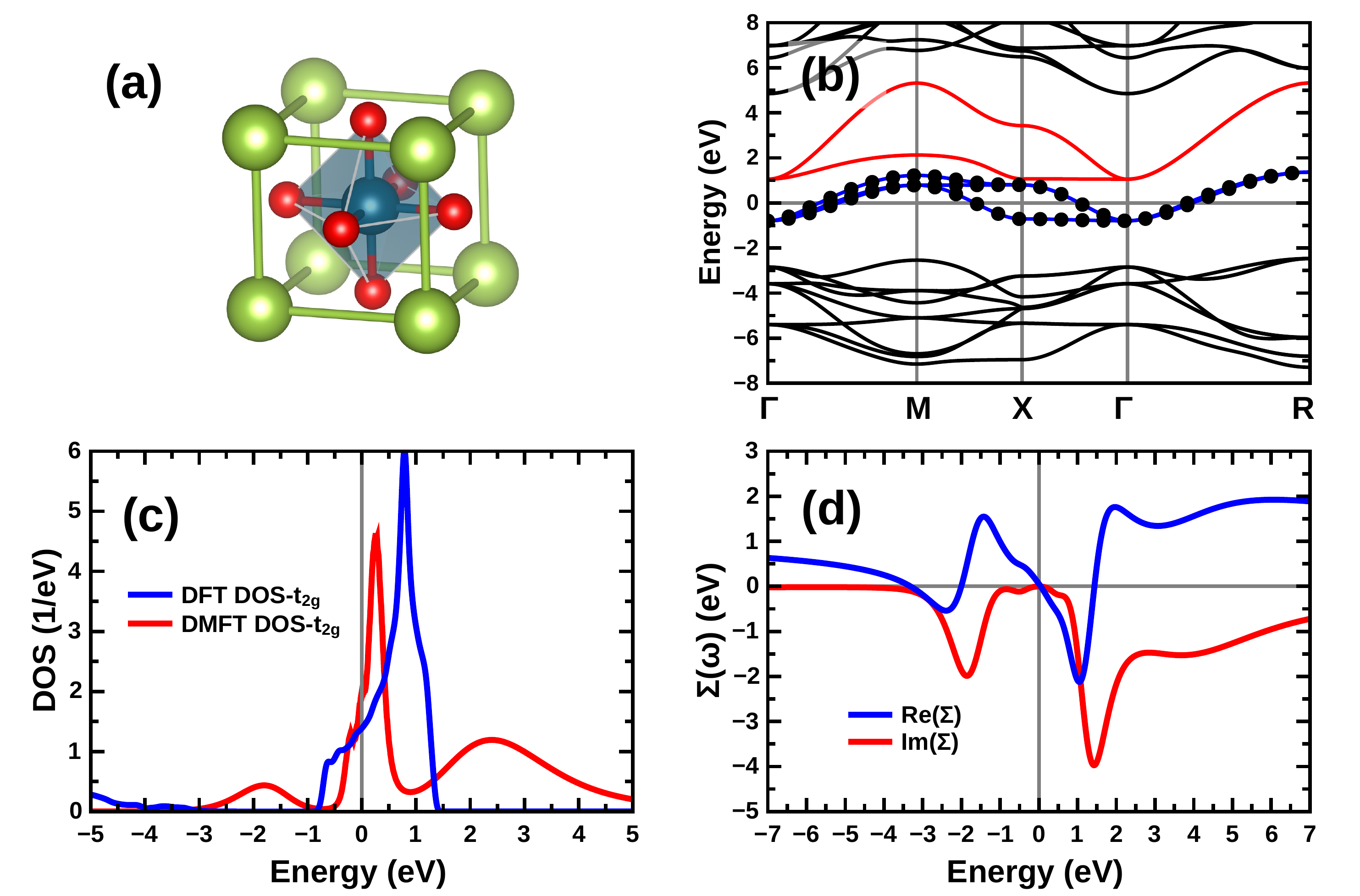}
\caption{DFT and DMFT electronic structures of SrVO$_3$.
(a) Crystal structure of SrVO$_3$: the green atoms indicate Sr; the center dark green V atom and the six oxygen atoms (red color) compose the VO$_6$ octahedron.
(b) DFT bands of SrVO$_3$; the bands between -8\,eV to -2\,eV derive from O-2$p$ orbitals, the V-$t_{2g}$ and $e_g$ bands are labeled by blue and red colors, the bands with black dots indicate Wannier projection onto the DFT $t_{2g}$ bands.
(c) The comparison between the DFT density of states (DOS) and the DMFT spectral functions $A(\omega)$ for the $t_{2g}$ orbitals.
(d) DMFT real-frequency self-energy $\Sigma(\omega)$ for the $t_{2g}$ orbitals.}
\label{Fig3}
\end{figure*}

\begin{figure*}[h]
\includegraphics[width=0.80\textwidth]{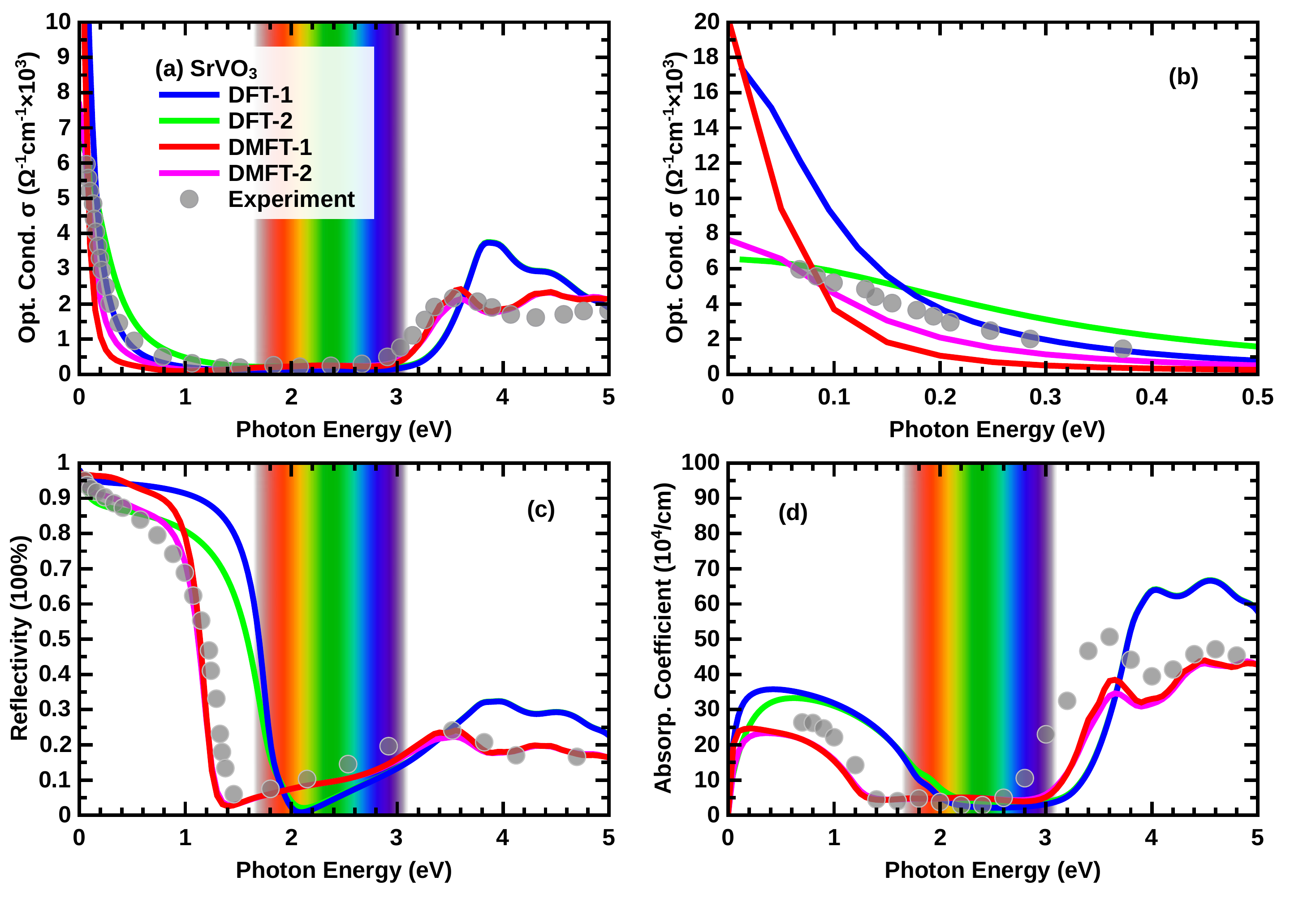}
\caption{Optical properties of SrVO$_3$ within DFT and DFT+DMFT (\textsc{Woptic}): (a) optical conductivity $\sigma(\omega)$, (b) zoom-in of (a), (c) reflectivity $R(\omega)$ and (d) absorption coefficient $A(\omega)$. 
  DFT-1 and DMFT-1 are with (impurity) scattering $\tau_0^{-1}=0.1\,eV$ and 0; DFT-2 and DMFT-2  with  $\tau_0^{-1}=0.27\,eV$ and  0.13\,eV. 
For comparison, the experimental data of SrVO$_3$ is also shown. The experimental $R$ is taken from Ref.\,\onlinecite{PhysRevB.58.4384}; the experimental $A$ is computed from the dielectric functions $\epsilon$ from Ref.~\onlinecite{zhang2016correlated}.}
\label{Fig4}
\end{figure*}

\clearpage

\subsection{Section 4: Additional DFT results for 3$d^1$ SrVO$_3$, 4$d^1$ SrNbO$_3$ and 5$d^1$ SrTaO$_3$ with hole doping}

As shown in Fig.~\ref{Fig2} in the main text and Fig.~\ref{Fig5} below, hole doping is a good way to shift the plasma frequency. Moreover, 4$d$ and 5$d$ TMOs are preferable because of their wide frequency range with lower reflectivity owing to their larger bandwidth and larger distance between $d$ and $p$ bands. The last two reduces the intra- and inter-band transition in the visible-light region, respectively.

In Fig.~\ref{Fig5}, we show the DFT absorption $A(\omega)$ of hole doped the 5$d$ transition metal oxide (TMO) SrTaO$_3$, using the VCA. As a limitation of hole doping, the result of 5$d^0$ SrHfO$_3$ is also shown. As a comparison the experimental data of the absorption coefficient of SrVO$_3$ is also shown. In Fig.~\ref{Fig2}(b) of the main text we have already shown that undoped SrTaO$_3$ ($n$=1.00) is not a perfect transparent conductor because of the high plasma frequency $\omega_p$: only the blue part of visible-light is allowed to pass through the surface without reflection. An effective way to reduce  $\omega_p$ is hole-doping. When $n$$\le$0.2, SrTaO$_3$ exhibits excellent transparency with only a tiny reflectivity for $E$$\sim$1.65\,eV-3.10\,eV (see Fig.~\ref{Fig2} of the main text). For the shown absorption, similarly promising behavior is obtained: undoped SrTaO$_3$ ($n$=1.00) has large absorption coefficient at $E$$<$3.00\,eV. With hole-doping, the absorption peak starting from $\omega\sim$3.0\,eV is reduced and shifted to $\omega\sim$0\,eV, finally reaching the non-absorption band insulator SrHfO$_3$ (5$d^0$). Excellent transparency is predicted for hole-doped SrTaO$_3$ ($n$$\le$0.20).

\begin{figure*}[h]
\includegraphics[width=0.50\textwidth]{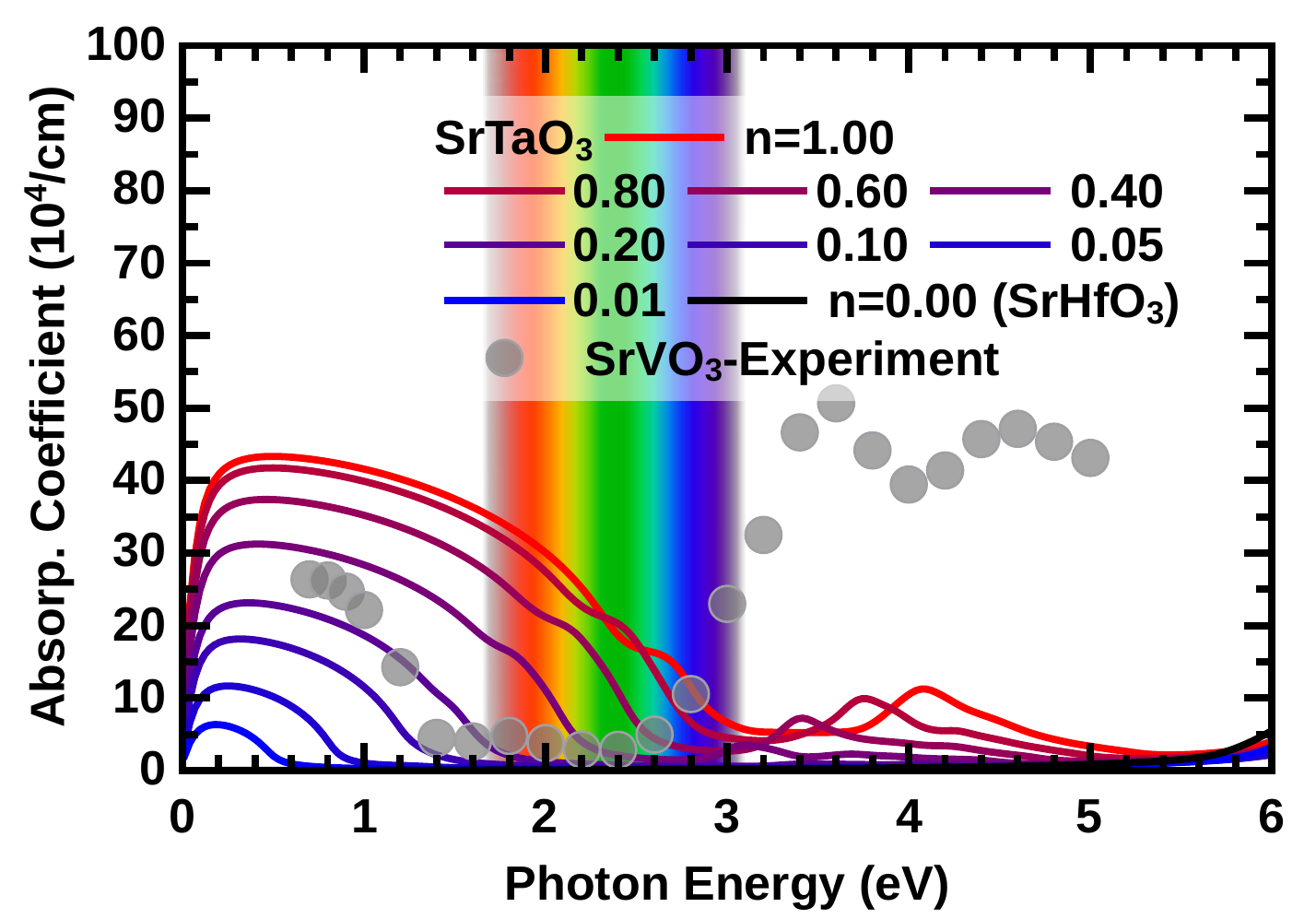}
\caption{DFT-derived absorption coefficient of SrTaO$_3$ with hole doping, with modified electron count $n$ from the original value $n$=1.00 down to 0.00 (for the 0.00, calculations of the reflectivity and the absorption curves refer to the band insulators 5$d^0$ SrHfO$_3$).}
\label{Fig5}
\end{figure*}

\begin{figure*}[h]
\includegraphics[width=0.80\textwidth]{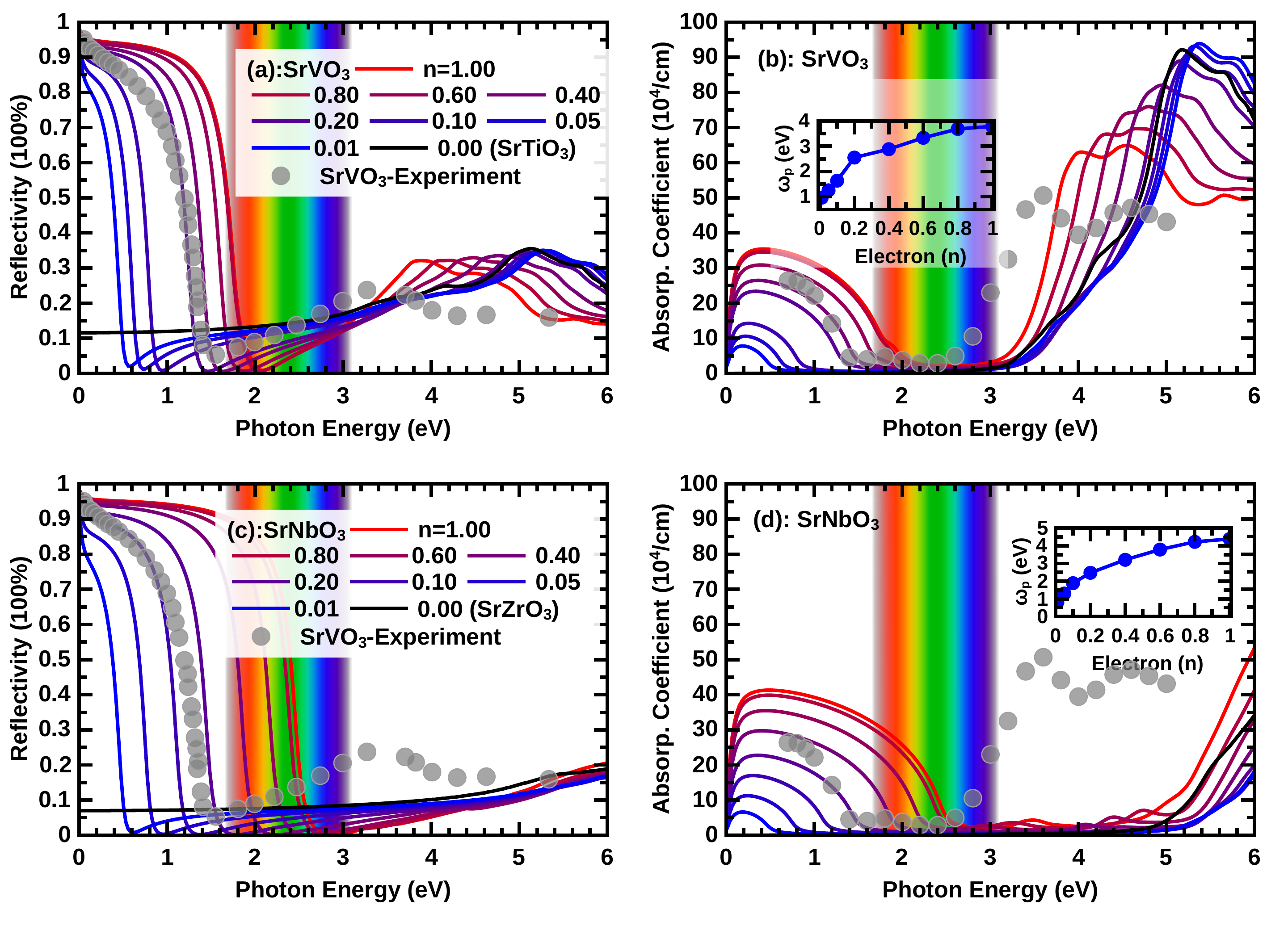}
\caption{DFT-derived optical properties of SrVO$_3$ (a-b) and SrNbO$_3$ (c-d) with hole doping, with modified electron number $n$ from the original value $n$=1.00 to 0.00 (for $n$=0.00, reflectivity and absorption curves are obtained by calculating the band insulators 3$d^0$ SrTiO$_3$ and 4$d^0$ SrZrO$_3$).
The insets of (b) and (d) show the relationship between the plasma frequency $\omega_p$ and $n$.
As a comparison the experimental data of the reflectivity and the absorption coefficient of SrVO$_3$ are also shown (gray dots).}
\label{Fig6}
\end{figure*}

The same hole doping is also applied to 3$d^1$ SrVO$_3$ and 4$d^1$ SrNbO$_3$, as shown in Fig.~\ref{Fig6}(a-d). As a comparison the experimental data of the reflectivity and the absorption coefficient of SrVO$_3$ are also shown (gray dots). For SrVO$_3$ [Fig.~\ref{Fig6}(a,b)], hole doping reduces  $\omega_p$ from 3.8\,eV ($n$=1.0) to 0.96\,eV ($n$=0.01) [inset of Fig.~\ref{Fig6}(b)]. Consequently,  $R_{min}$ and $A_{min}$ are shifted to lower frequencies, too. Finally, at $n$=0.0 (SrTiO$_3$), a constant $R$ and $A\sim$0 are obtained. Please note that these results were computed at the DFT-level, without including dynamical correlation effects. In realistic experiments, even without hole doping, the frequency realizing $R_{min}$ is already below the visible region $\omega\sim$1.65\,eV. Hence, we exclude that the performance of SrVO$_3$ could be further increased substantially by doping.

The 4$d^1$ perovskite SrNbO$_3$ [Fig.~\ref{Fig6}(c,d)] exhibits similar results as 5$d^1$ SrTaO$_3$. The insets of Fig.~\ref{Fig6}(d) show the relationship between the plasma frequency $\omega_p$ and $n$, indicating that the reduction of the band filling can effectively reduce $\omega_p$ from $\sim$4.6\,eV ($n$=1.0) to $\sim$0.8\,eV ($n$=0.01), crossing the red-light edge of the visible-light window (1.65\,eV). For SrVO$_3$, it naturally blocks and absorb the blue-light due to the high reflectivity and absorption at $\omega\sim$3.10\,eV. However, for hole-doped 4$d$ SrNbO$_3$ [Fig.~\ref{Fig6}(c,d)] and 5$d$ SrTaO$_3$ [Fig.~\ref{Fig2}(b) of the main text and Fig.~\ref{Fig4} in SM], such suppression of the blue-light transparency is eliminated. For SrNbO$_3$, our DMFT calculations predict that the renormalization factor $Z$=0.74,  consistent with the previously reported $Z$=0.72 \cite{Park2020}. A simple formula to estimate how $Z$ modifies $\omega_p$ is: $\omega_p$(DMFT)$=$$\omega_p$(DFT)/$\sqrt{Z^{-1}}$. Then  $\omega_p$ of SrNbO$_3$ is reduced from $\sim$4.4\,eV (DFT) to $\sim$3.8\,eV (DMFT).

\begin{figure*}[h]
\includegraphics[width=0.50\textwidth]{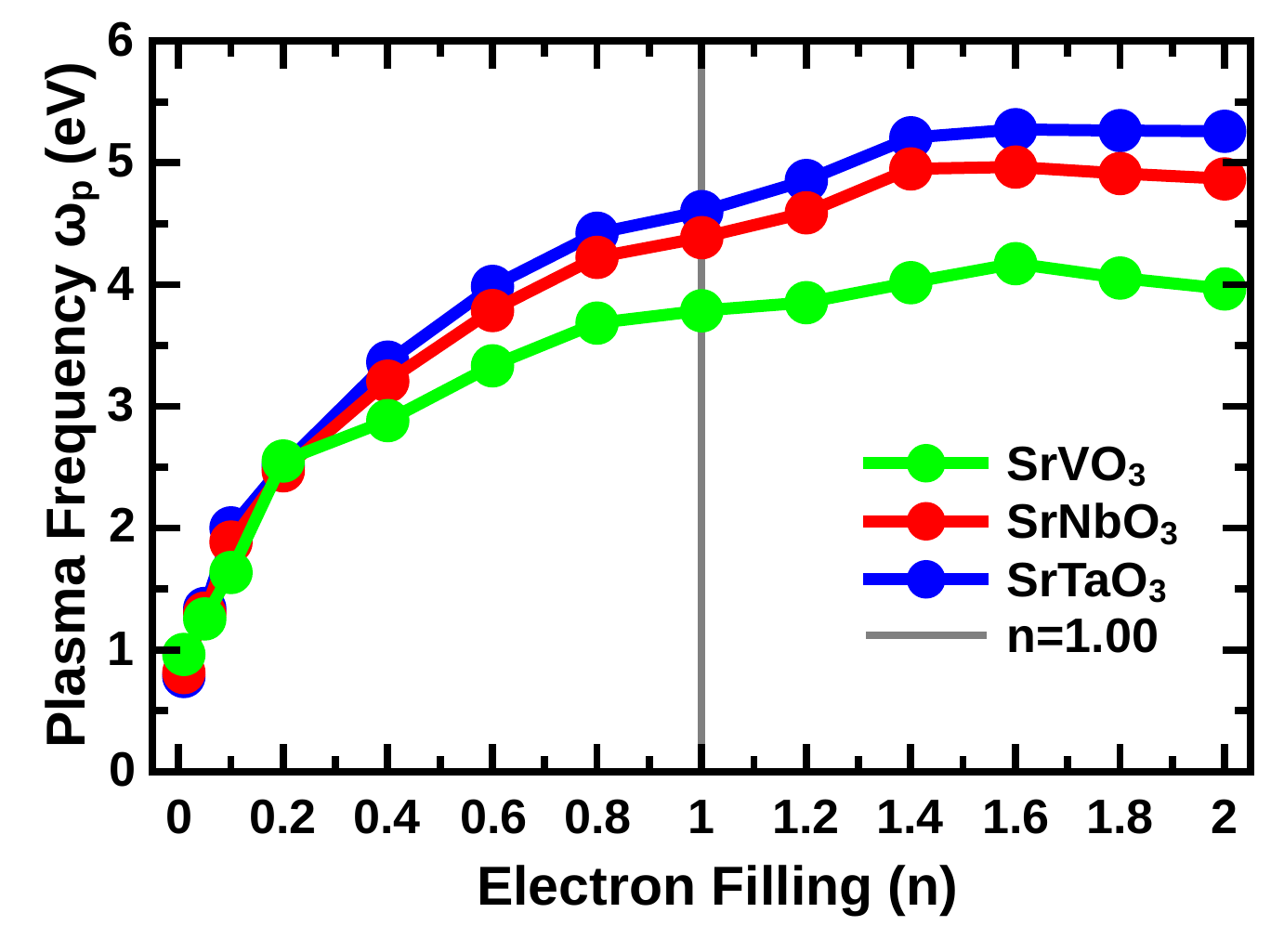}
\caption{DFT-derived plasma frequency $\omega_p$ of doped perovskite $AB$O$_3$ systems.}
\label{Fig7}
\end{figure*}

Fig.~\ref{Fig7} shows the $\omega_p$ of SrVO$_3$, SrNbO$_3$ and SrTaO$_3$ under hole doping. Again hole doping is achieved by the VCA as implemented in \textsc{Wien2k}. The electron numbers $n$ of V, Nb, Ta are changed from $n$=0.01 to $n$=2.0, passing the original value $n$=1. All three TMOs exhibit a similar behavior: the $\omega_p$ are reduced as $n$ shrinks. This can be explained by $\omega_p=e\sqrt{4\pi/\epsilon_{core}}\sqrt{n/m^*}$. This proves that hole doping is an effective way to reduce $\omega_p$ in TMOs and shifting $R_{min}$ and $A_{min}$ below the visible region.

\begin{figure*}[h]
\includegraphics[width=0.50\textwidth]{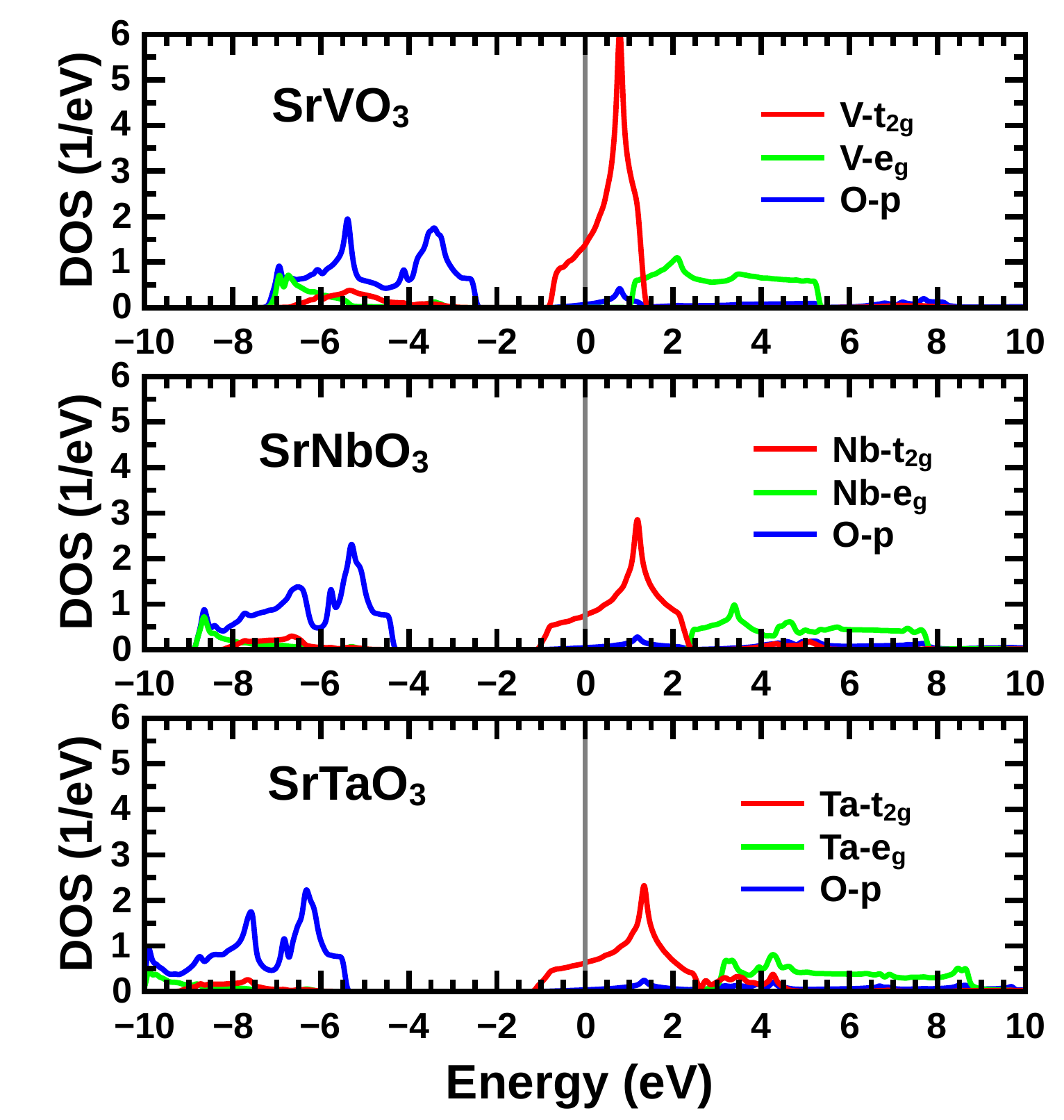}
\caption{DFT-derived DOS of 3$d^1$ SrVO$_3$, 4$d^1$ SrNbO$_3$ and 5$d^1$ SrTaO$_3$. The partial DOS of the $t_{2g}$, $e_g$ and O-$p$ orbitals are labeled by red, green and blue colors, respectively. 
}
\label{Fig8}
\end{figure*}

In Fig.~\ref{Fig8} we compare the orbital-resolved density of states (DOS) of SrVO$_3$, SrNbO$_3$ and SrTaO$_3$. In DFT there is no electron-electron scattering. Hence, the optical properties within the visible-light window (1.65\,eV-3.10\,eV) are mostly dominated by  interband transitions between $t_{2g}$ to $e_g$, and the transition between 0\,eV-1.65\,eV comes from the intra-band transitions within the $t_{2g}$ bands. The optical properties in the energy range of $E>$3.1\,eV originates from the transition between $t_{2g}$ and O-$p$ orbitals. In 4$d^1$ SrNbO$_3$ and 5$d^1$ SrTaO$_3$, the wider $d$-bands and larger distance between $d$ to O-$p$ lead to the reduced $R$ and $A$, as shown in Fig.~\ref{Fig2} of main text and Fig.~\ref{Fig5} and Fig.~\ref{Fig6} in this SM.

\clearpage

\subsection{Section 5: Materials proposed and DFT optical properties}

\begin{figure*}[h]
\includegraphics[width=0.50\textwidth]{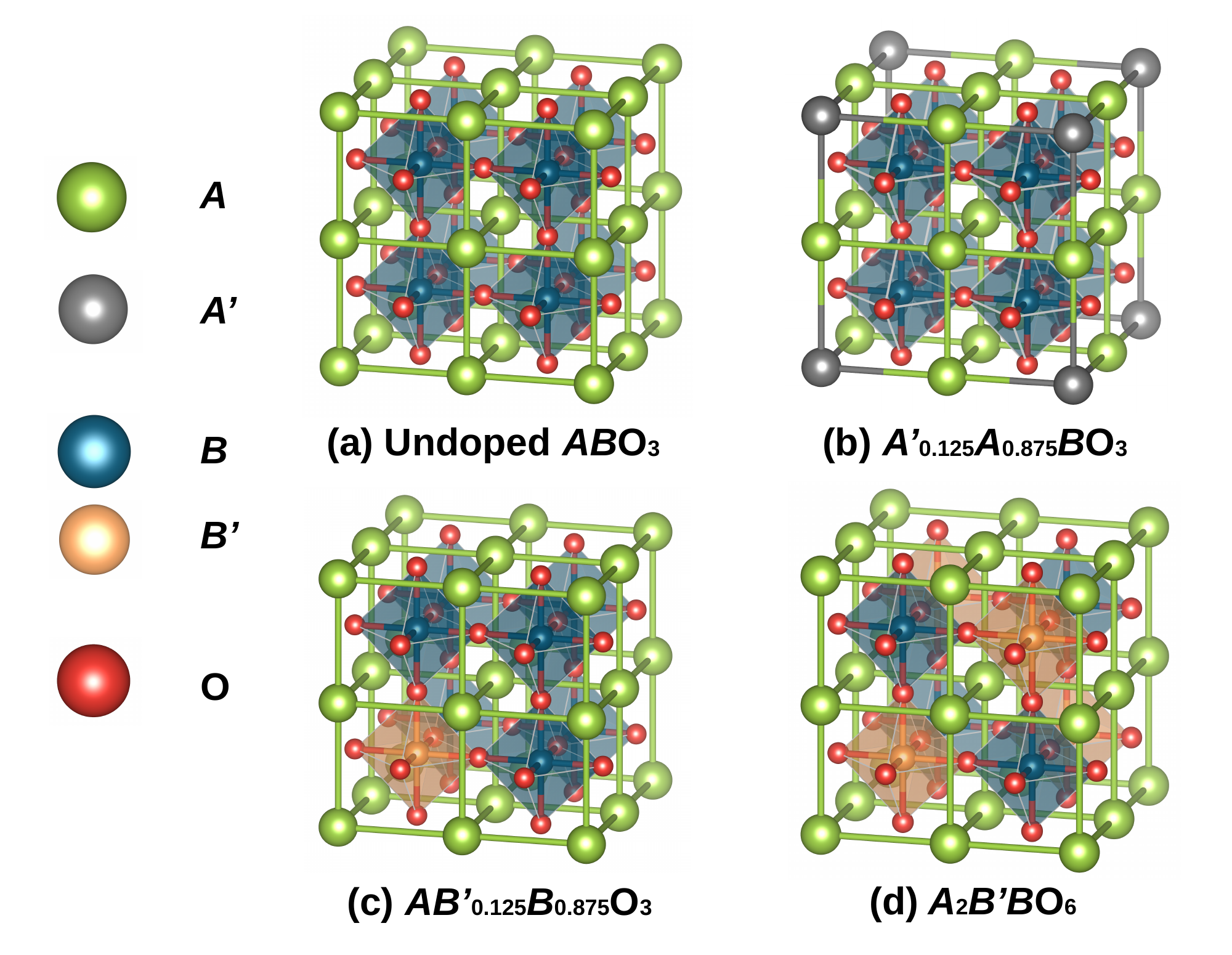}
\caption{Proposed ways to dope $AB$O$_3$ systems with holes: (a) undoped $AB$O$_3$, (b) 12.5\% $A$-site doping, (c) 12.5\% $B$-site doping and (d) double perovskite.}
\label{Fig9}
\end{figure*}

Based on the above results that still employed the VCA, we now turn to actual doped crystal structures and materials. We study three ways of inducing hole doping in Sr$B$O$_3$ [Fig.~\ref{Fig9}(a)]: $A$-site doping [Fig.~\ref{Fig9}(b)], $B$-site doping [Fig.~\ref{Fig9}(c)], and as  a special case of $B$-site doping  double perovskites [Fig.~\ref{Fig9}(d)]. This third way avoids structural disorder which might be problematic for $B$-site doping, as it will induce a strong disorder scattering potential. Double perovskites are similar to $B$-site doping but avoid this disorder scattering through a regular periodic structure. 
Hence we mainly focus on two alternative ways of engineering doped perovskites: $A$-site doping [Fig.~\ref{Fig9}(b)] and double perovskites  [Fig.~\ref{Fig9}(d)].

\begin{figure*}[h]
\includegraphics[width=0.80\textwidth]{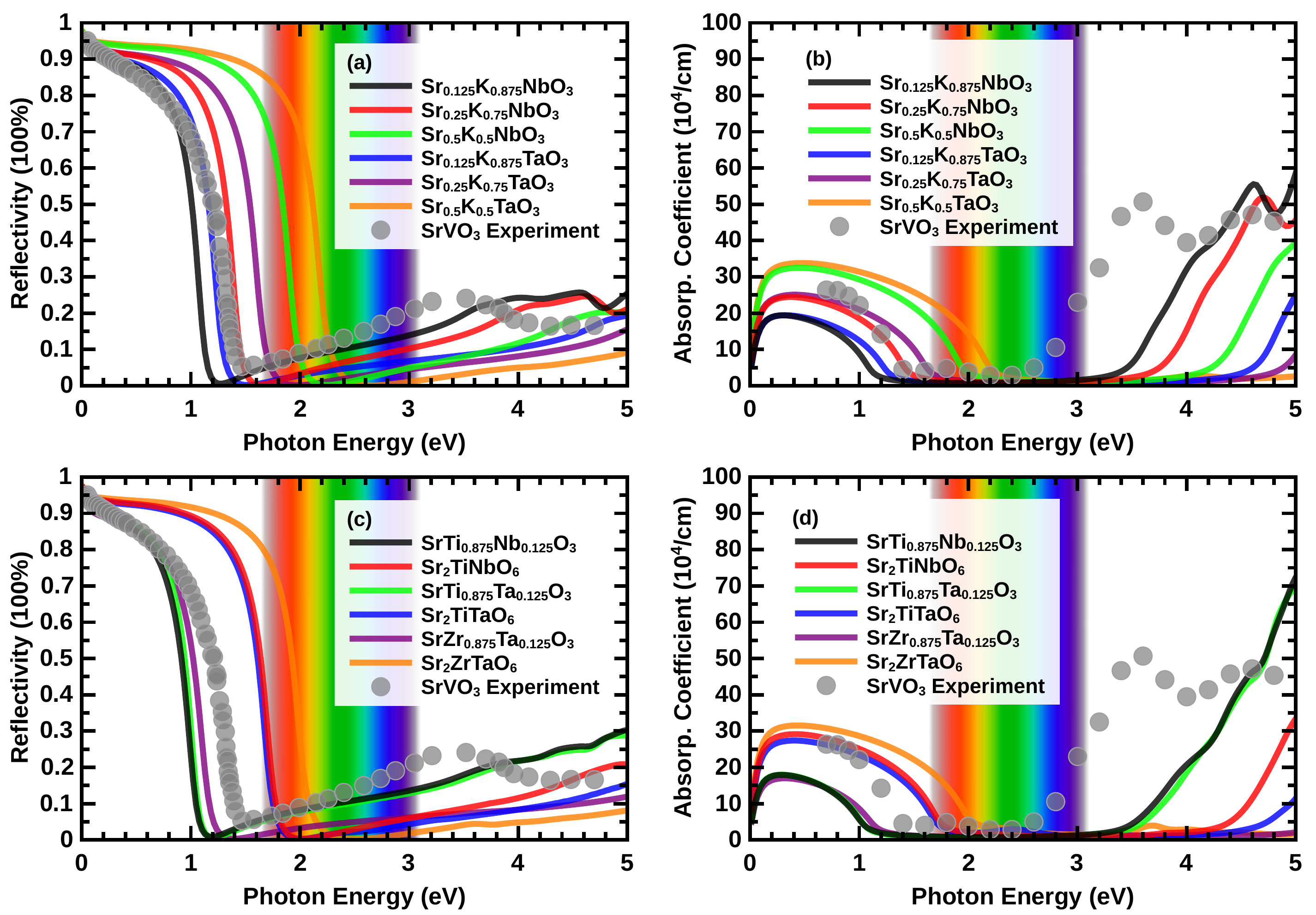}
\caption{DFT reflectivity $R$ (a) and absorption $A$ (b) of $A$-site hole doping (K-doping) of 4$d^1$ SrNbO$_3$ and 5$d^1$ SrTaO$_3$ with a K-concentration of 12.5\% and 50\%. $R$ (c) and $A$ (d) of $B$-site doped SrTi$_{0.875}$Nb$_{0.125}$O$_3$, SrZr$_{0.875}$Ta$_{0.125}$O$_3$, SrTi$_{0.875}$Nb$_{0.125}$O$_3$, and double perovskites Sr$_2$TiNbO$_6$, Sr$_2$ZrTaO$_6$ and Sr$_2$TiTaO$_6$.}
\label{Fig10}
\end{figure*}

For $A$-site doping in Fig.~\ref{Fig9}(b), we consider the series Sr$_{1-x}$K$_x$NbO$_3$ (4$d^{1-x}$) and Sr$_{1-x}$K$_x$TaO$_3$ (5$d^{1-x}$), i.e.\ doping those TMOs identified above as most promising. Please note that these ways of hole doping are effectively equivalent  to electron doping to the band insulators KNbO$_3$ and KTaO$_3$ by Sr-doping in K$_{1-x}$Sr$_x$NbO$_3$ and K$_{1-x}$Sr$_x$TaO$_3$. These two materials are intensively studied and doping with electrons is experimentally doable \cite{sudrajat2019electron,zhu2014ptcr}. The DFT optical properties $R$ and $A$ of Sr$_{1-x}$K$_x$NbO$_3$ and Sr$_{1-x}$K$_x$TaO$_3$ are shown in Fig.~\ref{Fig10}(a-d).  The difference to the previous Section is that we now calculate actual crystal structures of doped materials, instead of using the VCA.

In Fig.~\ref{Fig10}(c,d), we show three different double perovskites within $G$-type ordering of the transition metal $B$ sites as displayed in the inset of Fig.~\ref{Fig9}(d): Here, the combination of B sites induces an intrinsic charge transfer \cite{PhysRevX.7.011023},  in configurations $d^n$ (0$<$n$<$1). Our DFT calculations exclude Sr$_2$ZrTaO$_6$ as a candidate for a transparent conductor because of its high $\omega_p$=3.37\,eV. But both, Sr$_2$TiNbO$_6$ and Sr$_2$TiTaO$_6$, have a very appropriate $\omega_p= 3.12\,$eV and 2.94\,eV, so that  $R_{min}$ and $A_{min}$ are realized already close to the red-edge of the visible-window $\omega\sim$1.65\,eV [Fig.~\ref{Fig10}(c,d)]. These $\omega_p$'s will be further reduced by electronic correlation as we discussed above.

\begin{figure*}[h]
\includegraphics[width=0.80\textwidth]{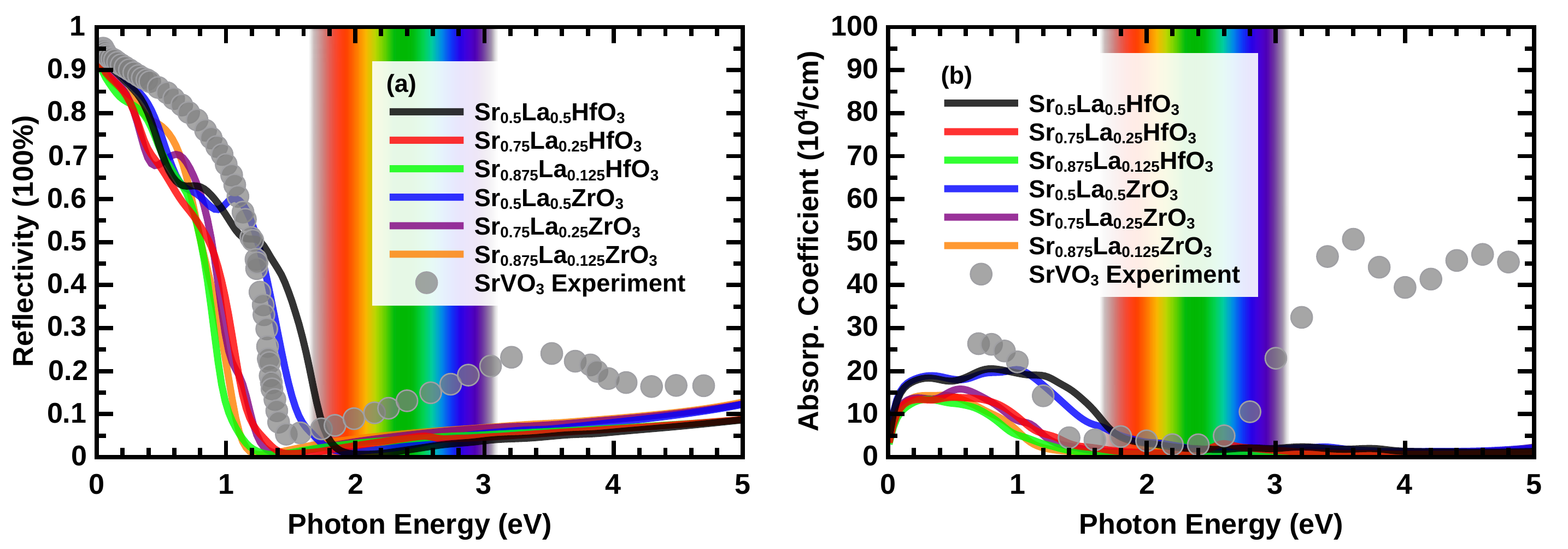}
\caption{DFT reflectivity $R$ (a) and absorption $A$ (b) of $A$-site electron doping (La-doping) of band insulators 4$d^0$ SrZrO$_3$ and 5$d^0$ SrHfO$_3$ with a Sr concentration of 12.5\% and 50\%.}
\label{Fig11}
\end{figure*}

\begin{figure*}[h]
\includegraphics[width=0.50\textwidth]{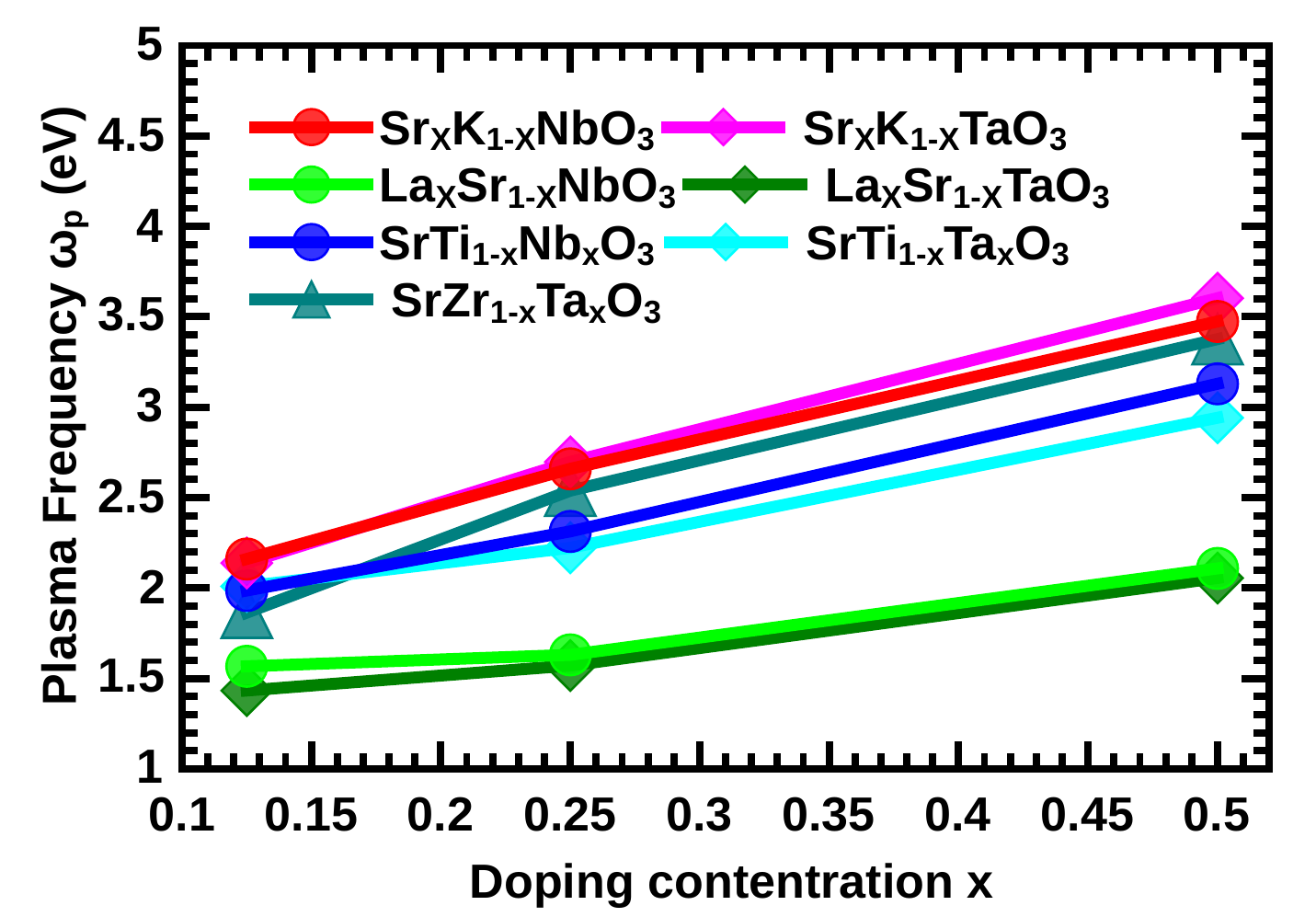}
\caption{DFT-derived plasma frequencies of $A$-site K-doped (hole doping) 4$d^1$ SrNbO$_3$ and 5$d^1$ SrTaO$_3$, and $B$-site doped SrTi$_{1-x}$Nb$_x$O$_3$, SrZr$_{1-x}$Ta$_x$O$_3$ and SrTi$_{1-x}$Ta$_x$O$_3$.}
\label{Fig12}
\end{figure*}

Besides hole doping of $d^1$ configurations, we also proposed another way of making $AB$O$_3$ perovskites transparent and conductive: electron doping towards $d^0$ configurations. As already mentioned,  this way is essentially the same as hole doping of $d^1$. Both aim at a $d^{0-1}$ electronic configuration. Considering the larger bandwidth and excellent optical transparency of 4$d$ and 5$d$ TMOs, we introduce electron doping to 4$d^0$ SrZrO$_3$ and 5$d^0$ SrHfO$_3$ by using La as dopant. Their optical properties, and $R$, $A$ are shown in Fig.~\ref{Fig11}(a,d). Similar trends are predicted here: as the electronic configurations change from 4$d^{0.5}$ (Sr$_{0.5}$La$_{0.5}$ZrO$_3$) and 5$d^{0.5}$ (Sr$_{0.5}$La$_{0.5}$HfO$_3$), to 4$d^{0.125}$ (Sr$_{0.875}$La$_{0.125}$ZrO$_3$) and 5$d^{0.125}$ (Sr$_{0.875}$La$_{0.125}$HfO$_3$), there is a remarkable reduction of $\omega_p$, and both $R$ and $A$ are close to zero for visible light (1.65-3.10\,eV). The shift of $R_{min}$ and $A_{min}$ can be explained by the reduced $\omega_p$ upon hole doping, as indicated by Fig.~\ref{Fig12}.


\clearpage

\subsection{Section 6: Additional DMFT results for the proposed materials}

\begin{figure*}[h]
\includegraphics[width=0.80\textwidth]{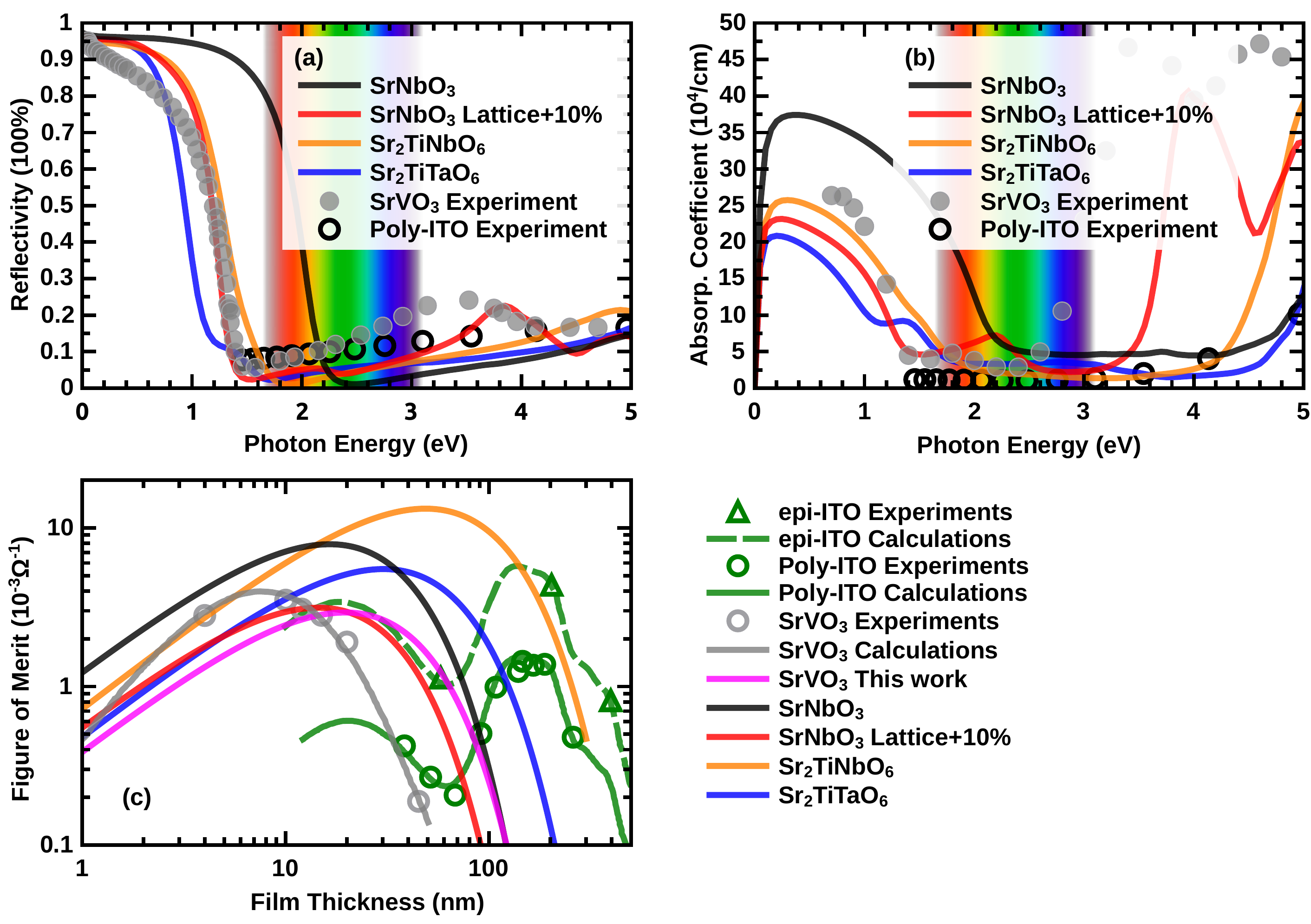}
\caption{Supplementary to Fig.~\ref{Fig3} in main text. Reflectivity $R$ (a) and absorption $A$ (b) of SrNbO$_3$ (DFT-relaxed), SrNbO$_3$ with expanding lattice (10\%), Sr$_2$TiNbO$_6$ and Sr$_2$TiTaO$_6$. As comparisons, experimental data of SrVO$_3$ \cite{PhysRevB.58.4384,zhang2016correlated} and poly(crystalline)-ITO are also shown \cite{polyITO}. (c) Figure of Merit of SrNbO$_3$ (DFT-relaxed), SrNbO$_3$ with expanding lattice (10\%), Sr$_2$TiNbO$_6$ and Sr$_2$TiTaO$_6$. The experimental results and computational curves of SrVO$_3$ films, poly(crystalline)-ITO films and epi(taxial)-ITO films,  are from Ref.\,\cite{ellmer2012past,polyITO,ohta2002surface,moyer2013highly,zhang2016correlated} as presented in Ref.\,\cite{zhang2016correlated}.}
\label{Fig13}
\end{figure*}

Besides the results shown in Fig.~\ref{Fig3} of the main text, we here show additionally  the DFT+DMFT (\textsc{Woptic}) results of SrNbO$_3$ (DFT-relaxed lattice) and SrNbO$_3$ with expanding lattice (+10\%), and another double perovskite Sr$_2$TiTaO$_6$, see Fig.~\ref{Fig13}.

Unlike SrVO$_3$, for which $R_{min}$ and $A_{min}$ are already above the red edge ($\omega\sim$ 1.65\,eV) of the visible region, SrNbO$_3$ at the bulk lattice constant exhibits strong reflection and absorption at $\omega\textless$2.4\,eV, indicating SrNbO$_3$ does not host good transparency at thicker films, unless it is deployed in thin films $\sim$10\,nm \cite{Park2020}. Lattice expansion effectively reduces its $\omega_p$ from 4.4\,eV (without lattice expansion) to 3.3\,eV (10\% expansion). This shifts  $R_{min}$ and $A_{min}$ below $\omega\sim$1.65\,eV, and most importantly, this does not reduce its $\sigma(0)$ too much, as shown in Table.~\ref{table1}.

The results of the FOM shown in Fig.~\ref{Fig3} of the main text are computed from $A$ and $R$ at the  wavelength 550\,nm ($\omega\sim$2.25\,eV). Here, we also compute $A_{average}$ and $R_{average}$ averaged over the range $\omega=$1.65-3.10\,eV; the results are shown in Table.~\ref{table1}. As one can see, for SrVO$_3$, Sr$_2$TiNbO$_6$ and Sr$_2$TiTaO$_6$,  $R_{average}$ and $A_{average}$ are basically equal to  $R_{550\,nm}$ and $A_{550\,nm}$; this is because these three materials host quite low and linear optical properties in the visible region.
However, for SrNbO$_3$ (without lattice expansion),  $R$ and $A$ at 1.65\,eV$\textless\omega
\textless$2.40\,eV are quite high, leading to larger  $R_{average}$ and $A_{average}$.


\begin{table*}[h]
\caption{Parameters for the calculation of the figure of merit (FOM).
$\sigma(0)$ is the zero frequency optical conductivity $\sigma$ from DMFT (without additional scattering) calculations at 300\,K.
$R_{550\,nm}$ and $A_{550\,nm}$ are the reflectivities and absorptions at wavelength 550\,nm ($\omega\sim$2.25\,eV).
$R_{average}$ and $A_{average}$ are the average reflectivities and absorptions in the interval $\omega=$1.65-3.10\,eV.
The units of $\sigma(0)$, $R$ and $A$ are 10$^{3} \Omega^{-1}$cm$^{-1}$, 100\% and 10$^{4}$cm$^{-1}$, respectively.
The DMFT $\sigma(0)$ of SrVO$_3$ is consistent with the experimental data of SrVO$_3$ films from 4\,nm to 45\,nm: 16.2-34.4 \cite{zhang2016correlated} (at 300\,K).
}
\begin{tabular}{ c|c|c|c|c|c }
\hline
\hline
 Materials & $\sigma(0)_{DMFT}$ & $R_{550\,nm}$ & $R_{average}$ &  $A_{550\,nm}$ & $A_{average}$   \\
\hline
SrVO$_3$  &   20.2  & 8.9\% & 10.1\% & 5.1 & 4.7  \\
SrNbO$_3$  &  28.4  & 3.9\% & 20.7\% & 6.1 & 9.0  \\
SrNbO$_3$ [lattice+10\%]    & 15.1 & 4.7\% & 5.5\% & 6.8 & 4.3  \\
Sr$_2$TiNbO$_6$   & 13.7 & 3.1\% & 4.2\% & 2.0 & 2.4  \\
Sr$_2$TiTaO$_6$   & 13.2 & 4.9\% & 5.0\% & 3.3 & 3.5  \\
\hline
\hline
\end{tabular}
\label{table1}
\end{table*}

\clearpage

\subsection{Section 7: Analytical continuation of the self-energy}

The basis for the analytic continuation is the Hilbert transform:

\begin{equation}
\label{eq:hilbert}
G(i\omega_n) = \int_{-\infty}^{\infty} d\nu \frac{A(\nu)}{i\omega_n - \nu} =: \int_{-\infty}^{\infty} d\nu K(i\omega_n, \nu) A(\nu),
\end{equation}

which holds for (impurity) Green's functions. The (impurity) self-energy $\Sigma$ is given by the Dyson equation:

\begin{equation}
\Sigma(i\omega_n) = [\mathcal{G}(i\omega_n)]^{-1} - [G(i\omega_n)]^{-1},
\end{equation}
where $\mathcal{G}$ is the non-interacting Green's function of the impurity, also called the Weiss field in the context of DMFT. It follows that the dynamical (i.e.~frequency-dependent) part of the self-energy
obeys Eq.~(\ref{eq:hilbert}) \cite{Luttinger61}.
In other words, the self-energy can be analytically continued by inversion of Eq.~(\ref{eq:hilbert})
after subtraction of the Hartree term:

\begin{equation}
\label{eq:hartree-se}
\Sigma(i\omega_n) - \Sigma_H = \int_{-\infty}^{\infty} d\nu K(i\omega_n, \nu) a(\nu),
\end{equation}

where  $a(\nu)$ is the \emph{spectrum of the self-energy}. It is a well-known fact that the above-mentioned inversion is not possible in a direct way, which has led to the development of several methods to get a spectral function that fulfills Eq.~(\ref{eq:hilbert}) \cite{Vidberg77,Bryan90,JarrellGubernatis96,Sandvik98,StochReg1,SpM}. Most common is the Maximum Entropy method (MEM), which is used also here in the implementation of Refs.\ \cite{Geffroy2019,kaufmannGithub}.

\begin{figure*}
  \centering
  \includegraphics[width=\textwidth]{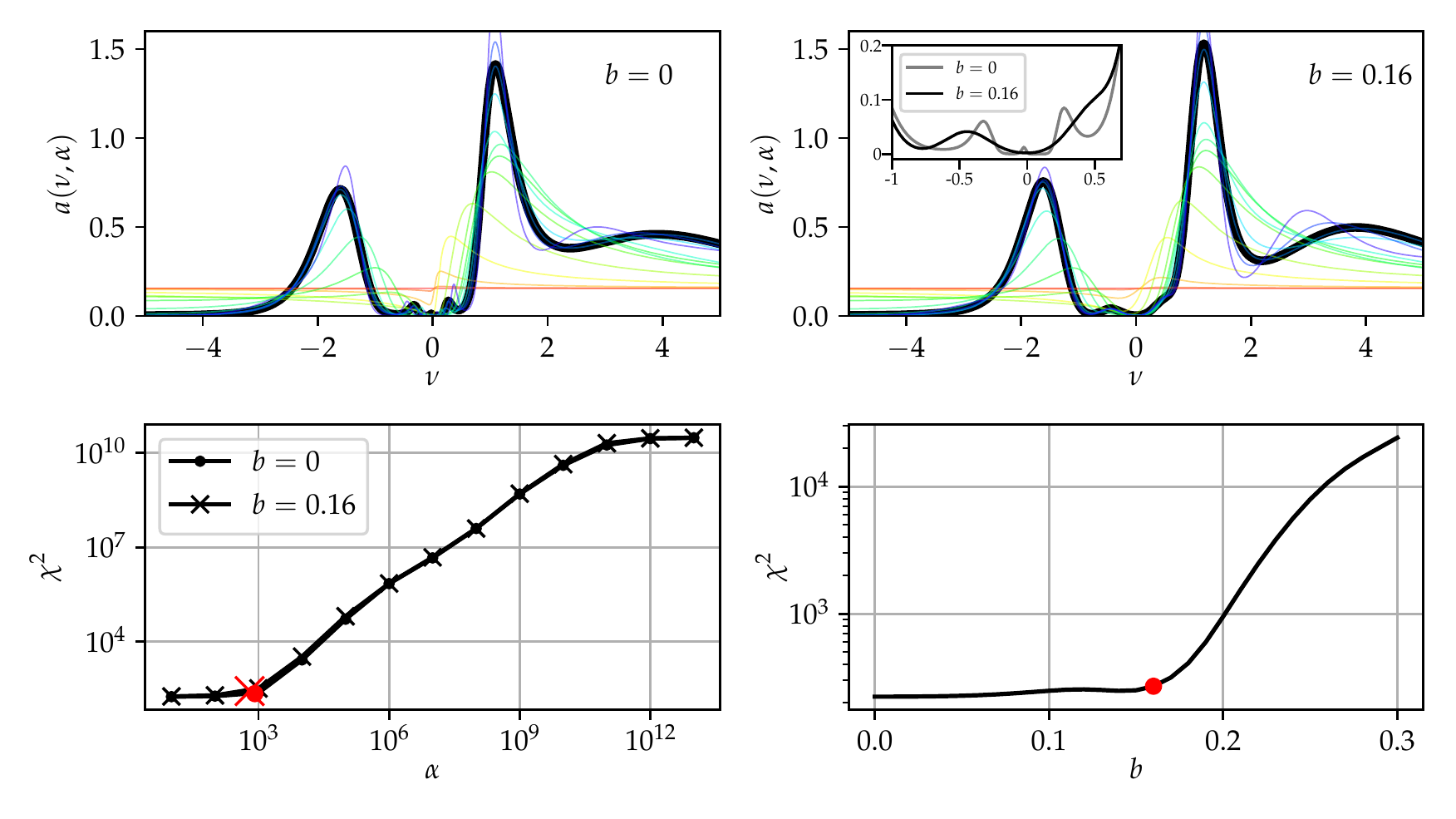}
  \caption{\label{Fig14}
    Analytic continuation of the DMFT self-energy of the V-$t_{2g}$ orbitals of SrVO$_3$, 
    which was obtained with symmetric improved estimators
    for extra precision \cite{Kaufmann2019}.
    \emph{Upper row}: Spectrum of the imaginary part of the self-energy spectrum $a(\omega)=- {\rm Im}\Sigma (\omega)/\pi$ 
    for SrVO$_3$ without [left] and with [right] preblur. The figures show optimization results
    for several different values of $\alpha$. 
    Red corresponds to $\alpha=10^{13}$, and as the color changes into blue, 
    the value is lowered by a factor of $10$ in every step. 
    The final result, at optimal $\alpha$, is drawn in black. 
    Clearly, for the highest values of $\alpha$, we recover the constant default model.
    \emph{Lower row}: Behavior of $\chi^2$ 
    as a function of $\alpha$ [left] and $b$ [right] 
    for the MEM analytical continuation of SrVO$_3$. 
    The values taken as optimal are highlighted in red.}
\end{figure*}

Not only the direct inversion of the kernel matrix $K(i\omega_n, \nu)$, but also
the minimization of the $\chi^2$-deviation $L[A] = \sum_n\big[G(i\omega_n)-\int d\nu K(i\omega_n, \nu)A(\nu)\big]^2/\sigma(i\omega_n)^2$
leads to unusable results. The same holds for $\Sigma(i\omega_n)$ and $a(\nu)$ which we actually use here. The MEM regularizes the optimization problem by adding an additional entropy term
$S[A] = -\int d\nu D(\nu) \mathrm{log} A(\nu) / D(\nu)$, with a default model $D(\nu)$. 
Thus, the optimization problem reads

\begin{equation}
  \label{eq:mem-opt}
  F[A] = L[A] - \alpha S[A],
\end{equation}

with the MEM hyperparameter $\alpha$ that balances the influence of the data and the default model.
Several methods for the determination of $\alpha$ have been proposed. The simplest and arguably most
elegant \cite{bergeron2016,Kraberger17} is described in the following.
We solve the optimization problem Eq.~(\ref{eq:mem-opt})  for several values of $\alpha$ on a logarithmic
scale, and then analyze $\mathrm{log}(L[A])$ as a function of $\mathrm{log}(\alpha)$.
Without doing any calculations, we anticipate the following behavior:
In the limit of infinite $\alpha$, $\mathrm{log}(L[A])$ will go to a constant value, 
namely $\mathrm{log}(L[D])$. As $\alpha$ goes to zero, also $L[A]$ will go to a (very small)
constant value, corresponding to the solution of the $\chi^2$ optimization without entropy term.
(Note that it can never be truly zero, since noise cannot be written in the form of Eq.~(\ref{eq:hilbert})
with positive spectrum.)
The optimal $\alpha$ is then taken to be close to where $L[A]$ reaches its small constant value.
For the case of SrVO$_3$, this is illustrated in the lower left panel of Fig.~\ref{Fig14}.
Further decreasing $\alpha$ is considered overfitting, since the quality of the fit does not improve,
although the spectrum $A$ undergoes considerable changes, which can be seen in the upper left panel of 
Fig.~\ref{Fig14} (for $b=0$). 

However, in the case of large noise $\sigma(i\omega_n)$ this procedure does not yet lead to satisfying results, 
which usually manifests itself in unphysical and sharp peaks in the spectrum. To tackle this well-known problem,
the \emph{preblur} variant of MEM has been introduced \cite{Skilling91,Kraberger17}. 
There, the spectral function is convoluted with a Gaussian (i.e.~blurred) 
before evaluating $L[A]$, and hence the blurred spectral function is the solution 
of the optimization problem. This essentially has the effect of only peaks 
with a certain minimal width being permitted in the spectrum.

Unfortunately, the preblur procedure introduces one more hyperparameter, 
namely the width $b$ of the Gaussian convolution kernel. 
Therefore it is necessary to perform the optimization for different values of $b$ 
and perform a similar analysis as in the search for the optimal $\alpha$. 
In the limit of $b=0$, standard MEM behavior is recovered and $L[A]$ 
should have a small value. Upon increasing $b$, also $L[A]$ will increase, 
since the search space for fit functions $A$ is gradually restricted to broader peaks. 
It however turns out that the increase of $L[A]$ is very small 
up to a certain value $b^\ast$, where it starts to rapidly increase. 
This behavior is illustrated in the lower right panel of Fig.~\ref{Fig14} . 
The value $b^\ast$ can be taken as optimal, since, apparently, 
the fit quality is not significantly decreased yet, 
although all peaks have a minimal width of $b^\ast$. 
Optimization results for several values of $\alpha$ at blur width 
$b^\ast$ are shown in the upper right panel of Fig.~\ref{Fig14}.

\begin{figure*}
  \centering
  \includegraphics[width=\textwidth]{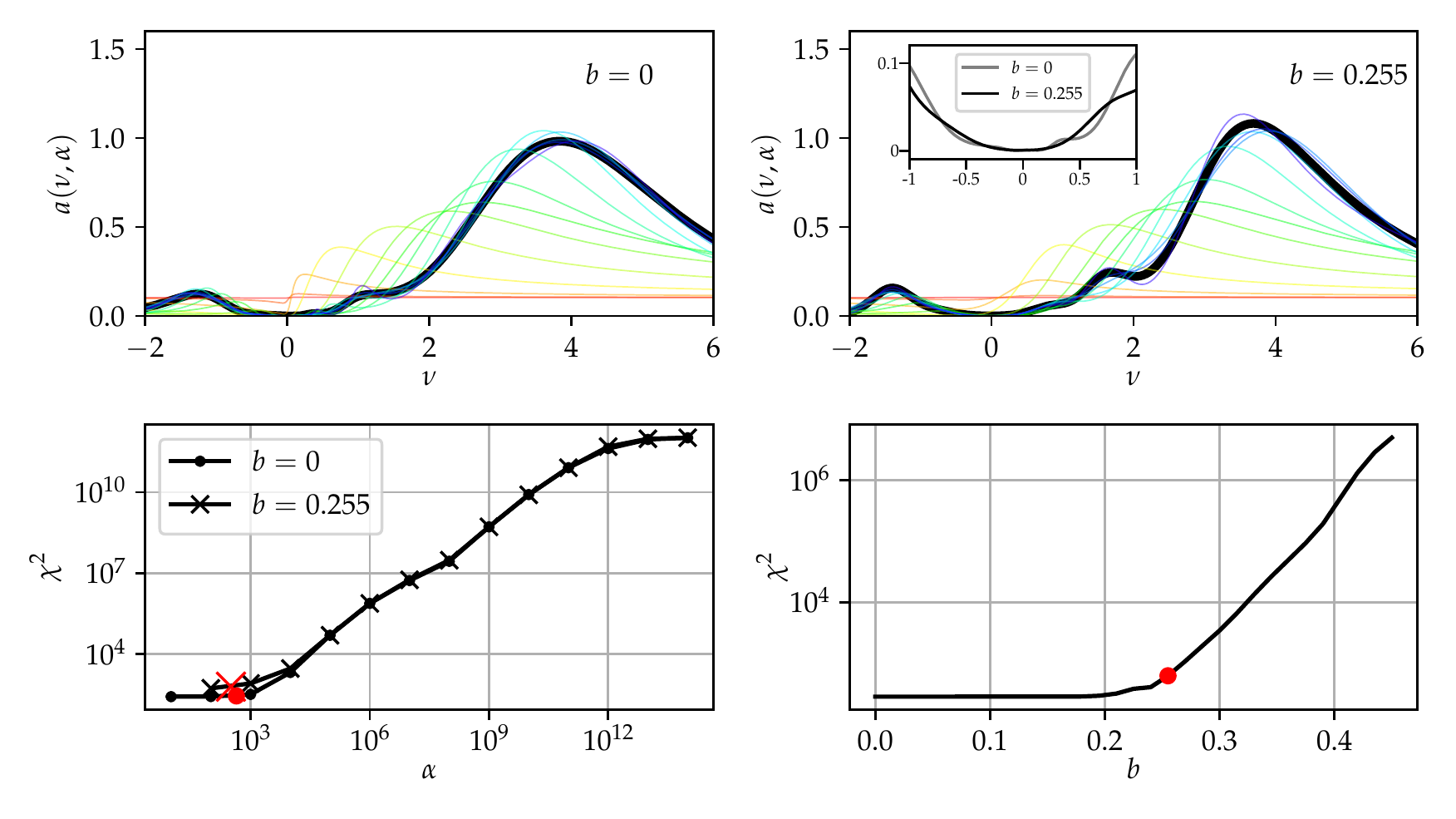}
  \caption{\label{Fig15}
    Analytic continuation of the DMFT self-energy of the Ti-$t_{2g}$-orbitals of Sr$_2$TiNbO$_6$.}
\end{figure*}

\begin{figure*}
  \centering
  \includegraphics[width=\textwidth]{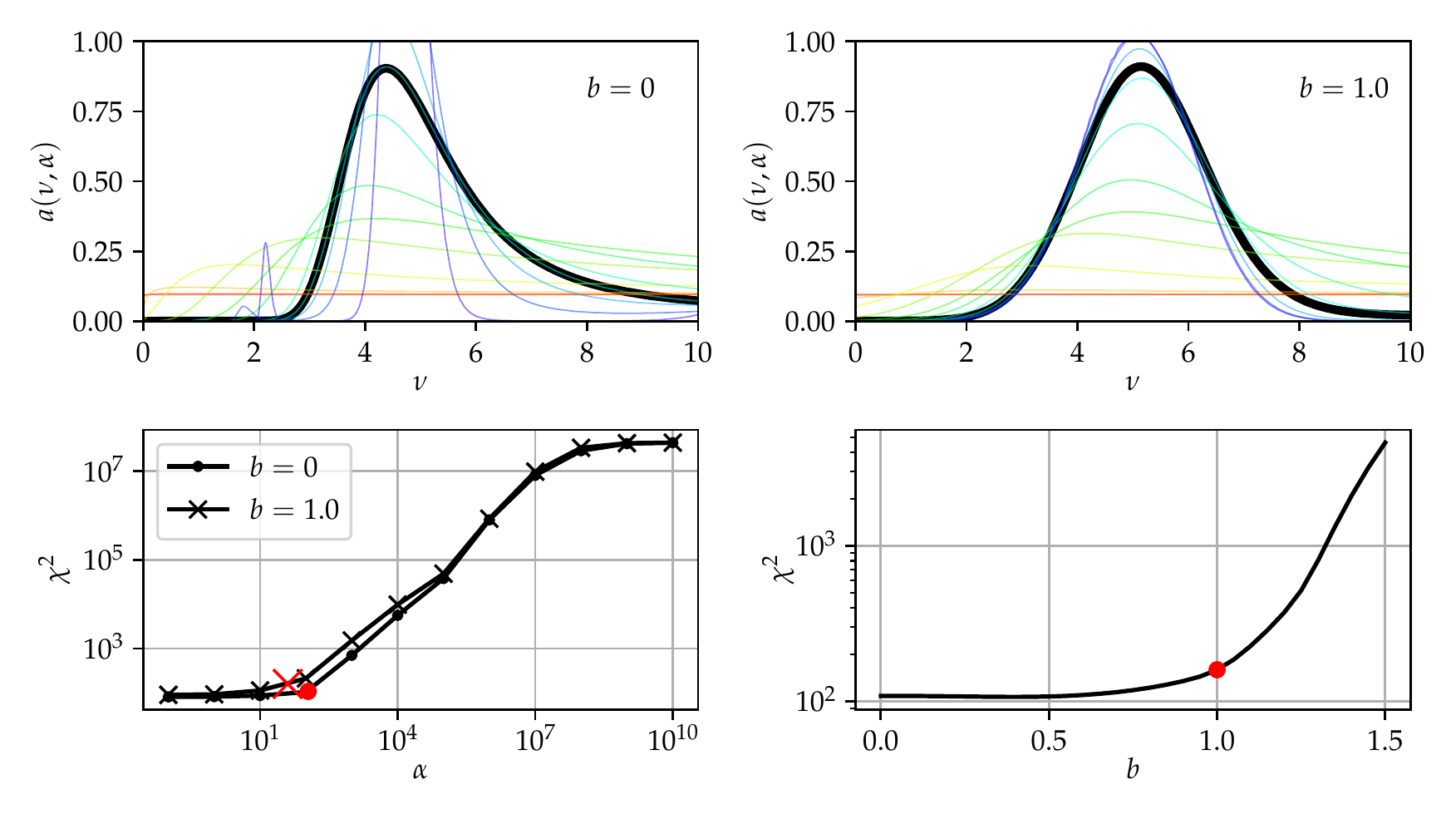}
  \caption{\label{Fig16}
    Analytic continuation of the DMFT self-energy of the Ti-$e_g$-orbitals of Sr$_2$TiNbO$_6$.}
\end{figure*}

\begin{figure*}
  \centering
  \includegraphics[width=\textwidth]{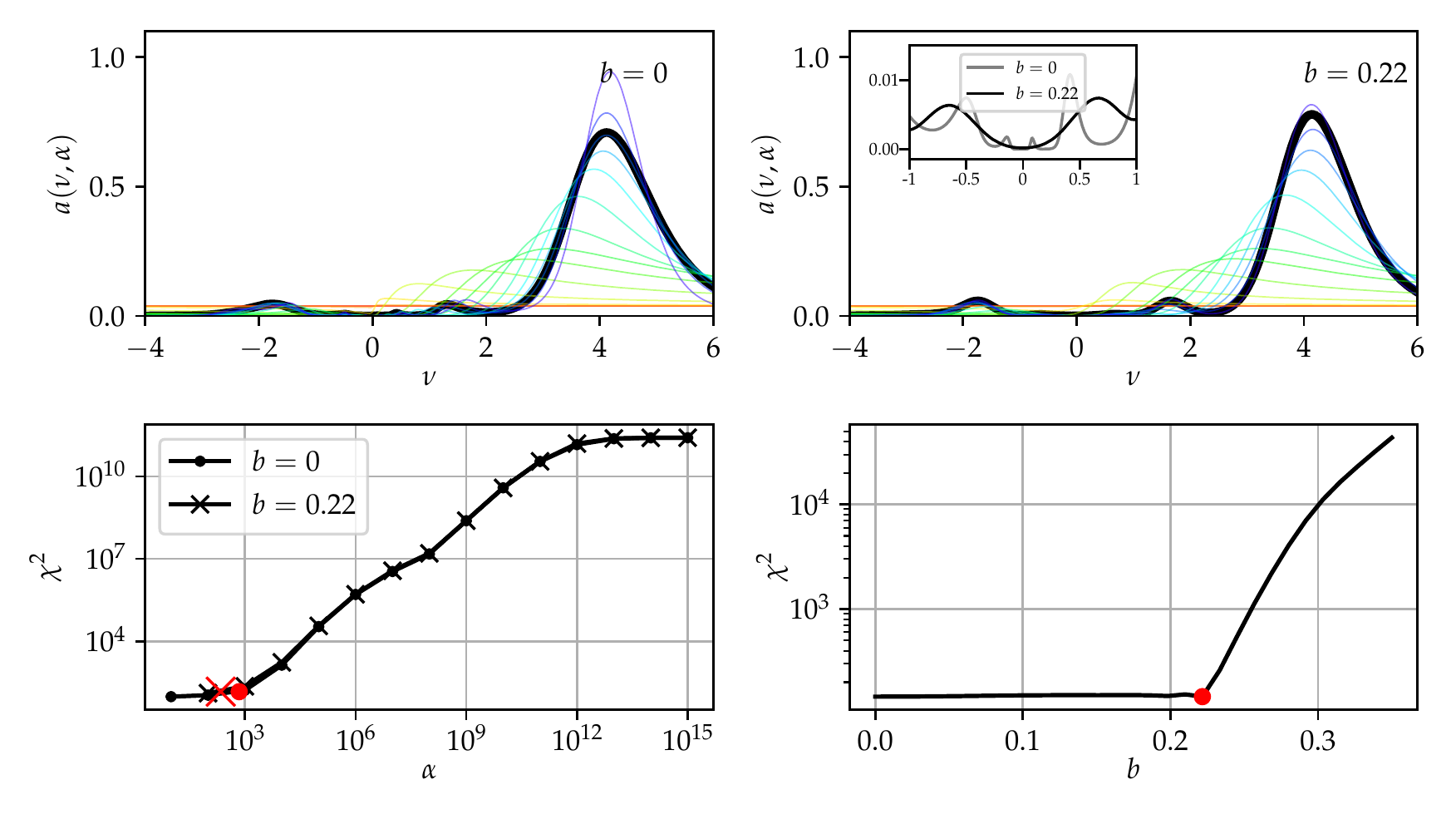}
  \caption{\label{Fig17}
    Analytic continuation of the DMFT self-energy of the Nb-$t_{2g}$-orbitals of Sr$_2$TiNbO$_6$.}
\end{figure*}

The same procedure is also applied for the analytic continuation
of Sr$_2$TiNbO$_6$ and illustrated in Figs.~\ref{Fig15}-\ref{Fig17}
for all inequivalent orbitals.

\begin{thebibliography}{70}%
\makeatletter
\providecommand \@ifxundefined [1]{%
 \@ifx{#1\undefined}
}%
\providecommand \@ifnum [1]{%
 \ifnum #1\expandafter \@firstoftwo
 \else \expandafter \@secondoftwo
 \fi
}%
\providecommand \@ifx [1]{%
 \ifx #1\expandafter \@firstoftwo
 \else \expandafter \@secondoftwo
 \fi
}%
\providecommand \natexlab [1]{#1}%
\providecommand \enquote  [1]{``#1''}%
\providecommand \bibnamefont  [1]{#1}%
\providecommand \bibfnamefont [1]{#1}%
\providecommand \citenamefont [1]{#1}%
\providecommand \href@noop [0]{\@secondoftwo}%
\providecommand \href [0]{\begingroup \@sanitize@url \@href}%
\providecommand \@href[1]{\@@startlink{#1}\@@href}%
\providecommand \@@href[1]{\endgroup#1\@@endlink}%
\providecommand \@sanitize@url [0]{\catcode `\\12\catcode `\$12\catcode
  `\&12\catcode `\#12\catcode `\^12\catcode `\_12\catcode `\%12\relax}%
\providecommand \@@startlink[1]{}%
\providecommand \@@endlink[0]{}%
\providecommand \url  [0]{\begingroup\@sanitize@url \@url }%
\providecommand \@url [1]{\endgroup\@href {#1}{\urlprefix }}%
\providecommand \urlprefix  [0]{URL }%
\providecommand \Eprint [0]{\href }%
\providecommand \doibase [0]{http://dx.doi.org/}%
\providecommand \selectlanguage [0]{\@gobble}%
\providecommand \bibinfo  [0]{\@secondoftwo}%
\providecommand \bibfield  [0]{\@secondoftwo}%
\providecommand \translation [1]{[#1]}%
\providecommand \BibitemOpen [0]{}%
\providecommand \bibitemStop [0]{}%
\providecommand \bibitemNoStop [0]{.\EOS\space}%
\providecommand \EOS [0]{\spacefactor3000\relax}%
\providecommand \BibitemShut  [1]{\csname bibitem#1\endcsname}%
\let\auto@bib@innerbib\@empty
\bibitem [{\citenamefont {Ginley}(2010)}]{Ginley2010}%
  \BibitemOpen
  \bibfield  {author} {\bibinfo {author} {\bibfnamefont {D.~S.}\ \bibnamefont
  {Ginley}},\ }\href {\doibase https://doi.org/10.1007/978-1-4419-1638-9}
  {\emph {\bibinfo {title} {Handbook of Transparent Conductors}}}\ (\bibinfo
  {publisher} {Springer, Boston, MA},\ \bibinfo {year} {2010})\BibitemShut
  {NoStop}%
\bibitem [{\citenamefont {Ginley}\ and\ \citenamefont
  {Bright}(2000)}]{ginley2000transparent}%
  \BibitemOpen
  \bibfield  {author} {\bibinfo {author} {\bibfnamefont {D.~S.}\ \bibnamefont
  {Ginley}}\ and\ \bibinfo {author} {\bibfnamefont {C.}~\bibnamefont
  {Bright}},\ }\href@noop {} {\bibfield  {journal} {\bibinfo  {journal} {MRS
  bulletin}\ }\textbf {\bibinfo {volume} {25}},\ \bibinfo {pages} {15}
  (\bibinfo {year} {2000})}\BibitemShut {NoStop}%
\bibitem [{Note1()}]{Note1}%
  \BibitemOpen
  \bibinfo {note} {Amporphous graphene has also been proposed as a transparent
  conductor \cite {PhysRevB.84.205414}}\BibitemShut {NoStop}%
\bibitem [{\citenamefont {Zhang}\ \emph {et~al.}(2016)\citenamefont {Zhang},
  \citenamefont {Zhou}, \citenamefont {Guo}, \citenamefont {Zhao},
  \citenamefont {Barnes}, \citenamefont {Zhang}, \citenamefont {Eaton},
  \citenamefont {Zheng}, \citenamefont {Brahlek}, \citenamefont {Haneef} \emph
  {et~al.}}]{zhang2016correlated}%
  \BibitemOpen
  \bibfield  {author} {\bibinfo {author} {\bibfnamefont {L.}~\bibnamefont
  {Zhang}}, \bibinfo {author} {\bibfnamefont {Y.}~\bibnamefont {Zhou}},
  \bibinfo {author} {\bibfnamefont {L.}~\bibnamefont {Guo}}, \bibinfo {author}
  {\bibfnamefont {W.}~\bibnamefont {Zhao}}, \bibinfo {author} {\bibfnamefont
  {A.}~\bibnamefont {Barnes}}, \bibinfo {author} {\bibfnamefont {H.-T.}\
  \bibnamefont {Zhang}}, \bibinfo {author} {\bibfnamefont {C.}~\bibnamefont
  {Eaton}}, \bibinfo {author} {\bibfnamefont {Y.}~\bibnamefont {Zheng}},
  \bibinfo {author} {\bibfnamefont {M.}~\bibnamefont {Brahlek}}, \bibinfo
  {author} {\bibfnamefont {H.~F.}\ \bibnamefont {Haneef}},  \emph {et~al.},\
  }\href@noop {} {\bibfield  {journal} {\bibinfo  {journal} {Nature materials}\
  }\textbf {\bibinfo {volume} {15}},\ \bibinfo {pages} {204} (\bibinfo {year}
  {2016})}\BibitemShut {NoStop}%
\bibitem [{\citenamefont {Park}\ \emph {et~al.}(2020)\citenamefont {Park},
  \citenamefont {Roth}, \citenamefont {Oka}, \citenamefont {Hirose},
  \citenamefont {Hasegawa}, \citenamefont {Paul}, \citenamefont {Pogrebnyakov},
  \citenamefont {Gopalan}, \citenamefont {Birol},\ and\ \citenamefont
  {Engel-Herbert}}]{Park2020}%
  \BibitemOpen
  \bibfield  {author} {\bibinfo {author} {\bibfnamefont {Y.}~\bibnamefont
  {Park}}, \bibinfo {author} {\bibfnamefont {J.}~\bibnamefont {Roth}}, \bibinfo
  {author} {\bibfnamefont {D.}~\bibnamefont {Oka}}, \bibinfo {author}
  {\bibfnamefont {Y.}~\bibnamefont {Hirose}}, \bibinfo {author} {\bibfnamefont
  {T.}~\bibnamefont {Hasegawa}}, \bibinfo {author} {\bibfnamefont
  {A.}~\bibnamefont {Paul}}, \bibinfo {author} {\bibfnamefont {A.}~\bibnamefont
  {Pogrebnyakov}}, \bibinfo {author} {\bibfnamefont {V.}~\bibnamefont
  {Gopalan}}, \bibinfo {author} {\bibfnamefont {T.}~\bibnamefont {Birol}}, \
  and\ \bibinfo {author} {\bibfnamefont {R.}~\bibnamefont {Engel-Herbert}},\
  }\href {\doibase https://doi.org/10.1038/s42005-020-0372-9} {\bibfield
  {journal} {\bibinfo  {journal} {Commun. Phys.}\ }\textbf {\bibinfo {volume}
  {3}},\ \bibinfo {pages} {102} (\bibinfo {year} {2020})}\BibitemShut {NoStop}%
\bibitem [{\citenamefont {Hohenberg}\ and\ \citenamefont
  {Kohn}(1964)}]{PhysRev.136.B864}%
  \BibitemOpen
  \bibfield  {author} {\bibinfo {author} {\bibfnamefont {P.}~\bibnamefont
  {Hohenberg}}\ and\ \bibinfo {author} {\bibfnamefont {W.}~\bibnamefont
  {Kohn}},\ }\href {\doibase 10.1103/PhysRev.136.B864} {\bibfield  {journal}
  {\bibinfo  {journal} {Phys. Rev.}\ }\textbf {\bibinfo {volume} {136}},\
  \bibinfo {pages} {B864} (\bibinfo {year} {1964})}\BibitemShut {NoStop}%
\bibitem [{\citenamefont {Georges}\ \emph {et~al.}(1996)\citenamefont
  {Georges}, \citenamefont {Kotliar}, \citenamefont {Krauth},\ and\
  \citenamefont {Rozenberg}}]{RevModPhys.68.13}%
  \BibitemOpen
  \bibfield  {author} {\bibinfo {author} {\bibfnamefont {A.}~\bibnamefont
  {Georges}}, \bibinfo {author} {\bibfnamefont {G.}~\bibnamefont {Kotliar}},
  \bibinfo {author} {\bibfnamefont {W.}~\bibnamefont {Krauth}}, \ and\ \bibinfo
  {author} {\bibfnamefont {M.~J.}\ \bibnamefont {Rozenberg}},\ }\href {\doibase
  10.1103/RevModPhys.68.13} {\bibfield  {journal} {\bibinfo  {journal} {Rev.
  Mod. Phys.}\ }\textbf {\bibinfo {volume} {68}},\ \bibinfo {pages} {13}
  (\bibinfo {year} {1996})}\BibitemShut {NoStop}%
\bibitem [{\citenamefont {Kotliar}\ and\ \citenamefont
  {Vollhardt}(2004)}]{kotliar2004strongly}%
  \BibitemOpen
  \bibfield  {author} {\bibinfo {author} {\bibfnamefont {G.}~\bibnamefont
  {Kotliar}}\ and\ \bibinfo {author} {\bibfnamefont {D.}~\bibnamefont
  {Vollhardt}},\ }\href@noop {} {\bibfield  {journal} {\bibinfo  {journal}
  {Physics Today}\ }\textbf {\bibinfo {volume} {57}},\ \bibinfo {pages} {53}
  (\bibinfo {year} {2004})}\BibitemShut {NoStop}%
\bibitem [{\citenamefont {Metzner}\ and\ \citenamefont
  {Vollhardt}(1989)}]{PhysRevLett.62.324}%
  \BibitemOpen
  \bibfield  {author} {\bibinfo {author} {\bibfnamefont {W.}~\bibnamefont
  {Metzner}}\ and\ \bibinfo {author} {\bibfnamefont {D.}~\bibnamefont
  {Vollhardt}},\ }\href {\doibase 10.1103/PhysRevLett.62.324} {\bibfield
  {journal} {\bibinfo  {journal} {Phys. Rev. Lett.}\ }\textbf {\bibinfo
  {volume} {62}},\ \bibinfo {pages} {324} (\bibinfo {year} {1989})}\BibitemShut
  {NoStop}%
\bibitem [{\citenamefont {Held}(2007)}]{held2007electronic}%
  \BibitemOpen
  \bibfield  {author} {\bibinfo {author} {\bibfnamefont {K.}~\bibnamefont
  {Held}},\ }\href@noop {} {\bibfield  {journal} {\bibinfo  {journal} {Advances
  in physics}\ }\textbf {\bibinfo {volume} {56}},\ \bibinfo {pages} {829}
  (\bibinfo {year} {2007})}\BibitemShut {NoStop}%
\bibitem [{\citenamefont {Paul}\ and\ \citenamefont {Birol}(2019)}]{Paul2019}%
  \BibitemOpen
  \bibfield  {author} {\bibinfo {author} {\bibfnamefont {A.}~\bibnamefont
  {Paul}}\ and\ \bibinfo {author} {\bibfnamefont {T.}~\bibnamefont {Birol}},\
  }\href {\doibase 10.1103/PhysRevMaterials.3.085001} {\bibfield  {journal}
  {\bibinfo  {journal} {Phys. Rev. Materials}\ }\textbf {\bibinfo {volume}
  {3}},\ \bibinfo {pages} {085001} (\bibinfo {year} {2019})}\BibitemShut
  {NoStop}%
\bibitem [{\citenamefont {Paul}\ and\ \citenamefont
  {Birol}(2020)}]{PhysRevResearch.2.033156}%
  \BibitemOpen
  \bibfield  {author} {\bibinfo {author} {\bibfnamefont {A.}~\bibnamefont
  {Paul}}\ and\ \bibinfo {author} {\bibfnamefont {T.}~\bibnamefont {Birol}},\
  }\href {\doibase 10.1103/PhysRevResearch.2.033156} {\bibfield  {journal}
  {\bibinfo  {journal} {Phys. Rev. Research}\ }\textbf {\bibinfo {volume}
  {2}},\ \bibinfo {pages} {033156} (\bibinfo {year} {2020})}\BibitemShut
  {NoStop}%
\bibitem [{\citenamefont {Makino}\ \emph {et~al.}(1998)\citenamefont {Makino},
  \citenamefont {Inoue}, \citenamefont {Rozenberg}, \citenamefont {Hase},
  \citenamefont {Aiura},\ and\ \citenamefont {Onari}}]{PhysRevB.58.4384}%
  \BibitemOpen
  \bibfield  {author} {\bibinfo {author} {\bibfnamefont {H.}~\bibnamefont
  {Makino}}, \bibinfo {author} {\bibfnamefont {I.~H.}\ \bibnamefont {Inoue}},
  \bibinfo {author} {\bibfnamefont {M.~J.}\ \bibnamefont {Rozenberg}}, \bibinfo
  {author} {\bibfnamefont {I.}~\bibnamefont {Hase}}, \bibinfo {author}
  {\bibfnamefont {Y.}~\bibnamefont {Aiura}}, \ and\ \bibinfo {author}
  {\bibfnamefont {S.}~\bibnamefont {Onari}},\ }\href {\doibase
  10.1103/PhysRevB.58.4384} {\bibfield  {journal} {\bibinfo  {journal} {Phys.
  Rev. B}\ }\textbf {\bibinfo {volume} {58}},\ \bibinfo {pages} {4384}
  (\bibinfo {year} {1998})}\BibitemShut {NoStop}%
\bibitem [{\citenamefont {Blaha}\ \emph {et~al.}(2001)\citenamefont {Blaha},
  \citenamefont {Schwarz}, \citenamefont {Madsen}, \citenamefont {Kvasnicka},\
  and\ \citenamefont {Luitz}}]{blaha2001wien2k}%
  \BibitemOpen
  \bibfield  {author} {\bibinfo {author} {\bibfnamefont {P.}~\bibnamefont
  {Blaha}}, \bibinfo {author} {\bibfnamefont {K.}~\bibnamefont {Schwarz}},
  \bibinfo {author} {\bibfnamefont {G.}~\bibnamefont {Madsen}}, \bibinfo
  {author} {\bibfnamefont {D.}~\bibnamefont {Kvasnicka}}, \ and\ \bibinfo
  {author} {\bibfnamefont {J.}~\bibnamefont {Luitz}},\ }\href@noop {}
  {\bibfield  {journal} {\bibinfo  {journal} {An augmented plane wave+ local
  orbitals program for calculating crystal properties}\ } (\bibinfo {year}
  {2001})}\BibitemShut {NoStop}%
\bibitem [{\citenamefont {Blaha}\ \emph {et~al.}(2020)\citenamefont {Blaha},
  \citenamefont {Schwarz}, \citenamefont {Tran}, \citenamefont {Laskowski},
  \citenamefont {Madsen},\ and\ \citenamefont {Marks}}]{wien2k2020}%
  \BibitemOpen
  \bibfield  {author} {\bibinfo {author} {\bibfnamefont {P.}~\bibnamefont
  {Blaha}}, \bibinfo {author} {\bibfnamefont {K.}~\bibnamefont {Schwarz}},
  \bibinfo {author} {\bibfnamefont {F.}~\bibnamefont {Tran}}, \bibinfo {author}
  {\bibfnamefont {R.}~\bibnamefont {Laskowski}}, \bibinfo {author}
  {\bibfnamefont {G.~K.}\ \bibnamefont {Madsen}}, \ and\ \bibinfo {author}
  {\bibfnamefont {L.~D.}\ \bibnamefont {Marks}},\ }\href {\doibase
  10.1063/1.5143061} {\bibfield  {journal} {\bibinfo  {journal} {The Journal of
  Chemical Physics}\ }\textbf {\bibinfo {volume} {152}},\ \bibinfo {pages}
  {074101} (\bibinfo {year} {2020})},\ \Eprint
  {http://arxiv.org/abs/https://doi.org/10.1063/1.5143061}
  {https://doi.org/10.1063/1.5143061} \BibitemShut {NoStop}%
\bibitem [{\citenamefont {Perdew}\ \emph {et~al.}(1996)\citenamefont {Perdew},
  \citenamefont {Burke},\ and\ \citenamefont
  {Ernzerhof}}]{PhysRevLett.77.3865}%
  \BibitemOpen
  \bibfield  {author} {\bibinfo {author} {\bibfnamefont {J.~P.}\ \bibnamefont
  {Perdew}}, \bibinfo {author} {\bibfnamefont {K.}~\bibnamefont {Burke}}, \
  and\ \bibinfo {author} {\bibfnamefont {M.}~\bibnamefont {Ernzerhof}},\ }\href
  {\doibase 10.1103/PhysRevLett.77.3865} {\bibfield  {journal} {\bibinfo
  {journal} {Phys. Rev. Lett.}\ }\textbf {\bibinfo {volume} {77}},\ \bibinfo
  {pages} {3865} (\bibinfo {year} {1996})}\BibitemShut {NoStop}%
\bibitem [{\citenamefont {Tran}\ and\ \citenamefont
  {Blaha}(2009)}]{PhysRevLett.102.226401}%
  \BibitemOpen
  \bibfield  {author} {\bibinfo {author} {\bibfnamefont {F.}~\bibnamefont
  {Tran}}\ and\ \bibinfo {author} {\bibfnamefont {P.}~\bibnamefont {Blaha}},\
  }\href {\doibase 10.1103/PhysRevLett.102.226401} {\bibfield  {journal}
  {\bibinfo  {journal} {Phys. Rev. Lett.}\ }\textbf {\bibinfo {volume} {102}},\
  \bibinfo {pages} {226401} (\bibinfo {year} {2009})}\BibitemShut {NoStop}%
\bibitem [{\citenamefont {Bellaiche}\ and\ \citenamefont
  {Vanderbilt}(2000)}]{PhysRevB.61.7877}%
  \BibitemOpen
  \bibfield  {author} {\bibinfo {author} {\bibfnamefont {L.}~\bibnamefont
  {Bellaiche}}\ and\ \bibinfo {author} {\bibfnamefont {D.}~\bibnamefont
  {Vanderbilt}},\ }\href {\doibase 10.1103/PhysRevB.61.7877} {\bibfield
  {journal} {\bibinfo  {journal} {Phys. Rev. B}\ }\textbf {\bibinfo {volume}
  {61}},\ \bibinfo {pages} {7877} (\bibinfo {year} {2000})}\BibitemShut
  {NoStop}%
\bibitem [{Note2()}]{Note2}%
  \BibitemOpen
  \bibinfo {note} {For Sr$_2$TiNbO$_6$, the Ti-$e_g$ bands are also projected
  onto the Wannier orbitals because they are energetically comparable to the
  Nb-$t_{2g}$ bands. However, DMFT predicts the Ti-$e_g$ orbitals to be
  unoccupied.}\BibitemShut {Stop}%
\bibitem [{\citenamefont {Wannier}(1937)}]{PhysRev.52.191}%
  \BibitemOpen
  \bibfield  {author} {\bibinfo {author} {\bibfnamefont {G.~H.}\ \bibnamefont
  {Wannier}},\ }\href {\doibase 10.1103/PhysRev.52.191} {\bibfield  {journal}
  {\bibinfo  {journal} {Phys. Rev.}\ }\textbf {\bibinfo {volume} {52}},\
  \bibinfo {pages} {191} (\bibinfo {year} {1937})}\BibitemShut {NoStop}%
\bibitem [{\citenamefont {Marzari}\ \emph {et~al.}(2012)\citenamefont
  {Marzari}, \citenamefont {Mostofi}, \citenamefont {Yates}, \citenamefont
  {Souza},\ and\ \citenamefont {Vanderbilt}}]{RevModPhys.84.1419}%
  \BibitemOpen
  \bibfield  {author} {\bibinfo {author} {\bibfnamefont {N.}~\bibnamefont
  {Marzari}}, \bibinfo {author} {\bibfnamefont {A.~A.}\ \bibnamefont
  {Mostofi}}, \bibinfo {author} {\bibfnamefont {J.~R.}\ \bibnamefont {Yates}},
  \bibinfo {author} {\bibfnamefont {I.}~\bibnamefont {Souza}}, \ and\ \bibinfo
  {author} {\bibfnamefont {D.}~\bibnamefont {Vanderbilt}},\ }\href {\doibase
  10.1103/RevModPhys.84.1419} {\bibfield  {journal} {\bibinfo  {journal} {Rev.
  Mod. Phys.}\ }\textbf {\bibinfo {volume} {84}},\ \bibinfo {pages} {1419}
  (\bibinfo {year} {2012})}\BibitemShut {NoStop}%
\bibitem [{\citenamefont {Mostofi}\ \emph {et~al.}(2008)\citenamefont
  {Mostofi}, \citenamefont {Yates}, \citenamefont {Lee}, \citenamefont {Souza},
  \citenamefont {Vanderbilt},\ and\ \citenamefont
  {Marzari}}]{mostofi2008wannier90}%
  \BibitemOpen
  \bibfield  {author} {\bibinfo {author} {\bibfnamefont {A.~A.}\ \bibnamefont
  {Mostofi}}, \bibinfo {author} {\bibfnamefont {J.~R.}\ \bibnamefont {Yates}},
  \bibinfo {author} {\bibfnamefont {Y.-S.}\ \bibnamefont {Lee}}, \bibinfo
  {author} {\bibfnamefont {I.}~\bibnamefont {Souza}}, \bibinfo {author}
  {\bibfnamefont {D.}~\bibnamefont {Vanderbilt}}, \ and\ \bibinfo {author}
  {\bibfnamefont {N.}~\bibnamefont {Marzari}},\ }\href@noop {} {\bibfield
  {journal} {\bibinfo  {journal} {Computer physics communications}\ }\textbf
  {\bibinfo {volume} {178}},\ \bibinfo {pages} {685} (\bibinfo {year}
  {2008})}\BibitemShut {NoStop}%
\bibitem [{\citenamefont {Kune{\v{s}}}\ \emph {et~al.}(2010)\citenamefont
  {Kune{\v{s}}}, \citenamefont {Arita}, \citenamefont {Wissgott}, \citenamefont
  {Toschi}, \citenamefont {Ikeda},\ and\ \citenamefont
  {Held}}]{kunevs2010wien2wannier}%
  \BibitemOpen
  \bibfield  {author} {\bibinfo {author} {\bibfnamefont {J.}~\bibnamefont
  {Kune{\v{s}}}}, \bibinfo {author} {\bibfnamefont {R.}~\bibnamefont {Arita}},
  \bibinfo {author} {\bibfnamefont {P.}~\bibnamefont {Wissgott}}, \bibinfo
  {author} {\bibfnamefont {A.}~\bibnamefont {Toschi}}, \bibinfo {author}
  {\bibfnamefont {H.}~\bibnamefont {Ikeda}}, \ and\ \bibinfo {author}
  {\bibfnamefont {K.}~\bibnamefont {Held}},\ }\href@noop {} {\bibfield
  {journal} {\bibinfo  {journal} {Computer Physics Communications}\ }\textbf
  {\bibinfo {volume} {181}},\ \bibinfo {pages} {1888} (\bibinfo {year}
  {2010})}\BibitemShut {NoStop}%
\bibitem [{\citenamefont {Nekrasov}\ \emph
  {et~al.}(2006{\natexlab{a}})\citenamefont {Nekrasov}, \citenamefont {Held},
  \citenamefont {Keller}, \citenamefont {Kondakov}, \citenamefont {Pruschke},
  \citenamefont {Kollar}, \citenamefont {Andersen}, \citenamefont {Anisimov},\
  and\ \citenamefont {Vollhardt}}]{PhysRevB.73.155112}%
  \BibitemOpen
  \bibfield  {author} {\bibinfo {author} {\bibfnamefont {I.~A.}\ \bibnamefont
  {Nekrasov}}, \bibinfo {author} {\bibfnamefont {K.}~\bibnamefont {Held}},
  \bibinfo {author} {\bibfnamefont {G.}~\bibnamefont {Keller}}, \bibinfo
  {author} {\bibfnamefont {D.~E.}\ \bibnamefont {Kondakov}}, \bibinfo {author}
  {\bibfnamefont {T.}~\bibnamefont {Pruschke}}, \bibinfo {author}
  {\bibfnamefont {M.}~\bibnamefont {Kollar}}, \bibinfo {author} {\bibfnamefont
  {O.~K.}\ \bibnamefont {Andersen}}, \bibinfo {author} {\bibfnamefont {V.~I.}\
  \bibnamefont {Anisimov}}, \ and\ \bibinfo {author} {\bibfnamefont
  {D.}~\bibnamefont {Vollhardt}},\ }\href {\doibase 10.1103/PhysRevB.73.155112}
  {\bibfield  {journal} {\bibinfo  {journal} {Phys. Rev. B}\ }\textbf {\bibinfo
  {volume} {73}},\ \bibinfo {pages} {155112} (\bibinfo {year}
  {2006}{\natexlab{a}})}\BibitemShut {NoStop}%
\bibitem [{\citenamefont {Si}\ \emph {et~al.}(2017)\citenamefont {Si},
  \citenamefont {Janson}, \citenamefont {Li}, \citenamefont {Zhong},
  \citenamefont {Liao}, \citenamefont {Koster},\ and\ \citenamefont
  {Held}}]{PhysRevLett.119.026402}%
  \BibitemOpen
  \bibfield  {author} {\bibinfo {author} {\bibfnamefont {L.}~\bibnamefont
  {Si}}, \bibinfo {author} {\bibfnamefont {O.}~\bibnamefont {Janson}}, \bibinfo
  {author} {\bibfnamefont {G.}~\bibnamefont {Li}}, \bibinfo {author}
  {\bibfnamefont {Z.}~\bibnamefont {Zhong}}, \bibinfo {author} {\bibfnamefont
  {Z.}~\bibnamefont {Liao}}, \bibinfo {author} {\bibfnamefont {G.}~\bibnamefont
  {Koster}}, \ and\ \bibinfo {author} {\bibfnamefont {K.}~\bibnamefont
  {Held}},\ }\href {\doibase 10.1103/PhysRevLett.119.026402} {\bibfield
  {journal} {\bibinfo  {journal} {Phys. Rev. Lett.}\ }\textbf {\bibinfo
  {volume} {119}},\ \bibinfo {pages} {026402} (\bibinfo {year}
  {2017})}\BibitemShut {NoStop}%
\bibitem [{\citenamefont {Okamoto}\ \emph {et~al.}(2014)\citenamefont
  {Okamoto}, \citenamefont {Zhu}, \citenamefont {Nomura}, \citenamefont
  {Arita}, \citenamefont {Xiao},\ and\ \citenamefont
  {Nagaosa}}]{PhysRevB.89.195121}%
  \BibitemOpen
  \bibfield  {author} {\bibinfo {author} {\bibfnamefont {S.}~\bibnamefont
  {Okamoto}}, \bibinfo {author} {\bibfnamefont {W.}~\bibnamefont {Zhu}},
  \bibinfo {author} {\bibfnamefont {Y.}~\bibnamefont {Nomura}}, \bibinfo
  {author} {\bibfnamefont {R.}~\bibnamefont {Arita}}, \bibinfo {author}
  {\bibfnamefont {D.}~\bibnamefont {Xiao}}, \ and\ \bibinfo {author}
  {\bibfnamefont {N.}~\bibnamefont {Nagaosa}},\ }\href {\doibase
  10.1103/PhysRevB.89.195121} {\bibfield  {journal} {\bibinfo  {journal} {Phys.
  Rev. B}\ }\textbf {\bibinfo {volume} {89}},\ \bibinfo {pages} {195121}
  (\bibinfo {year} {2014})}\BibitemShut {NoStop}%
\bibitem [{\citenamefont {Gull}\ \emph {et~al.}(2011)\citenamefont {Gull},
  \citenamefont {Millis}, \citenamefont {Lichtenstein}, \citenamefont
  {Rubtsov}, \citenamefont {Troyer},\ and\ \citenamefont
  {Werner}}]{RevModPhys.83.349}%
  \BibitemOpen
  \bibfield  {author} {\bibinfo {author} {\bibfnamefont {E.}~\bibnamefont
  {Gull}}, \bibinfo {author} {\bibfnamefont {A.~J.}\ \bibnamefont {Millis}},
  \bibinfo {author} {\bibfnamefont {A.~I.}\ \bibnamefont {Lichtenstein}},
  \bibinfo {author} {\bibfnamefont {A.~N.}\ \bibnamefont {Rubtsov}}, \bibinfo
  {author} {\bibfnamefont {M.}~\bibnamefont {Troyer}}, \ and\ \bibinfo {author}
  {\bibfnamefont {P.}~\bibnamefont {Werner}},\ }\href {\doibase
  10.1103/RevModPhys.83.349} {\bibfield  {journal} {\bibinfo  {journal} {Rev.
  Mod. Phys.}\ }\textbf {\bibinfo {volume} {83}},\ \bibinfo {pages} {349}
  (\bibinfo {year} {2011})}\BibitemShut {NoStop}%
\bibitem [{\citenamefont {Parragh}\ \emph {et~al.}(2012)\citenamefont
  {Parragh}, \citenamefont {Toschi}, \citenamefont {Held},\ and\ \citenamefont
  {Sangiovanni}}]{PhysRevB.86.155158}%
  \BibitemOpen
  \bibfield  {author} {\bibinfo {author} {\bibfnamefont {N.}~\bibnamefont
  {Parragh}}, \bibinfo {author} {\bibfnamefont {A.}~\bibnamefont {Toschi}},
  \bibinfo {author} {\bibfnamefont {K.}~\bibnamefont {Held}}, \ and\ \bibinfo
  {author} {\bibfnamefont {G.}~\bibnamefont {Sangiovanni}},\ }\href {\doibase
  10.1103/PhysRevB.86.155158} {\bibfield  {journal} {\bibinfo  {journal} {Phys.
  Rev. B}\ }\textbf {\bibinfo {volume} {86}},\ \bibinfo {pages} {155158}
  (\bibinfo {year} {2012})}\BibitemShut {NoStop}%
\bibitem [{\citenamefont {Wallerberger}\ \emph {et~al.}(2019)\citenamefont
  {Wallerberger}, \citenamefont {Hausoel}, \citenamefont {Gunacker},
  \citenamefont {Kowalski}, \citenamefont {Parragh}, \citenamefont {Goth},
  \citenamefont {Held},\ and\ \citenamefont
  {Sangiovanni}}]{wallerberger2019w2dynamics}%
  \BibitemOpen
  \bibfield  {author} {\bibinfo {author} {\bibfnamefont {M.}~\bibnamefont
  {Wallerberger}}, \bibinfo {author} {\bibfnamefont {A.}~\bibnamefont
  {Hausoel}}, \bibinfo {author} {\bibfnamefont {P.}~\bibnamefont {Gunacker}},
  \bibinfo {author} {\bibfnamefont {A.}~\bibnamefont {Kowalski}}, \bibinfo
  {author} {\bibfnamefont {N.}~\bibnamefont {Parragh}}, \bibinfo {author}
  {\bibfnamefont {F.}~\bibnamefont {Goth}}, \bibinfo {author} {\bibfnamefont
  {K.}~\bibnamefont {Held}}, \ and\ \bibinfo {author} {\bibfnamefont
  {G.}~\bibnamefont {Sangiovanni}},\ }\href@noop {} {\bibfield  {journal}
  {\bibinfo  {journal} {Computer Physics Communications}\ }\textbf {\bibinfo
  {volume} {235}},\ \bibinfo {pages} {388} (\bibinfo {year}
  {2019})}\BibitemShut {NoStop}%
\bibitem [{\citenamefont {Gubernatis}\ \emph {et~al.}(1991)\citenamefont
  {Gubernatis}, \citenamefont {Jarrell}, \citenamefont {Silver},\ and\
  \citenamefont {Sivia}}]{PhysRevB.44.6011}%
  \BibitemOpen
  \bibfield  {author} {\bibinfo {author} {\bibfnamefont {J.~E.}\ \bibnamefont
  {Gubernatis}}, \bibinfo {author} {\bibfnamefont {M.}~\bibnamefont {Jarrell}},
  \bibinfo {author} {\bibfnamefont {R.~N.}\ \bibnamefont {Silver}}, \ and\
  \bibinfo {author} {\bibfnamefont {D.~S.}\ \bibnamefont {Sivia}},\ }\href
  {\doibase 10.1103/PhysRevB.44.6011} {\bibfield  {journal} {\bibinfo
  {journal} {Phys. Rev. B}\ }\textbf {\bibinfo {volume} {44}},\ \bibinfo
  {pages} {6011} (\bibinfo {year} {1991})}\BibitemShut {NoStop}%
\bibitem [{\citenamefont {Sandvik}(1998{\natexlab{a}})}]{PhysRevB.57.10287}%
  \BibitemOpen
  \bibfield  {author} {\bibinfo {author} {\bibfnamefont {A.~W.}\ \bibnamefont
  {Sandvik}},\ }\href {\doibase 10.1103/PhysRevB.57.10287} {\bibfield
  {journal} {\bibinfo  {journal} {Phys. Rev. B}\ }\textbf {\bibinfo {volume}
  {57}},\ \bibinfo {pages} {10287} (\bibinfo {year}
  {1998}{\natexlab{a}})}\BibitemShut {NoStop}%
\bibitem [{\citenamefont {Ambrosch-Draxl}\ and\ \citenamefont
  {Sofo}(2006)}]{AmbroschDraxl20061}%
  \BibitemOpen
  \bibfield  {author} {\bibinfo {author} {\bibfnamefont {C.}~\bibnamefont
  {Ambrosch-Draxl}}\ and\ \bibinfo {author} {\bibfnamefont {J.~O.}\
  \bibnamefont {Sofo}},\ }\href {\doibase DOI: 10.1016/j.cpc.2006.03.005}
  {\bibfield  {journal} {\bibinfo  {journal} {Computer Physics Communications}\
  }\textbf {\bibinfo {volume} {175}},\ \bibinfo {pages} {1 } (\bibinfo {year}
  {2006})}\BibitemShut {NoStop}%
\bibitem [{\citenamefont {Assmann}\ \emph {et~al.}(2016)\citenamefont
  {Assmann}, \citenamefont {Wissgott}, \citenamefont {Kune{\v{s}}},
  \citenamefont {Toschi}, \citenamefont {Blaha},\ and\ \citenamefont
  {Held}}]{assmann2016woptic}%
  \BibitemOpen
  \bibfield  {author} {\bibinfo {author} {\bibfnamefont {E.}~\bibnamefont
  {Assmann}}, \bibinfo {author} {\bibfnamefont {P.}~\bibnamefont {Wissgott}},
  \bibinfo {author} {\bibfnamefont {J.}~\bibnamefont {Kune{\v{s}}}}, \bibinfo
  {author} {\bibfnamefont {A.}~\bibnamefont {Toschi}}, \bibinfo {author}
  {\bibfnamefont {P.}~\bibnamefont {Blaha}}, \ and\ \bibinfo {author}
  {\bibfnamefont {K.}~\bibnamefont {Held}},\ }\href@noop {} {\bibfield
  {journal} {\bibinfo  {journal} {Computer physics communications}\ }\textbf
  {\bibinfo {volume} {202}},\ \bibinfo {pages} {1} (\bibinfo {year}
  {2016})}\BibitemShut {NoStop}%
\bibitem [{\citenamefont {Ashcroft}\ and\ \citenamefont
  {Mermin}(1976)}]{ashcroft1976solid}%
  \BibitemOpen
  \bibfield  {author} {\bibinfo {author} {\bibfnamefont {N.~W.}\ \bibnamefont
  {Ashcroft}}\ and\ \bibinfo {author} {\bibfnamefont {N.~D.}\ \bibnamefont
  {Mermin}},\ }\href@noop {} {\emph {\bibinfo {title} {Solid state physics}}}\
  (\bibinfo  {publisher} {New York: Holt, Rinehart and Winston,},\ \bibinfo
  {year} {1976})\BibitemShut {NoStop}%
\bibitem [{\citenamefont {Haacke}(1976)}]{Haake1976}%
  \BibitemOpen
  \bibfield  {author} {\bibinfo {author} {\bibfnamefont {G.}~\bibnamefont
  {Haacke}},\ }\href {\doibase 10.1063/1.323240} {\bibfield  {journal}
  {\bibinfo  {journal} {Journal of Applied Physics}\ }\textbf {\bibinfo
  {volume} {47}},\ \bibinfo {pages} {4086} (\bibinfo {year} {1976})},\ \Eprint
  {http://arxiv.org/abs/https://doi.org/10.1063/1.323240}
  {https://doi.org/10.1063/1.323240} \BibitemShut {NoStop}%
\bibitem [{\citenamefont {Pavarini}\ \emph {et~al.}(2004)\citenamefont
  {Pavarini}, \citenamefont {Biermann}, \citenamefont {Poteryaev},
  \citenamefont {Lichtenstein}, \citenamefont {Georges},\ and\ \citenamefont
  {Andersen}}]{PhysRevLett.92.176403}%
  \BibitemOpen
  \bibfield  {author} {\bibinfo {author} {\bibfnamefont {E.}~\bibnamefont
  {Pavarini}}, \bibinfo {author} {\bibfnamefont {S.}~\bibnamefont {Biermann}},
  \bibinfo {author} {\bibfnamefont {A.}~\bibnamefont {Poteryaev}}, \bibinfo
  {author} {\bibfnamefont {A.~I.}\ \bibnamefont {Lichtenstein}}, \bibinfo
  {author} {\bibfnamefont {A.}~\bibnamefont {Georges}}, \ and\ \bibinfo
  {author} {\bibfnamefont {O.~K.}\ \bibnamefont {Andersen}},\ }\href {\doibase
  10.1103/PhysRevLett.92.176403} {\bibfield  {journal} {\bibinfo  {journal}
  {Phys. Rev. Lett.}\ }\textbf {\bibinfo {volume} {92}},\ \bibinfo {pages}
  {176403} (\bibinfo {year} {2004})}\BibitemShut {NoStop}%
\bibitem [{\citenamefont {Huang}\ and\ \citenamefont
  {Wang}(2012)}]{huang2012dynamical}%
  \BibitemOpen
  \bibfield  {author} {\bibinfo {author} {\bibfnamefont {L.}~\bibnamefont
  {Huang}}\ and\ \bibinfo {author} {\bibfnamefont {Y.}~\bibnamefont {Wang}},\
  }\href@noop {} {\bibfield  {journal} {\bibinfo  {journal} {EPL (Europhysics
  Letters)}\ }\textbf {\bibinfo {volume} {99}},\ \bibinfo {pages} {67003}
  (\bibinfo {year} {2012})}\BibitemShut {NoStop}%
\bibitem [{Note3()}]{Note3}%
  \BibitemOpen
  \bibinfo {note} {Here, the experimental absorption coefficient is calculated
  from the measured dielectric functions of Ref.\protect \tmspace +\thinmuskip
  {.1667em}\cite {zhang2016correlated} via $A$=2$\omega k$/c.}\BibitemShut
  {Stop}%
\bibitem [{\citenamefont {Filmetrics}(2015)}]{polyITO}%
  \BibitemOpen
  \bibfield  {author} {\bibinfo {author} {\bibnamefont {Filmetrics}},\ }\href
  {https://www.filmetrics.com/refractive-index-database/ITO/Indium-Tin-Oxide-InSnO}
  {\bibfield  {journal} {\bibinfo  {journal} {Refractive Index of ITO, Indium
  Tin Oxide, InSnO}\ } (\bibinfo {year} {2015})}\BibitemShut {NoStop}%
\bibitem [{\citenamefont {Ellmer}(2012)}]{ellmer2012past}%
  \BibitemOpen
  \bibfield  {author} {\bibinfo {author} {\bibfnamefont {K.}~\bibnamefont
  {Ellmer}},\ }\href@noop {} {\bibfield  {journal} {\bibinfo  {journal} {Nature
  Photonics}\ }\textbf {\bibinfo {volume} {6}},\ \bibinfo {pages} {809}
  (\bibinfo {year} {2012})}\BibitemShut {NoStop}%
\bibitem [{\citenamefont {Ohta}\ \emph {et~al.}(2002)\citenamefont {Ohta},
  \citenamefont {Orita}, \citenamefont {Hirano},\ and\ \citenamefont
  {Hosono}}]{ohta2002surface}%
  \BibitemOpen
  \bibfield  {author} {\bibinfo {author} {\bibfnamefont {H.}~\bibnamefont
  {Ohta}}, \bibinfo {author} {\bibfnamefont {M.}~\bibnamefont {Orita}},
  \bibinfo {author} {\bibfnamefont {M.}~\bibnamefont {Hirano}}, \ and\ \bibinfo
  {author} {\bibfnamefont {H.}~\bibnamefont {Hosono}},\ }\href@noop {}
  {\bibfield  {journal} {\bibinfo  {journal} {Journal of applied physics}\
  }\textbf {\bibinfo {volume} {91}},\ \bibinfo {pages} {3547} (\bibinfo {year}
  {2002})}\BibitemShut {NoStop}%
\bibitem [{\citenamefont {Moyer}\ \emph {et~al.}(2013)\citenamefont {Moyer},
  \citenamefont {Eaton},\ and\ \citenamefont
  {Engel-Herbert}}]{moyer2013highly}%
  \BibitemOpen
  \bibfield  {author} {\bibinfo {author} {\bibfnamefont {J.~A.}\ \bibnamefont
  {Moyer}}, \bibinfo {author} {\bibfnamefont {C.}~\bibnamefont {Eaton}}, \ and\
  \bibinfo {author} {\bibfnamefont {R.}~\bibnamefont {Engel-Herbert}},\
  }\href@noop {} {\bibfield  {journal} {\bibinfo  {journal} {Advanced
  Materials}\ }\textbf {\bibinfo {volume} {25}},\ \bibinfo {pages} {3578}
  (\bibinfo {year} {2013})}\BibitemShut {NoStop}%
\bibitem [{\citenamefont {Mizoguchi}\ \emph {et~al.}(2013)\citenamefont
  {Mizoguchi}, \citenamefont {Chen}, \citenamefont {Boolchand}, \citenamefont
  {Ksenofontov}, \citenamefont {Felser}, \citenamefont {Barnes},\ and\
  \citenamefont {Woodward}}]{mizoguchi2013electrical}%
  \BibitemOpen
  \bibfield  {author} {\bibinfo {author} {\bibfnamefont {H.}~\bibnamefont
  {Mizoguchi}}, \bibinfo {author} {\bibfnamefont {P.}~\bibnamefont {Chen}},
  \bibinfo {author} {\bibfnamefont {P.}~\bibnamefont {Boolchand}}, \bibinfo
  {author} {\bibfnamefont {V.}~\bibnamefont {Ksenofontov}}, \bibinfo {author}
  {\bibfnamefont {C.}~\bibnamefont {Felser}}, \bibinfo {author} {\bibfnamefont
  {P.~W.}\ \bibnamefont {Barnes}}, \ and\ \bibinfo {author} {\bibfnamefont
  {P.~M.}\ \bibnamefont {Woodward}},\ }\href@noop {} {\bibfield  {journal}
  {\bibinfo  {journal} {Chemistry of Materials}\ }\textbf {\bibinfo {volume}
  {25}},\ \bibinfo {pages} {3858} (\bibinfo {year} {2013})}\BibitemShut
  {NoStop}%
\bibitem [{Note4()}]{Note4}%
  \BibitemOpen
  \bibinfo {note} {The shift of the SrVO$_3$ FOM curve in DFT+DMFT compared to
  experiment is mainly because the absorption coefficient becomes larger for
  thin films while it is constant in our bulk calculations. Indeed ultrathin
  SrVO$_3$ films behave very different from bulk \cite
  {james2020quantum,PhysRevLett.114.246401,PhysRevLett.104.147601}}\BibitemShut
  {NoStop}%
\bibitem [{\citenamefont {Holmstr\"om}\ \emph {et~al.}(2011)\citenamefont
  {Holmstr\"om}, \citenamefont {Fransson}, \citenamefont {Eriksson},
  \citenamefont {Liz\'arraga}, \citenamefont {Sanyal}, \citenamefont
  {Bhandary},\ and\ \citenamefont {Katsnelson}}]{PhysRevB.84.205414}%
  \BibitemOpen
  \bibfield  {author} {\bibinfo {author} {\bibfnamefont {E.}~\bibnamefont
  {Holmstr\"om}}, \bibinfo {author} {\bibfnamefont {J.}~\bibnamefont
  {Fransson}}, \bibinfo {author} {\bibfnamefont {O.}~\bibnamefont {Eriksson}},
  \bibinfo {author} {\bibfnamefont {R.}~\bibnamefont {Liz\'arraga}}, \bibinfo
  {author} {\bibfnamefont {B.}~\bibnamefont {Sanyal}}, \bibinfo {author}
  {\bibfnamefont {S.}~\bibnamefont {Bhandary}}, \ and\ \bibinfo {author}
  {\bibfnamefont {M.~I.}\ \bibnamefont {Katsnelson}},\ }\href {\doibase
  10.1103/PhysRevB.84.205414} {\bibfield  {journal} {\bibinfo  {journal} {Phys.
  Rev. B}\ }\textbf {\bibinfo {volume} {84}},\ \bibinfo {pages} {205414}
  (\bibinfo {year} {2011})}\BibitemShut {NoStop}%
\bibitem [{\citenamefont {James}\ \emph {et~al.}(2020)\citenamefont {James},
  \citenamefont {Aichhorn},\ and\ \citenamefont {Laverock}}]{james2020quantum}%
  \BibitemOpen
  \bibfield  {author} {\bibinfo {author} {\bibfnamefont {A.}~\bibnamefont
  {James}}, \bibinfo {author} {\bibfnamefont {M.}~\bibnamefont {Aichhorn}}, \
  and\ \bibinfo {author} {\bibfnamefont {J.}~\bibnamefont {Laverock}},\
  }\href@noop {} {\bibfield  {journal} {\bibinfo  {journal} {arXiv preprint
  arXiv:2005.14329}\ } (\bibinfo {year} {2020})}\BibitemShut {NoStop}%
\bibitem [{\citenamefont {Zhong}\ \emph {et~al.}(2015)\citenamefont {Zhong},
  \citenamefont {Wallerberger}, \citenamefont {Tomczak}, \citenamefont
  {Taranto}, \citenamefont {Parragh}, \citenamefont {Toschi}, \citenamefont
  {Sangiovanni},\ and\ \citenamefont {Held}}]{PhysRevLett.114.246401}%
  \BibitemOpen
  \bibfield  {author} {\bibinfo {author} {\bibfnamefont {Z.}~\bibnamefont
  {Zhong}}, \bibinfo {author} {\bibfnamefont {M.}~\bibnamefont {Wallerberger}},
  \bibinfo {author} {\bibfnamefont {J.~M.}\ \bibnamefont {Tomczak}}, \bibinfo
  {author} {\bibfnamefont {C.}~\bibnamefont {Taranto}}, \bibinfo {author}
  {\bibfnamefont {N.}~\bibnamefont {Parragh}}, \bibinfo {author} {\bibfnamefont
  {A.}~\bibnamefont {Toschi}}, \bibinfo {author} {\bibfnamefont
  {G.}~\bibnamefont {Sangiovanni}}, \ and\ \bibinfo {author} {\bibfnamefont
  {K.}~\bibnamefont {Held}},\ }\href {\doibase 10.1103/PhysRevLett.114.246401}
  {\bibfield  {journal} {\bibinfo  {journal} {Phys. Rev. Lett.}\ }\textbf
  {\bibinfo {volume} {114}},\ \bibinfo {pages} {246401} (\bibinfo {year}
  {2015})}\BibitemShut {NoStop}%
\bibitem [{\citenamefont {Yoshimatsu}\ \emph {et~al.}(2010)\citenamefont
  {Yoshimatsu}, \citenamefont {Okabe}, \citenamefont {Kumigashira},
  \citenamefont {Okamoto}, \citenamefont {Aizaki}, \citenamefont {Fujimori},\
  and\ \citenamefont {Oshima}}]{PhysRevLett.104.147601}%
  \BibitemOpen
  \bibfield  {author} {\bibinfo {author} {\bibfnamefont {K.}~\bibnamefont
  {Yoshimatsu}}, \bibinfo {author} {\bibfnamefont {T.}~\bibnamefont {Okabe}},
  \bibinfo {author} {\bibfnamefont {H.}~\bibnamefont {Kumigashira}}, \bibinfo
  {author} {\bibfnamefont {S.}~\bibnamefont {Okamoto}}, \bibinfo {author}
  {\bibfnamefont {S.}~\bibnamefont {Aizaki}}, \bibinfo {author} {\bibfnamefont
  {A.}~\bibnamefont {Fujimori}}, \ and\ \bibinfo {author} {\bibfnamefont
  {M.}~\bibnamefont {Oshima}},\ }\href {\doibase
  10.1103/PhysRevLett.104.147601} {\bibfield  {journal} {\bibinfo  {journal}
  {Phys. Rev. Lett.}\ }\textbf {\bibinfo {volume} {104}},\ \bibinfo {pages}
  {147601} (\bibinfo {year} {2010})}\BibitemShut {NoStop}%
\bibitem [{\citenamefont {Wang}\ \emph {et~al.}(2011)\citenamefont {Wang},
  \citenamefont {Dang},\ and\ \citenamefont {Millis}}]{Wang11}%
  \BibitemOpen
  \bibfield  {author} {\bibinfo {author} {\bibfnamefont {X.}~\bibnamefont
  {Wang}}, \bibinfo {author} {\bibfnamefont {H.~T.}\ \bibnamefont {Dang}}, \
  and\ \bibinfo {author} {\bibfnamefont {A.~J.}\ \bibnamefont {Millis}},\
  }\href {\doibase 10.1103/PhysRevB.84.073104} {\bibfield  {journal} {\bibinfo
  {journal} {Phys. Rev. B}\ }\textbf {\bibinfo {volume} {84}},\ \bibinfo
  {pages} {073104} (\bibinfo {year} {2011})}\BibitemShut {NoStop}%
\bibitem [{\citenamefont {Pruschke}\ \emph {et~al.}(1993)\citenamefont
  {Pruschke}, \citenamefont {Cox},\ and\ \citenamefont {Jarrell}}]{Pruschke93}%
  \BibitemOpen
  \bibfield  {author} {\bibinfo {author} {\bibfnamefont {T.}~\bibnamefont
  {Pruschke}}, \bibinfo {author} {\bibfnamefont {D.~L.}\ \bibnamefont {Cox}}, \
  and\ \bibinfo {author} {\bibfnamefont {M.}~\bibnamefont {Jarrell}},\ }\href
  {\doibase 10.1103/PhysRevB.47.3553} {\bibfield  {journal} {\bibinfo
  {journal} {Phys. Rev. B}\ }\textbf {\bibinfo {volume} {47}},\ \bibinfo
  {pages} {3553} (\bibinfo {year} {1993})}\BibitemShut {NoStop}%
\bibitem [{\citenamefont {Bulla}(1999)}]{Bulla1999}%
  \BibitemOpen
  \bibfield  {author} {\bibinfo {author} {\bibfnamefont {R.}~\bibnamefont
  {Bulla}},\ }\href {\doibase 10.1103/PhysRevLett.83.136} {\bibfield  {journal}
  {\bibinfo  {journal} {Phys. Rev. Lett.}\ }\textbf {\bibinfo {volume} {83}},\
  \bibinfo {pages} {136} (\bibinfo {year} {1999})}\BibitemShut {NoStop}%
\bibitem [{\citenamefont {Nekrasov}\ \emph
  {et~al.}(2006{\natexlab{b}})\citenamefont {Nekrasov}, \citenamefont {Held},
  \citenamefont {Keller}, \citenamefont {Kondakov}, \citenamefont {Pruschke},
  \citenamefont {Kollar}, \citenamefont {Andersen}, \citenamefont {Anisimov},\
  and\ \citenamefont {Vollhardt}}]{Nekrasov05a}%
  \BibitemOpen
  \bibfield  {author} {\bibinfo {author} {\bibfnamefont {I.~A.}\ \bibnamefont
  {Nekrasov}}, \bibinfo {author} {\bibfnamefont {K.}~\bibnamefont {Held}},
  \bibinfo {author} {\bibfnamefont {G.}~\bibnamefont {Keller}}, \bibinfo
  {author} {\bibfnamefont {D.~E.}\ \bibnamefont {Kondakov}}, \bibinfo {author}
  {\bibfnamefont {T.}~\bibnamefont {Pruschke}}, \bibinfo {author}
  {\bibfnamefont {M.}~\bibnamefont {Kollar}}, \bibinfo {author} {\bibfnamefont
  {O.~K.}\ \bibnamefont {Andersen}}, \bibinfo {author} {\bibfnamefont {V.~I.}\
  \bibnamefont {Anisimov}}, \ and\ \bibinfo {author} {\bibfnamefont
  {D.}~\bibnamefont {Vollhardt}},\ }\href {\doibase 10.1103/PhysRevB.73.155112}
  {\bibfield  {journal} {\bibinfo  {journal} {Phys. Rev. B}\ }\textbf {\bibinfo
  {volume} {73}},\ \bibinfo {pages} {155112} (\bibinfo {year}
  {2006}{\natexlab{b}})}\BibitemShut {NoStop}%
\bibitem [{\citenamefont {Byczuk}\ \emph {et~al.}(2007)\citenamefont {Byczuk},
  \citenamefont {Kollar}, \citenamefont {Held}, \citenamefont {Yang},
  \citenamefont {Nekrasov}, \citenamefont {Pruschke},\ and\ \citenamefont
  {Vollhardt}}]{Byczuk2007}%
  \BibitemOpen
  \bibfield  {author} {\bibinfo {author} {\bibfnamefont {K.}~\bibnamefont
  {Byczuk}}, \bibinfo {author} {\bibfnamefont {M.}~\bibnamefont {Kollar}},
  \bibinfo {author} {\bibfnamefont {K.}~\bibnamefont {Held}}, \bibinfo {author}
  {\bibfnamefont {Y.-F.}\ \bibnamefont {Yang}}, \bibinfo {author}
  {\bibfnamefont {I.~A.}\ \bibnamefont {Nekrasov}}, \bibinfo {author}
  {\bibfnamefont {T.}~\bibnamefont {Pruschke}}, \ and\ \bibinfo {author}
  {\bibfnamefont {D.}~\bibnamefont {Vollhardt}},\ }\href {\doibase
  10.1038/nphys538} {\bibfield  {journal} {\bibinfo  {journal} {Nature
  Physics}\ ,\ \bibinfo {pages} {168}} (\bibinfo {year} {2007})}\BibitemShut
  {NoStop}%
\bibitem [{\citenamefont {Held}\ \emph {et~al.}(2013)\citenamefont {Held},
  \citenamefont {Peters},\ and\ \citenamefont {Toschi}}]{Held13}%
  \BibitemOpen
  \bibfield  {author} {\bibinfo {author} {\bibfnamefont {K.}~\bibnamefont
  {Held}}, \bibinfo {author} {\bibfnamefont {R.}~\bibnamefont {Peters}}, \ and\
  \bibinfo {author} {\bibfnamefont {A.}~\bibnamefont {Toschi}},\ }\href
  {\doibase 10.1103/PhysRevLett.110.246402} {\bibfield  {journal} {\bibinfo
  {journal} {Phys. Rev. Lett.}\ }\textbf {\bibinfo {volume} {110}},\ \bibinfo
  {pages} {246402} (\bibinfo {year} {2013})}\BibitemShut {NoStop}%
\bibitem [{\citenamefont {Sudrajat}\ \emph {et~al.}(2019)\citenamefont
  {Sudrajat}, \citenamefont {Dhakal}, \citenamefont {Kitta}, \citenamefont
  {Sasaki}, \citenamefont {Ozawa}, \citenamefont {Babel}, \citenamefont
  {Yoshida}, \citenamefont {Ichikuni},\ and\ \citenamefont
  {Onishi}}]{sudrajat2019electron}%
  \BibitemOpen
  \bibfield  {author} {\bibinfo {author} {\bibfnamefont {H.}~\bibnamefont
  {Sudrajat}}, \bibinfo {author} {\bibfnamefont {D.}~\bibnamefont {Dhakal}},
  \bibinfo {author} {\bibfnamefont {M.}~\bibnamefont {Kitta}}, \bibinfo
  {author} {\bibfnamefont {T.}~\bibnamefont {Sasaki}}, \bibinfo {author}
  {\bibfnamefont {A.}~\bibnamefont {Ozawa}}, \bibinfo {author} {\bibfnamefont
  {S.}~\bibnamefont {Babel}}, \bibinfo {author} {\bibfnamefont
  {T.}~\bibnamefont {Yoshida}}, \bibinfo {author} {\bibfnamefont
  {N.}~\bibnamefont {Ichikuni}}, \ and\ \bibinfo {author} {\bibfnamefont
  {H.}~\bibnamefont {Onishi}},\ }\href@noop {} {\bibfield  {journal} {\bibinfo
  {journal} {The Journal of Physical Chemistry C}\ }\textbf {\bibinfo {volume}
  {123}},\ \bibinfo {pages} {18387} (\bibinfo {year} {2019})}\BibitemShut
  {NoStop}%
\bibitem [{\citenamefont {Zhu}\ \emph {et~al.}(2014)\citenamefont {Zhu},
  \citenamefont {Wang}, \citenamefont {Zhong}, \citenamefont {Liu},
  \citenamefont {Shen}, \citenamefont {Jiang},\ and\ \citenamefont
  {Zhou}}]{zhu2014ptcr}%
  \BibitemOpen
  \bibfield  {author} {\bibinfo {author} {\bibfnamefont {X.-W.}\ \bibnamefont
  {Zhu}}, \bibinfo {author} {\bibfnamefont {S.-W.}\ \bibnamefont {Wang}},
  \bibinfo {author} {\bibfnamefont {S.-W.}\ \bibnamefont {Zhong}}, \bibinfo
  {author} {\bibfnamefont {Y.-Y.}\ \bibnamefont {Liu}}, \bibinfo {author}
  {\bibfnamefont {S.-Y.}\ \bibnamefont {Shen}}, \bibinfo {author}
  {\bibfnamefont {W.-Z.}\ \bibnamefont {Jiang}}, \ and\ \bibinfo {author}
  {\bibfnamefont {X.}~\bibnamefont {Zhou}},\ }\href@noop {} {\bibfield
  {journal} {\bibinfo  {journal} {Ceramics International}\ }\textbf {\bibinfo
  {volume} {40}},\ \bibinfo {pages} {12383} (\bibinfo {year}
  {2014})}\BibitemShut {NoStop}%
\bibitem [{\citenamefont {Zhong}\ and\ \citenamefont
  {Hansmann}(2017)}]{PhysRevX.7.011023}%
  \BibitemOpen
  \bibfield  {author} {\bibinfo {author} {\bibfnamefont {Z.}~\bibnamefont
  {Zhong}}\ and\ \bibinfo {author} {\bibfnamefont {P.}~\bibnamefont
  {Hansmann}},\ }\href {\doibase 10.1103/PhysRevX.7.011023} {\bibfield
  {journal} {\bibinfo  {journal} {Phys. Rev. X}\ }\textbf {\bibinfo {volume}
  {7}},\ \bibinfo {pages} {011023} (\bibinfo {year} {2017})}\BibitemShut
  {NoStop}%
\bibitem [{\citenamefont {Luttinger}(1961)}]{Luttinger61}%
  \BibitemOpen
  \bibfield  {author} {\bibinfo {author} {\bibfnamefont {J.~M.}\ \bibnamefont
  {Luttinger}},\ }\href {\doibase 10.1103/PhysRev.121.942} {\bibfield
  {journal} {\bibinfo  {journal} {Phys. Rev.}\ }\textbf {\bibinfo {volume}
  {121}},\ \bibinfo {pages} {942} (\bibinfo {year} {1961})}\BibitemShut
  {NoStop}%
\bibitem [{\citenamefont {Vidberg}\ and\ \citenamefont
  {Serene}(1977)}]{Vidberg77}%
  \BibitemOpen
  \bibfield  {author} {\bibinfo {author} {\bibfnamefont {H.~J.}\ \bibnamefont
  {Vidberg}}\ and\ \bibinfo {author} {\bibfnamefont {J.~W.}\ \bibnamefont
  {Serene}},\ }\href {\doibase 10.1007/BF00655090} {\bibfield  {journal}
  {\bibinfo  {journal} {Journal of Low Temperature Physics}\ }\textbf {\bibinfo
  {volume} {29}},\ \bibinfo {pages} {179} (\bibinfo {year} {1977})}\BibitemShut
  {NoStop}%
\bibitem [{\citenamefont {Bryan}(1990)}]{Bryan90}%
  \BibitemOpen
  \bibfield  {author} {\bibinfo {author} {\bibfnamefont {R.~K.}\ \bibnamefont
  {Bryan}},\ }\href {\doibase 10.1007/BF02427376} {\bibfield  {journal}
  {\bibinfo  {journal} {European Biophysics Journal}\ }\textbf {\bibinfo
  {volume} {18}},\ \bibinfo {pages} {165} (\bibinfo {year} {1990})}\BibitemShut
  {NoStop}%
\bibitem [{\citenamefont {Jarrell}\ and\ \citenamefont
  {Gubernatis}(1996)}]{JarrellGubernatis96}%
  \BibitemOpen
  \bibfield  {author} {\bibinfo {author} {\bibfnamefont {M.}~\bibnamefont
  {Jarrell}}\ and\ \bibinfo {author} {\bibfnamefont {J.}~\bibnamefont
  {Gubernatis}},\ }\href {\doibase
  https://doi.org/10.1016/0370-1573(95)00074-7} {\bibfield  {journal} {\bibinfo
   {journal} {Physics Reports}\ }\textbf {\bibinfo {volume} {269}},\ \bibinfo
  {pages} {133 } (\bibinfo {year} {1996})}\BibitemShut {NoStop}%
\bibitem [{\citenamefont {Sandvik}(1998{\natexlab{b}})}]{Sandvik98}%
  \BibitemOpen
  \bibfield  {author} {\bibinfo {author} {\bibfnamefont {A.~W.}\ \bibnamefont
  {Sandvik}},\ }\href {\doibase 10.1103/PhysRevB.57.10287} {\bibfield
  {journal} {\bibinfo  {journal} {Phys. Rev. B}\ }\textbf {\bibinfo {volume}
  {57}},\ \bibinfo {pages} {10287} (\bibinfo {year}
  {1998}{\natexlab{b}})}\BibitemShut {NoStop}%
\bibitem [{\citenamefont {Mishchenko}\ \emph {et~al.}(2000)\citenamefont
  {Mishchenko}, \citenamefont {Prokof'ev}, \citenamefont {Sakamoto},\ and\
  \citenamefont {Svistunov}}]{StochReg1}%
  \BibitemOpen
  \bibfield  {author} {\bibinfo {author} {\bibfnamefont {A.~S.}\ \bibnamefont
  {Mishchenko}}, \bibinfo {author} {\bibfnamefont {N.~V.}\ \bibnamefont
  {Prokof'ev}}, \bibinfo {author} {\bibfnamefont {A.}~\bibnamefont {Sakamoto}},
  \ and\ \bibinfo {author} {\bibfnamefont {B.~V.}\ \bibnamefont {Svistunov}},\
  }\href {\doibase 10.1103/PhysRevB.62.6317} {\bibfield  {journal} {\bibinfo
  {journal} {Phys. Rev. B}\ }\textbf {\bibinfo {volume} {62}},\ \bibinfo
  {pages} {6317} (\bibinfo {year} {2000})}\BibitemShut {NoStop}%
\bibitem [{\citenamefont {Otsuki}\ \emph {et~al.}(2017)\citenamefont {Otsuki},
  \citenamefont {Ohzeki}, \citenamefont {Shinaoka},\ and\ \citenamefont
  {Yoshimi}}]{SpM}%
  \BibitemOpen
  \bibfield  {author} {\bibinfo {author} {\bibfnamefont {J.}~\bibnamefont
  {Otsuki}}, \bibinfo {author} {\bibfnamefont {M.}~\bibnamefont {Ohzeki}},
  \bibinfo {author} {\bibfnamefont {H.}~\bibnamefont {Shinaoka}}, \ and\
  \bibinfo {author} {\bibfnamefont {K.}~\bibnamefont {Yoshimi}},\ }\href
  {\doibase 10.1103/PhysRevE.95.061302} {\bibfield  {journal} {\bibinfo
  {journal} {Phys. Rev. E}\ }\textbf {\bibinfo {volume} {95}},\ \bibinfo
  {pages} {061302} (\bibinfo {year} {2017})}\BibitemShut {NoStop}%
\bibitem [{\citenamefont {Geffroy}\ \emph {et~al.}(2019)\citenamefont
  {Geffroy}, \citenamefont {Kaufmann}, \citenamefont {Hariki}, \citenamefont
  {Gunacker}, \citenamefont {Hausoel},\ and\ \citenamefont
  {Kune\ifmmode~\check{s}\else \v{s}\fi{}}}]{Geffroy2019}%
  \BibitemOpen
  \bibfield  {author} {\bibinfo {author} {\bibfnamefont {D.}~\bibnamefont
  {Geffroy}}, \bibinfo {author} {\bibfnamefont {J.}~\bibnamefont {Kaufmann}},
  \bibinfo {author} {\bibfnamefont {A.}~\bibnamefont {Hariki}}, \bibinfo
  {author} {\bibfnamefont {P.}~\bibnamefont {Gunacker}}, \bibinfo {author}
  {\bibfnamefont {A.}~\bibnamefont {Hausoel}}, \ and\ \bibinfo {author}
  {\bibfnamefont {J.}~\bibnamefont {Kune\ifmmode~\check{s}\else \v{s}\fi{}}},\
  }\href {\doibase 10.1103/PhysRevLett.122.127601} {\bibfield  {journal}
  {\bibinfo  {journal} {Phys. Rev. Lett.}\ }\textbf {\bibinfo {volume} {122}},\
  \bibinfo {pages} {127601} (\bibinfo {year} {2019})}\BibitemShut {NoStop}%
\bibitem [{\citenamefont {Kaufmann}(2020)}]{kaufmannGithub}%
  \BibitemOpen
  \bibfield  {author} {\bibinfo {author} {\bibfnamefont {J.}~\bibnamefont
  {Kaufmann}},\ }\href@noop {} {\enquote {\bibinfo {title} {ana\_cont: Package
  for analytic continuation of many-body green's functions},}\ }\bibinfo
  {howpublished} {\url{https://github.com/josefkaufmann/ana\_cont}} (\bibinfo
  {year} {2020})\BibitemShut {NoStop}%
\bibitem [{\citenamefont {Bergeron}\ and\ \citenamefont
  {Tremblay}(2016)}]{bergeron2016}%
  \BibitemOpen
  \bibfield  {author} {\bibinfo {author} {\bibfnamefont {D.}~\bibnamefont
  {Bergeron}}\ and\ \bibinfo {author} {\bibfnamefont {A.-M.~S.}\ \bibnamefont
  {Tremblay}},\ }\href {\doibase 10.1103/PhysRevE.94.023303} {\bibfield
  {journal} {\bibinfo  {journal} {Phys. Rev. E}\ }\textbf {\bibinfo {volume}
  {94}},\ \bibinfo {pages} {023303} (\bibinfo {year} {2016})}\BibitemShut
  {NoStop}%
\bibitem [{\citenamefont {Kraberger}\ \emph {et~al.}(2017)\citenamefont
  {Kraberger}, \citenamefont {Triebl}, \citenamefont {Zingl},\ and\
  \citenamefont {Aichhorn}}]{Kraberger17}%
  \BibitemOpen
  \bibfield  {author} {\bibinfo {author} {\bibfnamefont {G.~J.}\ \bibnamefont
  {Kraberger}}, \bibinfo {author} {\bibfnamefont {R.}~\bibnamefont {Triebl}},
  \bibinfo {author} {\bibfnamefont {M.}~\bibnamefont {Zingl}}, \ and\ \bibinfo
  {author} {\bibfnamefont {M.}~\bibnamefont {Aichhorn}},\ }\href {\doibase
  10.1103/PhysRevB.96.155128} {\bibfield  {journal} {\bibinfo  {journal} {Phys.
  Rev. B}\ }\textbf {\bibinfo {volume} {96}},\ \bibinfo {pages} {155128}
  (\bibinfo {year} {2017})}\BibitemShut {NoStop}%
\bibitem [{\citenamefont {Skilling}(1991)}]{Skilling91}%
  \BibitemOpen
  \bibfield  {author} {\bibinfo {author} {\bibfnamefont {J.}~\bibnamefont
  {Skilling}},\ }\href@noop {} {\emph {\bibinfo {title} {Maximum Entropy in
  Action}}},\ edited by\ \bibinfo {editor} {\bibfnamefont {B.}~\bibnamefont
  {Buck}}\ and\ \bibinfo {editor} {\bibfnamefont {V.~A.}\ \bibnamefont
  {Macaulay}}\ (\bibinfo  {publisher} {Clarendon Press},\ \bibinfo {address}
  {Oxford},\ \bibinfo {year} {1991})\ Chap.~\bibinfo {chapter} {2}, pp.\
  \bibinfo {pages} {19--40}\BibitemShut {NoStop}%
\bibitem [{\citenamefont {Kaufmann}\ \emph {et~al.}(2019)\citenamefont
  {Kaufmann}, \citenamefont {Gunacker}, \citenamefont {Kowalski}, \citenamefont
  {Sangiovanni},\ and\ \citenamefont {Held}}]{Kaufmann2019}%
  \BibitemOpen
  \bibfield  {author} {\bibinfo {author} {\bibfnamefont {J.}~\bibnamefont
  {Kaufmann}}, \bibinfo {author} {\bibfnamefont {P.}~\bibnamefont {Gunacker}},
  \bibinfo {author} {\bibfnamefont {A.}~\bibnamefont {Kowalski}}, \bibinfo
  {author} {\bibfnamefont {G.}~\bibnamefont {Sangiovanni}}, \ and\ \bibinfo
  {author} {\bibfnamefont {K.}~\bibnamefont {Held}},\ }\href {\doibase
  10.1103/PhysRevB.100.075119} {\bibfield  {journal} {\bibinfo  {journal}
  {Phys. Rev. B}\ }\textbf {\bibinfo {volume} {100}},\ \bibinfo {pages}
  {075119} (\bibinfo {year} {2019})}\BibitemShut {NoStop}%
\end{thebibliography}

\clearpage
\end{document}